\documentclass[aps,prd,groupedaddress,preprintnumbers,%
              showpacs,showkeys,floatfix]{revtex4}       % Documentstyle for
\usepackage[latin1]{inputenc}                    % Codepage latin1
\usepackage{graphicx}                            % Graphics package
\usepackage{epsfig}
\usepackage{epsf}
\usepackage{latexsym}                            % Load additional symbols
\usepackage{amsfonts}                            % Use AMS fonts,
\usepackage{amssymb}                             % symbols,
\usepackage{amsmath}                             % and definitions
\usepackage[mathscr]{eucal}                      % Euler font symbols
\usepackage{dcolumn}                             % Align table columns
\usepackage{multirow}
\usepackage{bm}                                  % Bold math
\usepackage{hyperref}                            % References as links
\usepackage[usenames]{color}

\graphicspath{{.}{Graphics/}}                    % Path for graphics
\newcommand{\be}{\begin{equation}}
\newcommand{\ee}{\end{equation}}
\newcommand{\beq}{\begin{eqnarray}}
\newcommand{\eeq}{\end{eqnarray}}
\newcommand{\bea}{\begin{eqnarray}}
\newcommand{\eea}{\end{eqnarray}}
\newcommand{\tendto}{\mathop{\longrightarrow}}

\usepackage{ifpdf}
\usepackage{graphicx}
\def\eq#1{Eq.~(\ref{#1})}

\def \3{\ss }

\newcommand{\tr}{\operatorname{Tr}}
\newcommand{\re}{\operatorname{Re}}

\newcommand{\beqn}{\begin{eqnarray}}
\newcommand{\eeqn}{\end{eqnarray}}

\def\cyp{a}
\def\cyi{b}
\def\sac{c}
\def\nic{d}
\def\gre{e}
\def\uam{f}

\begin{document}

\begin{titlepage}
  {\vspace{-0.5cm} \normalsize
  \hfill \parbox{60mm}{DESY 12-069\\
SFB/CPP-12-25%LPT-ORSAY 08-32 \\ 
                       %IRFU-08-29\\ 
                       % DESY 08-032 \\
                       %SFB/CPP-08-19\\
                       % ROM2F/2008/06\\
                       %    HU-EP-08/09\\
                       %     MS-TP-08-4.\\
                        %RM3-TH/, ROM2F/2007/\\
}}\\[10mm]
  \begin{center}
    \begin{LARGE}
      \textbf{Strange and charm baryon masses with two flavors of dynamical twisted mass fermions} \\
    \end{LARGE}
  \end{center}

 \vspace{.5cm}

\begin{figure*}[h!]
\begin{center}
\epsfxsize=2.5truecm
\epsfysize=3truecm
% \mbox{\epsfbox{ETMC.eps}}
    \end{center}
\end{figure*}

%  \begin{figure}[h]
%    \begin{center}
%      \includegraphics[draft=false]{ETMC_logo.ps}
%    \end{center}
%  \end{figure}

 \vspace{-0.8cm}
  \baselineskip 20pt plus 2pt minus 2pt
  \begin{center}
    \textbf{
      C.~Alexandrou$^{(\cyp, \cyi)}$,
      J.~Carbonell$^{(\sac)}$,
      D.~Christaras$^{(\cyp)}$, 
      V.~Drach$^{(\nic)}$,
      M. Gravina$^{(\cyp)}$,
      M. Papinutto$^{(\gre, \uam)}$
}
  \end{center}
  
  \begin{center}
    \begin{footnotesize}
      \noindent 
	
 	$^{(\cyp)}$ Department of Physics, University of Cyprus, P.O. Box 20537,
 	1678 Nicosia, Cyprus\\	
 	$^{(\cyi)}$ Computation-based Science and Technology Research Center, Cyprus Institute,20 Kavafi Str., Nicosia 2121, Cyprus 
	\vspace{0.2cm}
	
	$^{(\sac)}$CEA-Saclay, IRFU/Service de Physique Nucl\'eaire, 91191 Gif-sur-Yvette, France 
	\vspace{0.2cm}

      $^{(\nic)}$ NIC, DESY, Platanenallee 6, D-15738 Zeuthen, Germany\\
      \vspace{0.2cm}

	$^{(\gre)}$ Laboratoire de Physique Subatomique et Cosmologie,
               UJF/CNRS/IN2P3, 53 avenue des Martyrs, 38026 Grenoble, France\\
        $^{(\uam)}$ Dpto. de F\'isica Te\'orica and Instituto de F\'isica Te\'orica UAM/CSIC, 
               Universidad Aut\'onoma de Madrid,
               Cantoblanco, E-28049 Madrid, Spain
	 \vspace{0.2cm}

%            $^{(\rmii)}$ Dip. di Fisica, Universit{\`a} di Roma Tor Vergata and INFN,
 %     Sez. di Roma Tor Vergata, Via della Ricerca Scientifica, I-00133 Roma, Italy
 %     \vspace{0.2cm}

 %     $^{(\mns)}$ Universit\"at M\"unster, Institut f\"ur Theoretische Physik,
 %     Wilhelm-Klemm-Strasse 9, D-48149 M\"unster, Germany
 %     \vspace{0.2cm}

%      $^{(\ors)}$ Laboratoire de Physique Th\'eorique (B\^at.~210), Universit\'e
%      de Paris XI,CNRS-UMR8627,  Centre d'Orsay, 91405 Orsay-Cedex, France\\
%      \vspace{0.2cm}
      
%      $^{(\liv)}$ Theoretical Physics Division, Dept. of Mathematical Sciences,
%      University of Liverpool, Liverpool L69 7ZL, UK\\
%     \vspace{0.2cm}
%
% $^{(\ber)}$ Humboldt  Universit\"at  zu Berlin, Fachbereich Physik,
%             Inst. fur Elementarteilchenphysik, Newtonstr. 15, D-12489 Berlin,
%             Germany
%      \vspace{0.2cm}
%      
%      $^{(\zur)}$ Institute for Theoretical Physics, ETH Z{\"u}rich, CH-8093 Z{\"u}rich,
%      Switzerland\\
      
    \end{footnotesize}
  \end{center}

  \begin{abstract}

The masses of the low-lying strange and charm baryons are evaluated 
using two degenerate flavors of twisted
mass sea quarks for pion masses in the range of about  260~MeV to 450~MeV.
The strange and charm valence quark masses are tuned to reproduce 
the mass of the kaon and D-meson at the physical point.
%when the light pseudo scalar mass assumes its physical value. 
The tree-level Symanzik improved gauge action is employed.
We use  three  values of 
the lattice spacing, corresponding to $\beta=3.9$, $\beta=4.05$ and $\beta=4.2$ 
with $r_0/a=5.22(2)$, $r_0/a=6.61(3)$ and $r_0/a=8.31(5)$ respectively.
%spacings $a=0.0855(5)$ and $a=0.0667(3)$ determined from the pion decay constant.
We examine the dependence of the strange and charm baryons on the
lattice spacing and strange and charm quark masses.  The pion mass dependence is studied and physical results are obtained using heavy baryon chiral perturbation theory  to extrapolate to the physical point.

\begin{center}
\today
\end{center}
 \end{abstract}
\maketitle 
\end{titlepage}

%\tableofcontents

%\newpage
\section{Introduction}

Lattice QCD simulations with two light degenerate 
sea quarks $(N_f = 2$) as well as with a strange sea quark ($N_f = 2 + 1$) 
close to physical values of the pion mass
are  being carried out. 
Masses of low-lying hadrons are primary quantities that 
can be extracted using these simulations. Comparing the lattice
and experimental  values 
provides a check of lattice discretization effects. 
Such a comparison is necessary before one can use the lattice approach
to study hadron structure. 
The European Twisted Mass Collaboration (ETMC) 
has generated a number of $N_f = 2$
 ensembles at four values of the lattice spacing, ranging from 0.1~fm to about
0.05~fm, at several values of the light sea quark mass 
and for several physical volumes  with maximally twisted mass fermions.
 We will use
ensembles generated  at the three smallest lattice spacings
to evaluate the masses of strange and charm baryons. 
The strange and charm quarks 
are added  as valence quarks.

For heavy quarks the Compton wavelength of the associated heavy-light meson
is comparable to presently attainable lattice spacings, 
which means that cut-off effects maybe large. 
The charm quark mass is at the upper limit of the range of masses 
that can be directly  simulated at present. In order to obtain
values for the masses that can be compared to experiment,  it is 
 important to assess
 the size of lattice artifacts.
A first study of cut-off effects was carried out for light and strange baryons
in Refs.~\cite{Alexandrou:2008tn,Alexandrou:2009qu}. In this work
we extend the study  by including a finer lattice spacing
and calculate besides the mass of  strange baryons the masses of
 charm baryons. Having three lattice spacings the
continuum extrapolation can be better assessed.

 In this work we compare our results in the strange baryon sector 
with recent results obtained  with Clover-improved Wilson fermions
with different levels of
smearing.
 The PACS-CS~\cite{Aoki:2008sm} and  
BMW~\cite{Durr:2008zz} collaborations evaluated  the octet 
spectrum using two degenerate flavors of light quarks and
a strange quark with mass tuned to its physical value.
The PACS-CS has also computed the decuplet baryon masses.
In addition, we compare with the LHPC 
 that computed the octet and decuplet spectrum using a hybrid
action with domain wall valence 
fermions 
on Kogut-Susskind sea quarks~\cite{WalkerLoud:2008bp}
%Preliminary results on
%the nucleon mass are also  computed using $N_f=2+1$ domain wall fermions by the% RBC-UKQCD
%collaboration~\cite{Antonio:2006px,Antonio:2006zz}.
   
%The last two works
%were performed with quark masses very close to the physical point. 

Besides the strange baryons,  we also 
study the ground state spectrum of  charm baryons with spin $J=1/2^+$ 
  and spin $J=3/2^+$.  
Experimental searches  of charm  hadrons have
 received significant attention, mainly due to the experimental observation
for candidates of the doubly charm baryons $\Xi_{cc}^+ (3520)$ and $\Xi^{++}_{cc}(3460)$ by the  SELEX collaboration~\cite{Mattson:2002vu,Russ:2002bw,Ocherashvili:2004hi}. 
The 60~MeV mass difference between the singly and doubly charged  states 
is difficult to understand since it is an order of magnitude larger compared to what is expected.
No evidence was found for these states by the BABAR experiment~\cite{Aubert:2006qw} and FOCUS Collaboration~\cite{Ratti:2003ez}.
The BELLE Collaboration~\cite{Chistov:2006zj} finds $\Xi$-states lower in mass, that can be candidates of excited states 
of $\Xi_c$ but no doubly charm $\Xi$.
Additional experiments are planned at  the new Beijing Spectrometer (BES-III) and at the antiProton ANnihilation at DArmstadt (PANDA) experiment at GSI, that 
can shed light on these charm baryon states.
Several lattice QCD studies have been carried out to study charm baryons. 
We will compare the results of the current work with recent lattice QCD
results all computed in a hybrid action approach where the charm valence quark was introduced on gauge configurations produced with staggered sea fermions  by the MILC collaboration~\cite{Bernard:2001av,Aubin:2004wf,Bazavov:2010ru}.

As in the case of the other lattice
QCD studies of heavy baryons, also in this work we use a mixed action approach.
 For the strange and charm sector we use an Osterwalder-Seiler 
valence  quark, following the approach employed in the 
study of the pseudo scalar meson decay constants~\cite{Blossier:2007vv,Blossier:2009bx}. 
 The bare strange and charm valence quark mass is tuned by requiring
that 
 the physical values of the mass of the  kaon and D-meson  are reproduced
after the lattice results are extrapolated at the physical value of the pion mass.
The ETMC $N_f=2$ configurations~\cite{Boucaud:2007uk,Boucaud:2008xu}
analyzed in this work  correspond to  pion masses in the range 
of 260 to 450 MeV
and  three values of the 
lattice spacing  corresponding to $\beta=3.9,\,4.05$ and $4.2$ 
with $r_0/a=5.22(2), 6.61(3)$ and $8.31(5)$, respectively. 
The Sommer parameter $r_0$ is determined from the force between two static quarks, the continuum
value of which is determined to be 0.462(5)~fm.
At $\beta=4.2$ we use two ensembles, one corresponding to the lowest
value of the  pion mass considered in this work and one to the upper pion mass range.
We find that the baryon masses, in general,  show a very weak
dependence on the lattice spacing and are fully compatible with an ${\cal O}(a^2)$ behaviour with an almost vanishing coefficient of the $a^2$ term.
This justifies neglecting the ${\cal O}(a^2)$
term in extrapolating results to the continuum limit.

An important issue raised by the twisted mass fermion formulation
 is isospin symmetry breaking. 
This symmetry, although exact in the continuum limit,
is broken at a non-vanishing lattice spacing to $\mathcal{O}(a^2)$. 
There are, however, theoretical arguments \cite{Frezzotti:2007qv} and numerical evidences
\cite{Dimopoulos:2008sy,Jansen:2008vs} that these isospin breaking effects are only
sizable for the neutral pseudo scalar mass whereas for other quantities studied 
so far by ETMC they are compatible with zero.   
In this paper we  demonstrate that also in the baryon sector these isospin 
breaking
effects are in general small or even compatible with zero. 
Small isospin breaking effects
decrease as the lattice spacing decreases and they vanish at the continuum limit.
This corroborates our previous findings~\cite{Alexandrou:2008tn, Alexandrou:2009qu}.
The isospin breaking effects are relevant not only for neutral pions but also
for other particles, e.g. the kaons. However, since the mass of the kaon is
higher, the relative splitting (between $K^0$ and $K^+$) is less drastic.

The paper is organized as follows:
The details of our lattice formulation, namely those concerning 
the twisted mass action, 
the parameters of the simulations, the interpolating fields used and the tuning
of the strange and charm quark masses 
are given in Section~II. 
Section~III contains the numerical results of the baryon masses computed
 for different 
lattice volumes, lattice spacings and bare quark masses.
%
%as well as the Gell-Mann Okubo relations that are supposed to be 
%fulfilled in the exact SU(3) limit.
Lattice artifacts, including finite volume and discretization errors,
  and continuum 
extrapolation are also discussed in Section~III, with special 
emphasis on the $\mathcal{O}(a^2)$ isospin breaking effects inherent
 to the twisted mass formulation
of lattice QCD.
The chiral extrapolations are analyzed in Section~IV.
Section~V  contains a comparison with other existing calculations. Our
conclusions are finally  drawn in Section~VI.

\section{Lattice formulation}
\subsection{The lattice action}
%\subsection{The light quark sector}

For the gauge fields  we use the  tree-level Symanzik improved
gauge action~\cite{Weisz:1982zw}, which includes besides the
plaquette term $U^{1\times1}_{x,\mu,\nu}$ also rectangular $(1\times2)$ Wilson 
loops $U^{1\times2}_{x,\mu,\nu}$
\begin{equation}
  \label{eq:Sg}
    S_g =  \frac{\beta}{3}\sum_x\Biggl(  b_0\sum_{\substack{
      \mu,\nu=1\\1\leq\mu<\nu}}^4\left \{1-\re\tr(U^{1\times1}_{x,\mu,\nu})\right \}\Bigr. 
     \Bigl.+
    b_1\sum_{\substack{\mu,\nu=1\\\mu\neq\nu}}^4\left \{1
    -\re\tr(U^{1\times2}_{x,\mu,\nu})\right \}\Biggr)\,  
\end{equation}
with  $b_1=-1/12$ and the
(proper) normalization condition $b_0=1-8b_1$. Note that at $b_1=0$ this
action becomes the usual Wilson plaquette gauge action.

The fermionic action for two degenerate flavors of quarks
 in twisted mass QCD is given by
\be
S_F= a^4\sum_x  \bar{\chi}(x)\bigl(D_W[U] + m_0 
+ i \mu \gamma_5\tau^3  \bigr ) \chi(x)
\label{S_tm}
\ee
with   $\tau^3$ the Pauli matrix acting in
the isospin space, $\mu$ the bare twisted mass 
and the massless Wilson-Dirac operator given by 
\be
D_W[U] = \frac{1}{2} \gamma_{\mu}(\nabla_{\mu} + \nabla_{\mu}^{*})
-\frac{ar}{2} \nabla_{\mu}
\nabla^*_{\mu} 
\ee
where
\be
\nabla_\mu \psi(x)= \frac{1}{a}\biggl[U^\dagger_\mu(x)\psi(x+a\hat{\mu})-\psi(x)\biggr]
\hspace*{0.5cm} {\rm and}\hspace*{0.5cm} 
\nabla^*_{\mu}\psi(x)=-\frac{1}{a}\biggl[U_{\mu}(x-a\hat{\mu})\psi(x-a\hat{\mu})-\psi(x)\biggr]
\quad .
\ee
Maximally twisted Wilson quarks are obtained by setting the untwisted quark mass $m_0$ to its critical value $m_{\rm cr}$,
 while the twisted
quark mass parameter $\mu$ is kept non-vanishing in order to work away from the chiral limit.
In \eq{S_tm} the quark fields $\chi$
are in the so-called "twisted basis". The "physical basis" is obtained for
maximal twist by the simple transformation
\be
\psi(x)=\exp\left(\frac {i\pi} 4\gamma_5\tau^3\right) \chi(x),\qquad
\overline\psi(x)=\overline\chi(x) \exp\left(\frac {i\pi} 4\gamma_5\tau^3\right)
\quad.
 \ee
In terms of the physical fields the action is given by
\be
S_F^{\psi}= a^4\sum_x  \bar{\psi}(x)\left(\frac 12 \gamma_\mu 
[\nabla_\mu+\nabla^*_\mu]-i \gamma_5\tau^3 \left(- 
\frac{ar}{2} \;\nabla_\mu\nabla^*_\mu+ m_{\rm cr}\right ) 
+  \mu \right ) \psi(x)\quad.
\label{S_ph}
\ee
In this paper, unless otherwise stated, the quark fields will be understood as ``physical fields'',
 $\psi$, in particular when we define the baryonic interpolating fields. 

A crucial advantage of the twisted mass formulation is
the fact that, by tuning the bare untwisted quark mass $m_0$ to its critical value
 $m_{\rm cr}$, all physical observables are automatically 
${\cal O}(a)$ improved. 
In practice, we implement
maximal twist of Wilson quarks by tuning to zero the bare untwisted current
quark mass, commonly called PCAC mass, $m_{\rm PCAC}$~\cite{ETMClong}, which is proportional to
$m_0 - m_{\rm cr}$ up to ${\cal O}(a)$ corrections. 
The value of $m_{\rm cr}$ is determined at each $\beta$ value at the lowest 
twisted mass used in our simulations, a procedure that preserves ${\cal O}(a)$ improvement
and keeps ${\cal O}(a^2)$ small~\cite{Boucaud:2008xu,Frezzotti:2005gi}.
The twisted mass fermionic action breaks parity and isospin at 
non-vanishing lattice spacing, as it is apparent from the form of the Wilson term in 
 Eq.~(\ref{S_ph}).
In particular,  the isospin breaking in physical observables is a 
cut-off effect of ${\cal O}(a^2)$~\cite{Frezzotti:2004}.
To simulate the strange quark in the valence sector several choices are possible. 

The strange and charm
quarks are added 
as 
Osterwalder-Seiler valence quarks  and their action reads
\beq
S_{\rm heavy}^{\rm{OS}} = a^4 \sum_{x} \sum_{h=s}^{c} \bar{\chi}_h(x) \,
\Big( \frac{\gamma_\mu}{2}(\nabla_\mu+\nabla_\mu^*) - \frac{a}{2}
\nabla_\mu^{*}\nabla_\mu + M_{cr} + i \gamma_5 ~\mu_h  \Big) \, \chi_h(x)
\label{S_OS}
\eeq
where $\mu_s$ and $\mu_c$ are the strange and charm valence quark masses.
This is naturally realized in the twisted mass approach by introducing two additional 
doublets of strange and charm quarks and keeping only the positive diagonal component of $\tau_3$.
The  $m_0$ value is taken to be equal to the critical mass determined in the light sector, 
thus guaranteeing the $O(a)$  improvement in any observable. 
The reader interested in the advantage of this mixed action in the mesonic sector is referred to the Refs~\cite{Frezzotti:2004wz, AbdelRehim:2006ra, AbdelRehim:2006ve, Blossier:2007vv, Blossier:2009bx}.

\subsection{Simulation details}

The input parameters of the calculation, namely $\beta$, $L/a$ and $a\mu$ 
are summarized in Table~\ref{Table:params}. The corresponding lattice spacing $a$ 
and the pion mass values, spanning a mass range 
from 260 to 450~MeV, are taken 
from Ref.~\cite{Alexandrou:2010hf}.
%%
% {\bf MARIO: Here I substituted Urbach:2007 by Alexandrou:2010hf.
% It was the only time that Urbach:2007 was cited!}  
%%
At $m_{\pi}\approx 300$ MeV we have simulations 
for lattices of spatial size $L=2.1$~fm and $L=2.7$~fm at $\beta=3.9$ 
allowing to investigate finite size effects. 
Finite lattice spacing effects are investigated using three sets of 
results at $\beta=3.9$, $\beta=4.05$ and $\beta=4.2$.
These sets of gauge ensembles allow us to estimate all the systematic 
errors in order to have reliable predictions for the baryon spectrum.
%%
% {\bf MARIO: Here I put the entries found in same table in Alexandrou:2010hf, 
%namely lattice spacings and pion masses. Also I put a "-" for the statistics 
%in the charme case at beta=3.9 and volume $32^3$, 
%where we don't have determinations. 
%It was requested by the referee, but I don't know if it is understandable.
%With these entries the problems addressed by the referee in the first part
%of the report are solved.}
%%
\begin{table}[h]
\begin{center}
\begin{tabular}{c|lllll}
\hline\hline
\multicolumn{5}{c}{ $\beta=4.2$, $a=0.056(1)$~fm  ${r_0/a}=8.31(5)$}\\\hline
$32^3\times 64$, $L=1.8$~fm &$a\mu_{\rm sea}$         & 0.0065     &     &      \\
 & Statistics      &  \multicolumn{2}{l}{ 240, 76} &   \\
                               &$m_\pi$~(GeV) & 0.4698(18) &  &    \\
                               &$m_\pi L$     & 4.24     &   &           \\ \hline
$48^3\times 92$, $L=2.7$~fm &$a\mu_{\rm sea}$         & 0.0020     &    &       \\
 & Statistics      &  \multicolumn{2}{l}{ 458, 456} &   \\
                               &$m_\pi$~(GeV) & 0.262(1) &   &  \\
                               &$m_\pi L$     & 3.55         & &    \\ 

\hline\hline
\multicolumn{5}{c}{ $\beta=4.05$, $a=0.070(1)$~fm, ${r_0/a}=6.61(3)$ }\\
\hline
$32^3\times 64$, $L=2.13$~fm &$a\mu_{\rm sea}$         & 0.0030     & 0.0060     & 0.0080     \\
 & Statistics      & 144, 144 & 194, 193 & 201, 201      \\
                               &$m_\pi$~(GeV) & 0.2925(18) & 0.4035(18) & 0.4653(15)  \\
                               &$m_\pi L$     & 3.31       &   4.57     & 5.27            \\ \hline\hline
\multicolumn{6}{c}{$\beta=3.9$, $a=0.089(1)$~fm,   ${r_0/a}=5.22(2)$}\\\hline 
$24^3\times 48$, $L=2.05$~fm &$a\mu_{\rm sea}$            &    0.0040      &   0.0064     &  0.0085     &   0.010 \\ 
                               & Statistics & 4112, 310 & 545, 278 & 1817, 369  & 477, 475\\ 
                               &$m_\pi$~(GeV) &    0.3032(16) & 0.3770(9) & 0.4319(12) & 0.4675(12)\\
                               &$m_\pi L$     &    3.25       & 4.05      & 4.63       & 5.03     \\
$32^3\times 64$, $L=2.74$~fm  &$a\mu_{\rm sea}$ & 0.0030 & 0.0040 & & \\
                               & Statistics & 659,     NA  &  242,     NA  \\
                               & $m_\pi$~(GeV)& 0.2600(9)   & 0.2978(6) & &   \\
                               & $m_\pi L$    & 3.74        & 4.28      && \\\hline \hline
\end{tabular}
\caption{Input parameters ($\beta,L,\mu$) of our lattice simulations and corresponding lattice spacing ($a$) and pion mass ($m_{\pi}$).
The  statistics refer to the number of configurations used in calculation 
of the masses of the strange and charm baryons. The first entry gives the
number used for the tuned value of the strange quark and the second for the tuned value of the charm. An entry NA indicates that  no
masses were computed.
  The lattice spacing was determined using the nucleon mass~~\cite{Alexandrou:2010hf}.}
\label{Table:params}
\end{center}
\vspace*{-.0cm}
\end{table}

\subsection{Tuning of the bare strange and charm quark masses}

% Using the experimental value of 
%the mass ratio of the kaon to the pion, $m_K/m_\pi$, the 
%bare strange quark mass can be set. 
The dependence of the pseudoscalar meson mass on the valence and sea quarks
can be written as a  polynomial of the form~\cite{Blossier:2007vv} 
\begin{equation}
a^2 M^2_{PS}(a \mu_{\rm sea},a \mu_1,\mu_2)=B_0 (a \mu_1+a \mu_2 ) [ 1+a_V \xi_{12} 
+a_{\rm sea} \xi_{\rm sea} + a_{VV} \xi^2_{12} + 
a^\prime_{\rm sea} \xi^2_{\rm sea} +a_{V,\rm sea} \xi_{12} \xi_{\rm sea} + a_{VD} \xi^2_{D12}]\quad, 
\label{extrapolation}
\end{equation}
where $\mu_{\rm sea}$ is the sea quark mass and $a \mu_1 , a \mu_2$ are the valence quark masses,
$\xi_i= B_0 a \mu_i / (4 \pi f)^2$, $\xi_{ij}=2 B_0 (a \mu_i +a \mu_j) / (4 \pi f)^2$ and
$\xi_{Dij}= B_0 (a \mu_i - a \mu_j)/ (4 \pi f)^2$.
For the $\beta=3.9$ ensembles we consider in total 164 
pseudoscalar meson masses using  all possible
combinations of sea and valence quark masses. Namely, we have considered 150
combinations obtained from $a\mu_{\rm sea}$ and $a\mu_1$ independently  taking the
values

\begin{equation}
\{a\mu_{\rm sea} , a\mu_1\}=\{ 0.0040 \quad 0.0064 \quad 0.0085 \quad 0.0100 \quad 0.0150 \} \nonumber
\end{equation}
whereas  $a\mu_2$ takes the values
\begin{equation}
a\mu_2=\{ 0.0040 \quad 0.0064 \quad 0.0085 \quad 0.0100 \quad 0.0150 \quad 0.0220 \quad 0.0270 \quad 0.0320\}. \nonumber
\end{equation}
We have an additional 12 combinations coming from the combinations

\begin{equation}
a \mu_{\rm sea}=\mu_1=\{ 0.0040 \quad 0.0064 \quad 0.0085 \quad 0.0100 \} \,,\quad \quad \mu_2=\{  0.24  \quad 0.27 \quad 0.30 \}, \nonumber
\end{equation}
plus two extra combinations from 
\begin{equation}
a \mu_{\rm sea}=a \mu_1= 0.0040\,, \hspace*{1cm} a \mu_2= \{0.0217 \quad 0.25\}\,.\nonumber
\end{equation}
For the tuning at $\beta=4.05$  we use the following 20 combinations

\bea
a \mu_{\rm sea}&=a \mu_1= 0.0030 \hspace*{1cm} a \mu_2&= \{0.0030 \quad 0.014 \quad 0.0166 \quad 0.020 
\quad 0.17 \quad 0.20 \quad 0.23 \quad 0.26 \}\nonumber \\
a \mu_{\rm sea}&=a \mu_1= 0.0060 \hspace*{1cm} a \mu_2&= \{0.0060 \quad 0.0166 \quad 0.019 \quad 0.025 \} \nonumber\\
a \mu_{\rm sea}&=a \mu_1= 0.0080 \hspace*{1cm} a \mu_2&= \{0.0080 \quad 0.014 \quad 0.0166 \quad 0.020
\quad 0.17 \quad 0.20 \quad 0.23 \quad 0.26 \}\,.\nonumber
\eeq
 
\noindent
For the tuning at $\beta=4.2$  we consider 10 pseudoscalar
meson masses:

\bea
a \mu_{\rm sea}&=a \mu_1= 0.0065 \hspace*{1cm} a \mu_2&= \{0.0065 \quad 0.14 \quad 0.16 \quad 0.185 \quad 0.21 \}\nonumber \\
a \mu_{\rm sea}&=a \mu_1= 0.0020 \hspace*{1cm} a \mu_2&= \{0.0020 \quad 0.012 \quad 0.015 \quad 0.136 \quad 0.17 \} . \nonumber
\eea

\begin{figure}[h!]
\centering
\begin{tabular}{cc}
\includegraphics[width=.3\textwidth, angle=270]{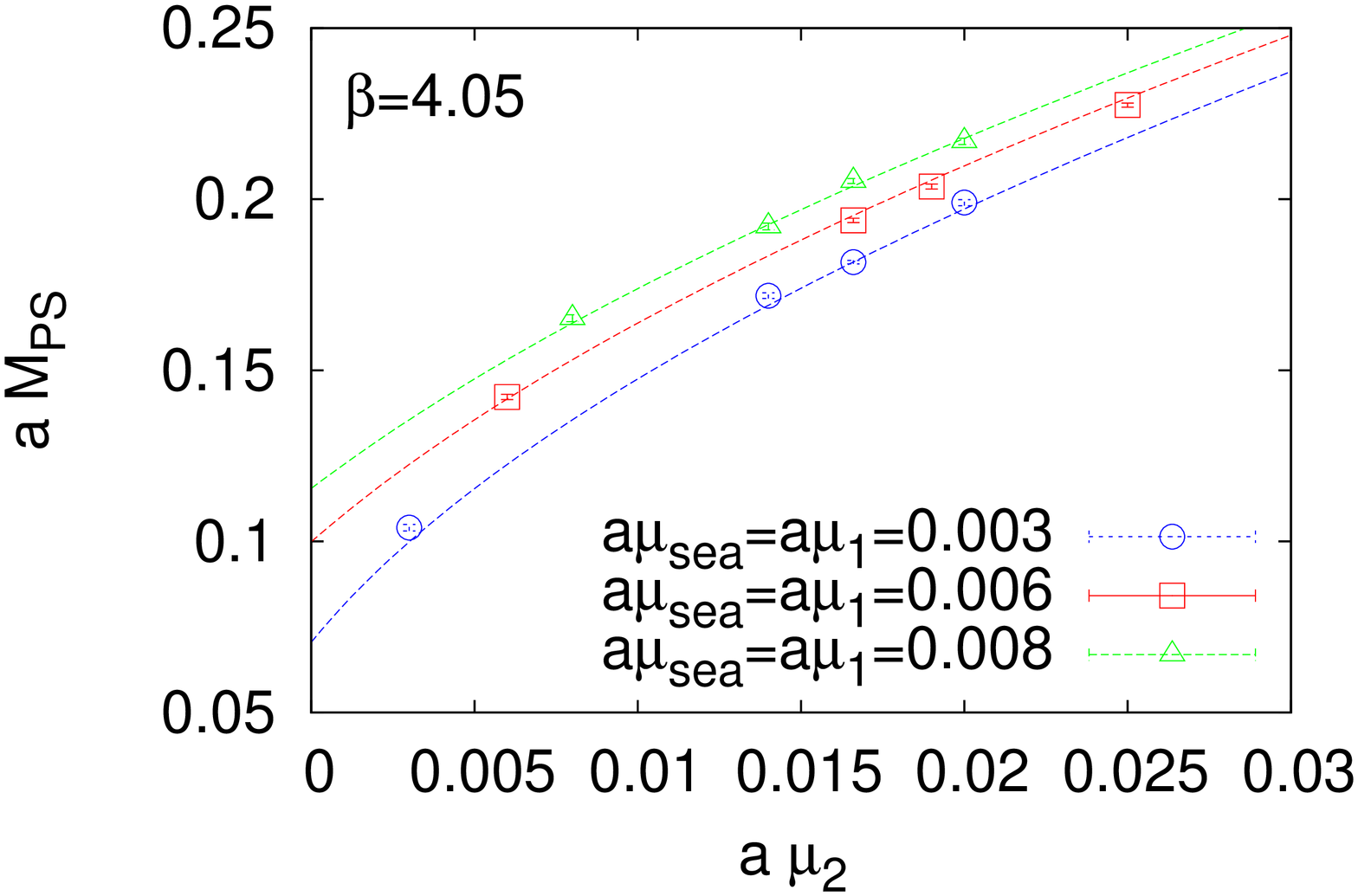} &
\includegraphics[width=.3\textwidth, angle=270]{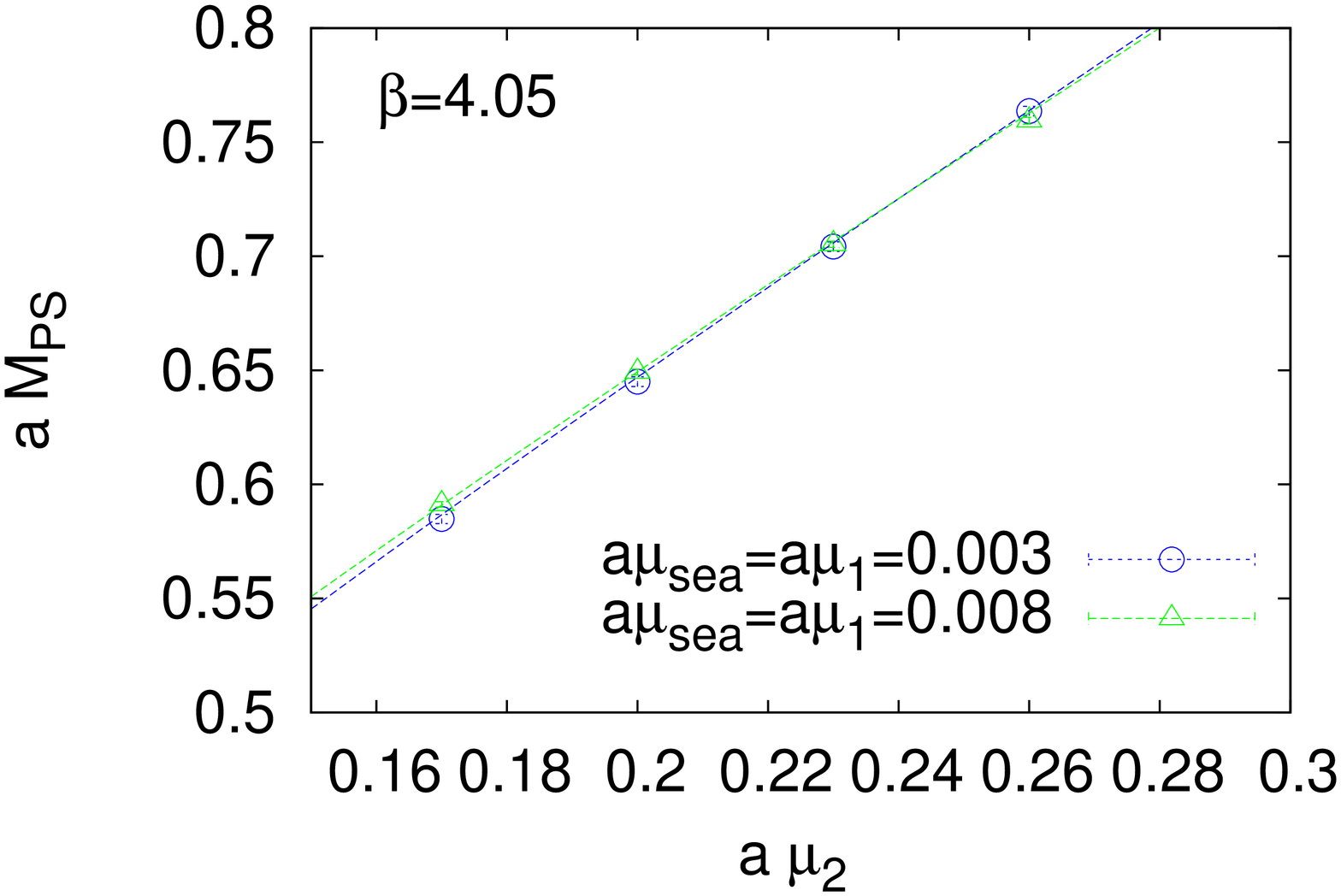} \\
\end{tabular}
\caption{Pseudo-scalar meson masses for $\beta=4.05$ as a function of the heavy valence 
quark mass $a \mu_2$ in the relevant
mass range for the strange quark (left) and charm quark (right).
In all the examples shown the sea quark mass $a \mu_{\rm sea}$ is set 
equal to the light valence quark mass $a \mu_1$.}
\label{fig:fits}
\end{figure}

\begin{figure}[h!]
\centering
\begin{tabular}{cc}
\includegraphics[width=.3\textwidth, angle=270]{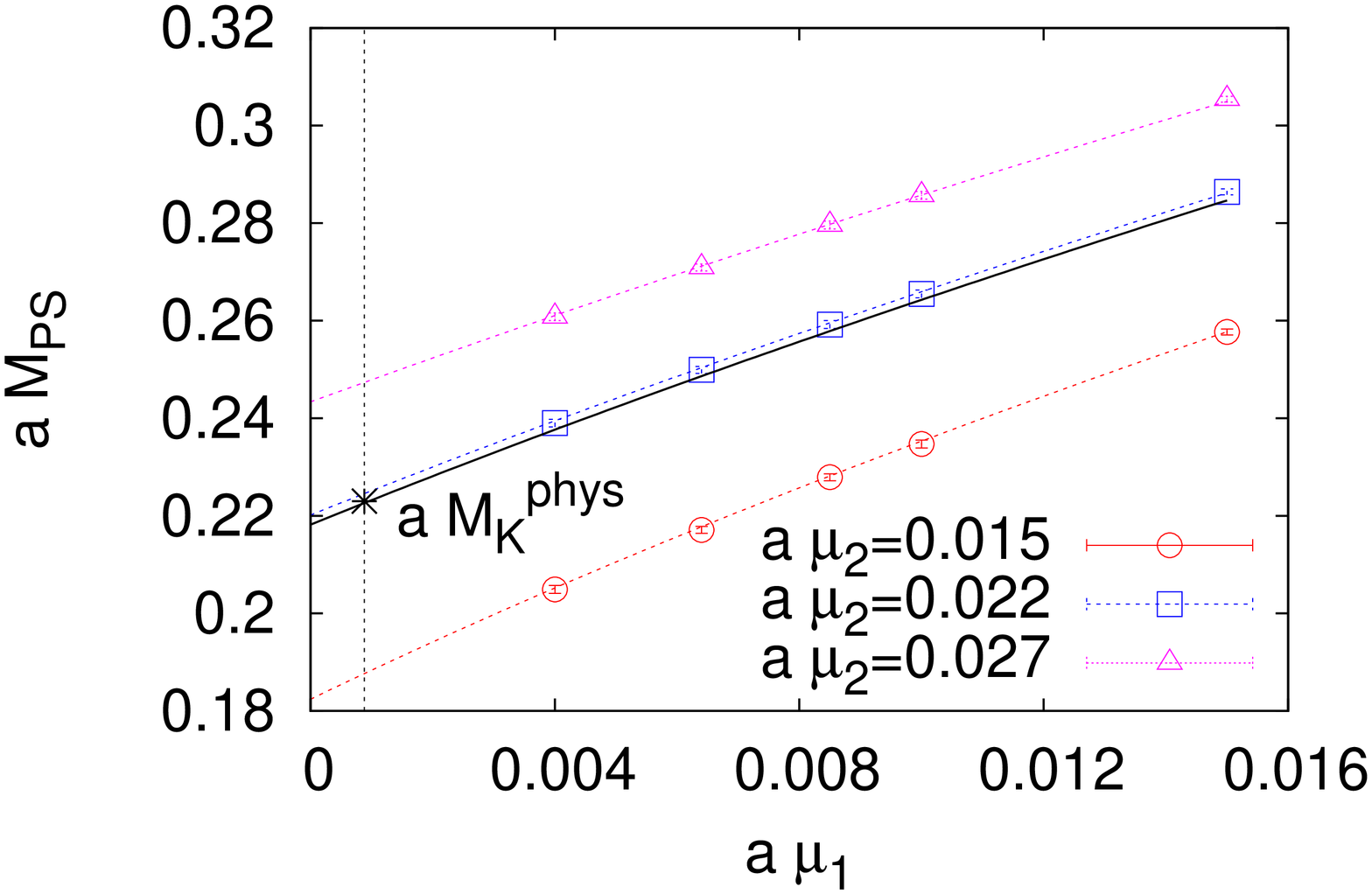} &
\includegraphics[width=.3\textwidth, angle=270]{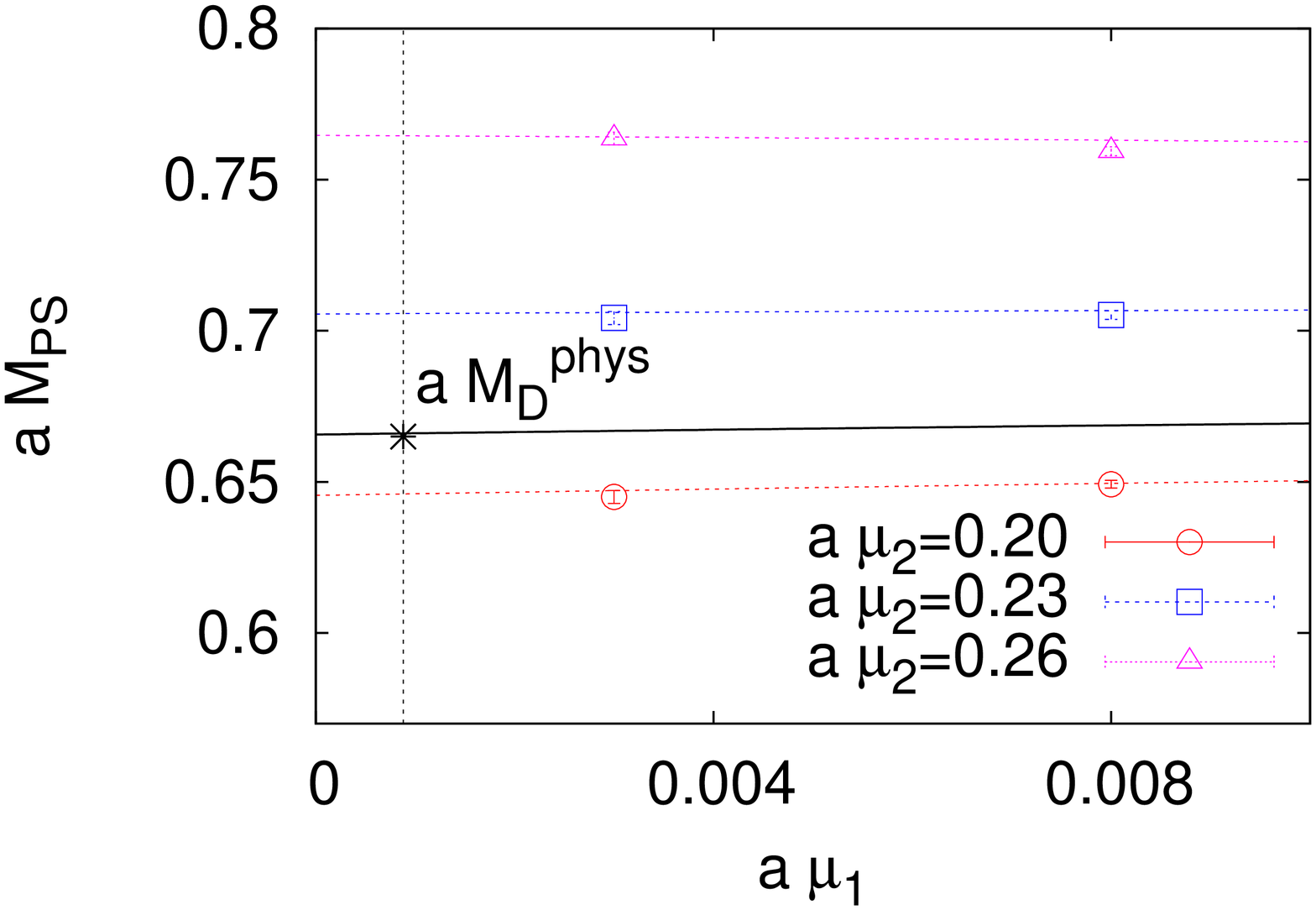} \\
\end{tabular}
\caption{The dependence of the pseudo-scalar
masses on the strange quark mass at $\beta=3.9$ (left) and 
charm quark mass at $\beta=4.05$ (right). The solid line 
shows the variation of
the pseudoscalar mass at the tuned strange (left) and charm  (right)
 quark masses. The dashed vertical 
line corresponds to the value of $a \mu_1$ at which the physical pion 
mass is recovered. The asterisk denotes the
physical value of the kaon and D-meson mass in lattice units. The 
lattice spacing is determined from the nucleon mass.}
\label{fig:fits2}
\end{figure}

In Figs.~\ref{fig:fits} and \ref{fig:fits2} we show representative fits to the pseudo-scalar masses
in the range of the kaon and D-meson 
masses using the expression given in Eq.~(\ref{extrapolation}).
 The values of the strange and charm quark masses 
are varied until the resulting  kaon and D-meson masses
are matched to their physical values.
The resulting fit parameters are listed in Table~\ref{tab:fit parameters}. 
We note that for $\beta=3.9$ two fitting ranges are used, one range
spanning the 
 the strange quark mass and one the charm quark mass,
For $\beta=4.05$ and $\beta=4.2$ we fit all data together since we do not have enough
mass combinations in order to apply Eq.~(\ref{extrapolation}). If one
does the same for $\beta=3.9$ then the tuned value for the strange quark mass
is  $a\mu_s=0.0216(7)$ compatible with the value of $a\mu_s=0.0217(5)$ if we restrict
the fit to the strange region. 
In addition, at each $\beta$-value we can restrict the fit in the charm region  using
the Ansatz
\be
 m_B= m_B^0+ b \mu_h+c/\mu_h.
\ee
The tuned charm quark value is found to be compatible with the one
extracted using Eq.~(\ref{extrapolation}). This procedure can be carried out
using either the lattice spacing determined from the nucleon mass or form
$f_\pi$. The difference in the tuned masses reflects the systematic error in
setting the scale. 

\begin{table}[h]
\begin{center}
\begin{tabular}{|l|l|l|l|l|}
\hline
 & $\beta=3.9$ (strange quark) & $\beta=3.9$ (charm quark) & $\beta=4.05$ & $\beta=4.2$ \\ \hline
$B_0$ & 2.252(5) & 2.38(6) & 1.652(5)  & 1.295(5) \\ \hline
$f$   & 0.077(2) & 0.112(2) & 0.093(4) & 0.069(4) \\ \hline
$a_V$ & -0.45(2) & 0.3(1) & 0.85(5)   & 0.56(5)  \\ \hline
$a_{\rm sea}$ &  0.0 & 0.0   & 0.0     & 0.0      \\ \hline
$a_{VV}$ & 3.0(1)  & 1.8(5) & -4.0(3)   & 2.6(2)   \\ \hline
$a_{V. \rm sea}$ &  0.0  & 0.0   & 0.0       & 0.0      \\ \hline
$a^\prime_{\rm sea}$ &  0.0  & 0.0   & 0.0       & 0.0      \\ \hline
$a_{VD}$ &  -2.25(3)  & -1.4(6)  & 4.94(5)       & -1.8(3)      \\ \hline
$\chi^2/{\rm d.of}$ & 0.51 & 1.33  &  4.52 & 4.40 \\ \hline
\end{tabular}
\caption{The values of the fit parameters.}
\label{tab:fit parameters}
\end{center}
\end{table}

\begin{table}[h]
\begin{tabular}{|l|l|l|l|l||l|l|}
\hline
$\beta$ & $a \mu_s$ & $a \mu_c $ & $\mu_s$ (GeV)& $\mu_c$ (GeV)& $\mu_s^{f_{\pi}} (GeV)$ & $\mu_c^{f_{\pi}}$ (GeV) \\ \hline
3.9     &  0.0216(7) & 0.27(3)  & 0.0478(16) & 0.598(66) & 0.0431(17) & 0.64(12)  \\ \hline
4.05     &  0.0178(5) & 0.21(1)  & 0.0501(14) & 0.591(28) & 0.0451(12) & 0.556(31)  \\ \hline
4.2     &  0.014(1) & 0.17(2)  & 0.0493(35) & 0.598(70) & 0.0464(15) & 0.575(38)  \\ \hline
\end{tabular}
\caption{The  strange and charm quark 
masses at each value of $\beta$ tuned using the kaon and D-meson masses and
the lattice spacing determined from the nucleon mass
are given in lattice units in the second and third columns respectively.  The tuned strange and charm quark masses in GeV are given in the fourth and fifth columns.  In the sixth and seventh columns  
we determined the corresponding masses in GeV using the lattice determined from $f_{\pi}$. }
\label{tab:tuned quarks}
\end{table}

%\begin{table}[h]
%\begin{tabular}{|l|l|l|l|}
%\hline
% & $\beta=3.9$ & $\beta=4.05$ & $\beta=4.2$ \\ \hline
%$a\mu_s$ & 0.0175(7) & 0.0146(4)  & 0.0121(4) \\ \hline
%$a\mu_c$ & 0.26(5) & 0.18(1)  & 0.15(1) \\ \hline
%\end{tabular}
%\caption{Tuned strange and charm quark masses in lattice units determined using% the scale from $f_\pi$. }
%\label{tab:tuned quarks fpi}
%\end{table}

The tuned values of the strange and charm quark masses $m_s$ and $m_c$ obtained
at the the physical pion are given in Table~\ref{tab:tuned quarks}.
%%
%{\bf MARIO: the referee had problems in comparing the tuned values in physical
%units found using the lattice spacing fixed with the nucleon and with $f_pi$.
%His mistake is that he tries to convert using different lattice spacings but the 
%same tuned quark masses, the ones listed in the table, which are found using
%the nucleon. Maybe we should specify better in the text that 
%$(a \mu_s)^N \ne (a \mu_s)^(f_pi)$}
%%
In a previous paper,  the ETMC computed  pseudo-scalar meson 
masses  for a number of sea and valence quark masses using the 
$\beta=3.9$ gauge configurations. Matching the experimental value of 
the mass ratio of the kaon to the pion, $m_K/m_\pi$, the 
bare strange quark mass was determined~\cite{Blossier:2007vv}.  Depending
on the polynomial fit used the values for $am_s$ at $\beta=3.9$ varied from $0.0243(5)$ to $0.0218(10)$. 
%We give the
%tuned values in Table~\ref{tab:tuned quarks fpi}.
Thus,  our value of $a\mu_s=0.0216(7)$ from
matching the physical value of the kaon mass in combination with the 
lattice spacing determined from the nucleon mass  is
compatible 
with the value determined
in Ref.~\cite{Blossier:2007vv}. 
%%
%{\bf MARIO: read this statement where the referee asked to be more precise about 
%the distance in standard deviations of the results}
%%
Such an agreement is satisfactory and shows
that the two procedures lead to the same 
 determination within the uncertainties associated with the extrapolation.
The systematic error introduced from the way the lattice scale  is fixed
 can
be assessed by comparing the
tuned values extracted using the lattice spacing determined from
 the nucleon mass
and from the pion decay constant $f_\pi$. The values of the lattice spacing
determined using $f_\pi$, taken from Ref.~\cite{Baron:2009wt}, are  $a_{\beta=3.9} = 0.0801(14)$~fm, $a_{\beta=4.05}=0.0638(10)$~fm and  $a_{\beta=4.2}= 0.05142(38)$~fm. In Table~\ref{tab:tuned quarks} we give 
 the tuned values for the charm and strange quark masses expressed in physical units. As can be seen, the values for the charm quark masses are in agreement,
 whereas for strange quark masses 
the differences are  within about two standard deviations.

%OLD TABLE
%\begin{table}[h]
%\begin{tabular}{|l|l|l|l|}
%\hline
% & $\beta=3.9$ & $\beta=4.05$ & $\beta=4.2$ \\ \hline
%%$\mu_l$ & 0.00088(2) & 0.00075(2)  & 0.00059(4) \\ \hline
%$\mu_s$ (GeV) & 0.0216(7) & 0.0178(5)  & 0.014(1) \\ \hline
%$\mu_c$ (GeV) & 0.27(3) & 0.21(1)  & 0.17(2) \\ \hline
%\end{tabular}
%\caption{Tuned strange and charm quark mass. The scale is set using th enucleon mass. {\bf Dina: are the units in GeV?}}
%\label{tab:tuned quarks}
%\end{table}

%OLD TABLE

\subsection{Interpolating fields}

\begin{figure}[h!]
\begin{minipage}{8cm}
\epsfxsize=8truecm
\epsfysize=6truecm
 \mbox{\epsfbox{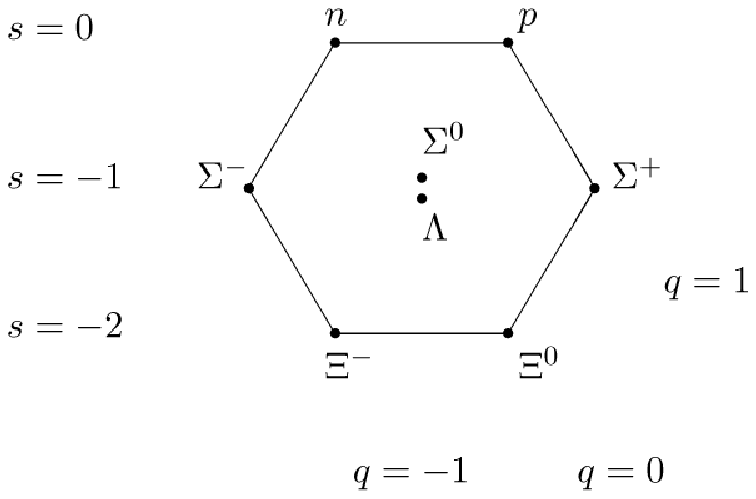}}
\caption{The low lying baryons belonging to the octet representation labeled by the value of $I_3$ and hyper-charge.} 
\label{fig:interp. octet}
\end{minipage}
\hfill
\begin{minipage}{8cm}
\epsfxsize=8truecm
\epsfysize=6truecm
 \mbox{\epsfbox{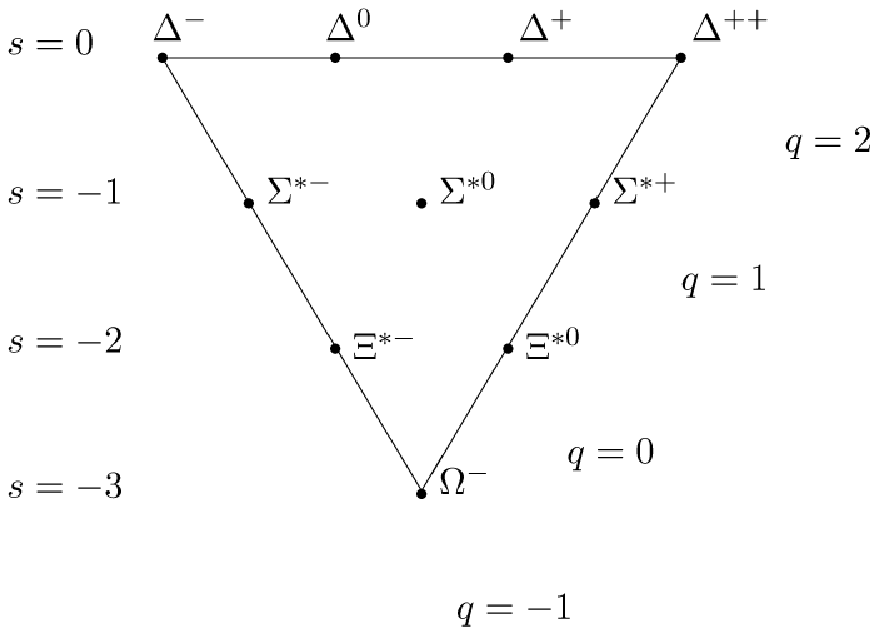}}
\caption{The low lying baryons belonging to the decuplet representation labeled by the value of $I_3$ and hyper-charge.}
\label{fig:interp. decuplet}
\end{minipage}
\end{figure}

The low-lying baryons belonging to the octet and decuplet representations 
 of $SU(3)$ are given in Figs.~\ref{fig:interp. octet}
 and \ref{fig:interp. decuplet} respectively.
They are classified by giving the isospin, $I$, the third component of the isospin, $I_3$, the strangeness (S), the spin and the parity.
In order to extract their masses in lattice QCD, we evaluate two-point correlators. We use interpolating fields to create these states from the vacuum
that have the correct quantum numbers and reduce to the quark model wave functions in the non-relativistic limit.
The interpolating fields used in this work are collected in Tables~\ref{Table:interpolating_octet}~\cite{Ioffe:1981kw, Leinweber:1990dv} and  ~\ref{Table:interpolating_decuplet}~\cite{Ioffe:1981kw, Leinweber:1992hy} for
the octet and decuplet respectively. 

Charm baryons with no strange quarks are obtained from the 
interpolating fields of strange baryons by
 replacing the strange with the charm quark. 
There are additional charm baryons containing strange  quarks, 
giving a 20-plet of
spin-1/2 and a 20-plet of spin-3/2. In most of  this work we 
do not consider the  particles that contained both a
strange and charm quarks. For the lattice with the smallest lattice spacing
and at the smallest pion mass we also consider  the spin-1/2 $\Xi_{c}$, $\Xi^{\prime}_{c}$, $\Omega_c$ and
 $\Omega_{cc}$ and the spin-3/2 $\Xi^*_{c}$, $\Omega^*_{c}$ and $\Omega^*_{cc}$. The interpolating fields for these baryons are given in Table~\ref{Table:interpolating_extra}.

\begin{table}[h!]
\begin{center}
\renewcommand{\arraystretch}{1.5}
\begin{tabular}{c|c c c c}
\hline 
 Strangeness  & Baryon  & Interpolating field & $I$  & $I_z$ \\ \hline
\multirow{2}{*}{$S=0$} & $p$ & $~\chi^{p} = \epsilon_{abc}(u_a^T C\gamma_5 d_b) u_c~$ & $1/2$ & $+1/2$\\
 & $n$ & $~\chi^{n} = \epsilon_{abc}(d_a^T C\gamma_5 u_b) d_c~$ &  $1/2$ & $-1/2$\\
\hline
 \multirow{4}{*}{$S=1$} & $\Lambda$ &$~\chi^{\Lambda^8}= \frac{1}{\sqrt{6}}\epsilon_{abc}\big\{2(u_a^T C\gamma_5 d_b )s_c +(u_a^T C\gamma_5 s_b)d_c -(d_a^T C\gamma_5 s_b)u_c\big\}~$ & $0$ & $0$\\
 & $\Sigma^+$ & $~\chi^{\Sigma^+} = \epsilon_{abc}(u_a^T C\gamma_5 s_b )u_c ~$  & $1$ & $+1$\\
 & $\Sigma^0$ & $~\chi^{\Sigma^0} = \frac{1}{\sqrt{2}}\epsilon_{abc}\big\{(u_a^T C\gamma_5 s_b )d_c +(d_a^T C\gamma_5 s_b)u_c \big\}~$ & $1$ & $+0$\\
 & $\Sigma^-$ & $~\chi^{\Sigma^-} = \epsilon_{abc}(d_a^T C\gamma_5 s_b )d_c ~$ & $1$ & $-1$\\
 \hline
\multirow{2}{*}{$S=2$} & $\Xi^0$ & $~\chi^{\Xi^{0}} = \epsilon_{abc}(s_a^T C\gamma_5 u_b) s_c~$ & $1/2$ & $+1/2$ \\
 & $\Xi^-$ & $~\chi^{\Xi^{-}} = \epsilon_{abc}(s_a^T C\gamma_5 d_b) s_c~$& $1/2$ & $-1/2$\\
 \hline \hline
\end{tabular}
\caption{Interpolating fields and quantum numbers for the baryons in the octet representation.}
\label{Table:interpolating_octet}
\end{center}
\end{table}

\begin{table}[h!]
\begin{center}
\renewcommand{\arraystretch}{1.5}
\begin{tabular}{c|c c c c}
\hline\hline
 Strangeness  & Baryon  & Interpolating field & $I$ & $I_z$ \\ \hline
\multirow{4}{*}{$S=0$} &$\Delta^{++}$ & $~\chi^{\Delta^{++}}_{\mu} = \epsilon_{abc}(u_a^T C\gamma_{\mu} u_b) u_c~$ &$3/2$ & $+3/2$\\
 &$\Delta^{+}$   & $~\chi^{\Delta^{+}}_{\mu} =  \frac{1}{\sqrt{3}}\epsilon_{abc}\big\{2 (u_a^T C\gamma_{\mu} d_b) u_c + (u_a^T C\gamma_{\mu} u_b) d_c\big\} ~$ &$3/2$ & $+1/2$ \\
 & $\Delta^{0}$  & $~\chi^{\Delta^{0}}_{\mu} =  \frac{1}{\sqrt{3}}\epsilon_{abc}\big\{2 (d_a^T C\gamma_{\mu} u_b) d_c + (d_a^T C\gamma_{\mu} d_b) u_c\big\}   ~$ &$3/2$ & $-1/2$\\
 & $\Delta^{-}$  & $~\chi^{\Delta^{-}}_{\mu} = \epsilon_{abc}(d_a^T C\gamma_{\mu} d_b) d_c~$ &$3/2$ & $-3/2$ \\
\hline
 \multirow{3}{*}{$S=1$}  & $\Sigma^{\ast +}$ & $~\chi^{\Sigma^{\ast +}}_{\mu}= \sqrt{\frac{2}{3}}\epsilon^{abc}\big\{ (u^{Ta}  C \gamma_{\mu}u^b )s^c + 2(u^{Ta}  C \gamma_{\mu}s^b) u^c  \big\}  ~$ & $1$ & $+1$\\
 & $\Sigma^{\ast 0}$ & $~\chi^{\Sigma^{\ast 0}}_{\mu} = \sqrt{\frac{2}{3}}\epsilon^{abc}\big\{ (u^{Ta}  C \gamma_{\mu}d^b )s^c + (d^{Ta}  C \gamma_{\mu}s^b) u^c +(s^{Ta}  C \gamma_{\mu}u^b) d^c \big\}~$ &$1$ & $+0$\\
 & $\Sigma^{\ast -}$ & $~\chi^{\Sigma^{\ast -}}_{\mu} = \sqrt{\frac{2}{3}}\epsilon^{abc}\big\{ (d^{Ta}  C \gamma_{\mu}d^b )s^c + 2(d^{Ta}  C \gamma_{\mu}s^b) d^c \big\} ~$ &$1$ & $-1$\\
 \hline
\multirow{2}{*}{$S=2$} & $\Xi^{\ast 0}$ & $~\chi^{\Xi^{\ast 0}}_{\mu} = \epsilon_{abc}(s_a^T C\gamma_\mu u_b) s_c~$  &$1/2$ & $+1/2$\\
 & $\Xi^{\ast -}$  & $~\chi^{\Xi^{\ast -}}_{\mu} = \epsilon_{abc}(s_a^T C\gamma_\mu d_b) s_c~$&$1/2$ & $-1/2$\\
 \hline
\multirow{1}{*}{$S=3$} & $\Omega^{-}$ & $~\chi^{\Omega^{-}}_{\mu} = \epsilon_{abc}(s_a^T C\gamma_{\mu} s_b) s_c~$  &$0$ & $+0$\\
 \hline \hline
\end{tabular}
\caption{Interpolating fields and quantum numbers for baryons in the decuplet representation.}
\label{Table:interpolating_decuplet}
\end{center}
\end{table}

%%
%{\bf MARIO: Indeed there was an error in the interpolating field of $\Omega_c$ %in the 
%following table. I used now the same interpolating field form as in you previous%
%paper on baryons. Now it should be ok.}
%%
\begin{table}[h!]
\begin{center}
\renewcommand{\arraystretch}{1.5}
\begin{tabular}{c|c}
\hline
 $J=1/2$  & $J=3/2$  \\ \hline
$\chi^{\Xi_c}= \frac{1}{\sqrt{6}}\epsilon_{abc}\big\{2(s_a^T C\gamma_5 d_b )c_c +(s_a^T C\gamma_5 c_b)d_c -(d_a^T C\gamma_5 c_b)s_c\big\}$   & 
$\chi^{\Xi_c^*}_{\mu} = \sqrt{\frac{2}{3}}\epsilon^{abc}\big\{ (s^{Ta}  C \gamma_{\mu}d^b )c^c + (d^{Ta}  C \gamma_{\mu}c^b) s^c +(c^{Ta}  C \gamma_{\mu}s^b) d^c \big\}$ \\
$\chi^{\Xi_c'} = \frac{1}{\sqrt{2}}\epsilon_{abc}\big\{(s_a^T C\gamma_5 c_b )d_c +(d_a^T C\gamma_5 c_b)s_c \big\}$ &  \\
$\chi^{\Omega_c}=\epsilon_{abc}(s_a^T C\gamma_5 c_b) s_c$ & $\chi^{\Omega_c^*}_{\mu}= \sqrt{\frac{2}{3}}\epsilon^{abc}\big\{ (s^{Ta}  C \gamma_{\mu}s^b )c^c + 2(s^{Ta}  C \gamma_{\mu}c^b) s^c  \big\}$ \\
$\chi^{\Omega_{cc}}=\epsilon_{abc}(c_a^T C\gamma_5 s_b) c_c$ & $\chi^{\Omega_{cc}^*}_{\mu} = \epsilon_{abc}(c_a^T C\gamma_{\mu} s_b) c_c $ \\
\hline
\end{tabular}
\caption{Interpolating fields  for the spin-1/2 $\Xi_{c}$, $\Xi^{\prime}_{c}$, $\Omega_c$ and
 $\Omega_{cc}$, and the spin-3/2 $\Xi^*_{c}$, $\Omega^*_{c}$ and $\Omega^*_{cc}$ baryons.}
\label{Table:interpolating_extra}
\end{center}
\end{table}

Local interpolating fields  are not optimal for suppressing excited state contributions. We instead apply
 Gaussian smearing to each  quark field,  $q({\bf x},t)$: $q^{\rm smear}({\bf x},t) = \sum_{\bf y} F({\bf x},{\bf y};U(t)) q({\bf y},t)$
using the gauge invariant smearing function  
\be 
F({\bf x},{\bf y};U(t)) = (1+\alpha H)^ n({\bf x},{\bf y};U(t)),
\ee
constructed from the hopping matrix,
\be
H({\bf x},{\bf y};U(t))= \sum_{i=1}^3 \biggl( U_i({\bf x},t)\delta_{{\bf x,y}-i} +  U_i^\dagger({\bf x}-i,t)\delta_{{\bf x,y}+i}\biggr).
\ee
 Furthermore we apply APE smearing to the spatial links  that enter the hopping matrix.
The parameters of the Gaussian and APE smearing are the same as those  used in our previous work devoted to the nucleon and $\Delta$
masses~\cite{Alexandrou:2008tn}.

\subsection{Two-point correlators}

To extract masses in the rest frame we consider two-point correlators defined by 
\bea
C^\pm_X(t,\vec{p}=\vec{0}) = \frac 1 2 {\rm Tr}(1 \pm \gamma_0) \sum_{\bf x_{\rm sink}}
\langle J_X( {\bf x}_{\rm sink}, t_{\rm sink}) \bar J_X({\bf x}_{\rm source}, t_{\rm source})\rangle,\qquad 
t=t_{\rm sink}-t_{\rm source} \quad.
\label{C_X}\eea
Space-time reflection symmetries of the action and the anti-periodic boundary conditions in the temporal direction for the quark fields
imply, for zero three-momentum correlators, that $C_X^+(t) = -C_X^-(T-t)$. So,  In order to decrease errors 
we average correlators in the forward and backward direction and define:
\be
   C_X(t) = C_X^+(t) - C_X^-(T-t) \, .
\ee 
In order to decrease correlation between measurement, we choose the source location  randomly on the whole lattice 
for each configuration.
Masses are then extracted from the so called effective mass which is defined by
\be
m_{\rm eff}^X(t)=-\log(C_X(t)/C_X(t-1))= m_X+\log\left(\frac{1+\sum_{i=1}^\infty c_ie^{\Delta_i t}}{1+\sum_{i=1}^\infty c_ie^{\Delta_i (t-1)}}\right)
\tendto_{t\rightarrow \infty}  m_X \quad,
\label{meff}
\ee
where $\Delta_i= m_i-m_X$ is the mass difference of the excited state $i$ with 
respect to the ground mass $m_X$.

%\begin{figure}[h!]
%{\bf Dina: Please show some representative effective masses}
%\begin{minipage}{0.46\linewidth}
%\includegraphics[width=\linewidth]{}
%\caption{Effective mass for ...}
%\label{fig:meff strange}
%\end{minipage}
%\begin{minipage}{0.46\linewidth}
%\includegraphics[width=\linewidth]{}
%\caption{Effective mass for ...}
%\label{fig:meff charm}
%\end{minipage}
%\end{figure}

\begin{figure}[h!]
%\centering
\begin{tabular}{cc}
\includegraphics[width=.33\textwidth, angle=270]{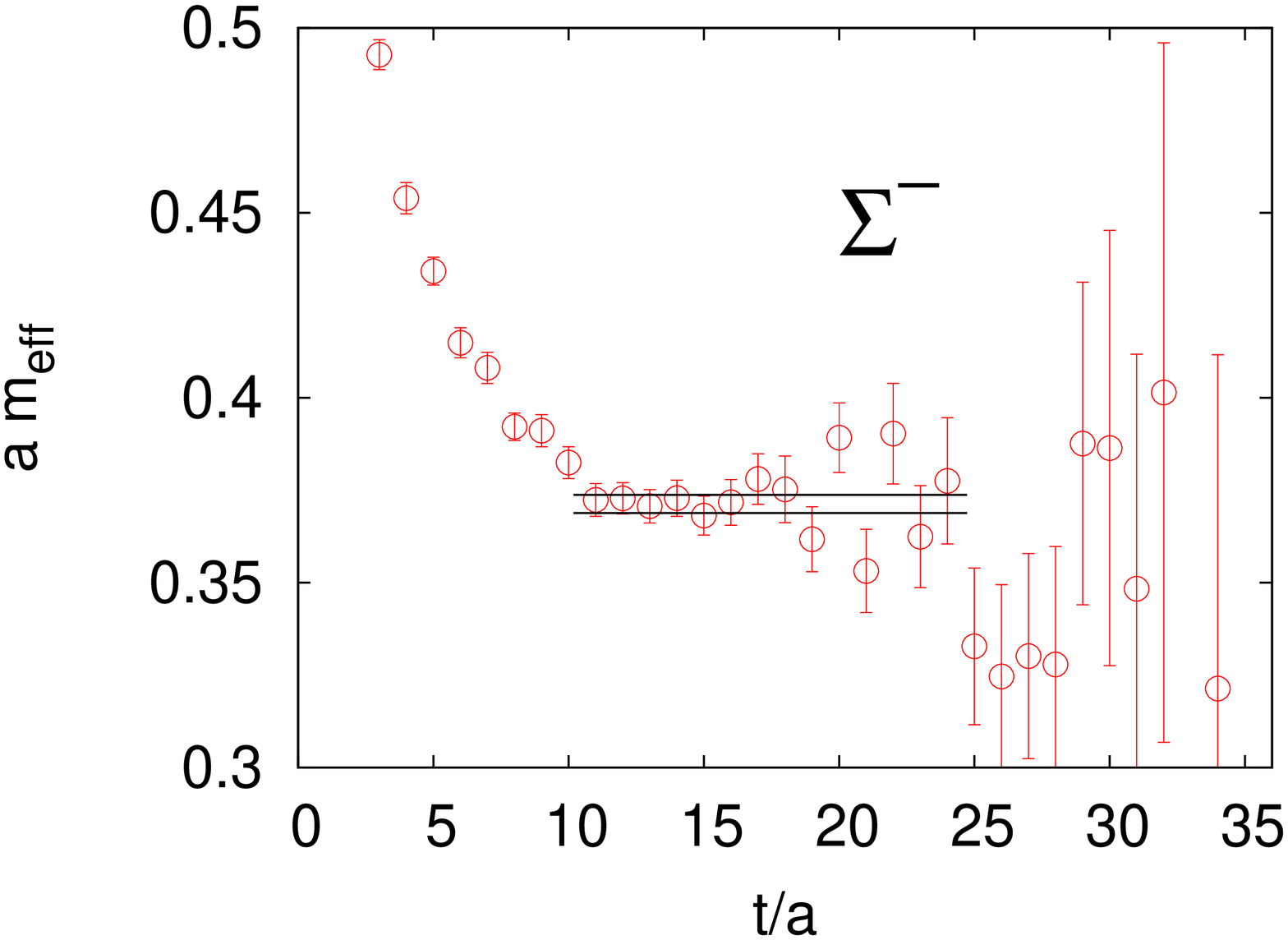} &
\includegraphics[width=.33\textwidth, angle=270]{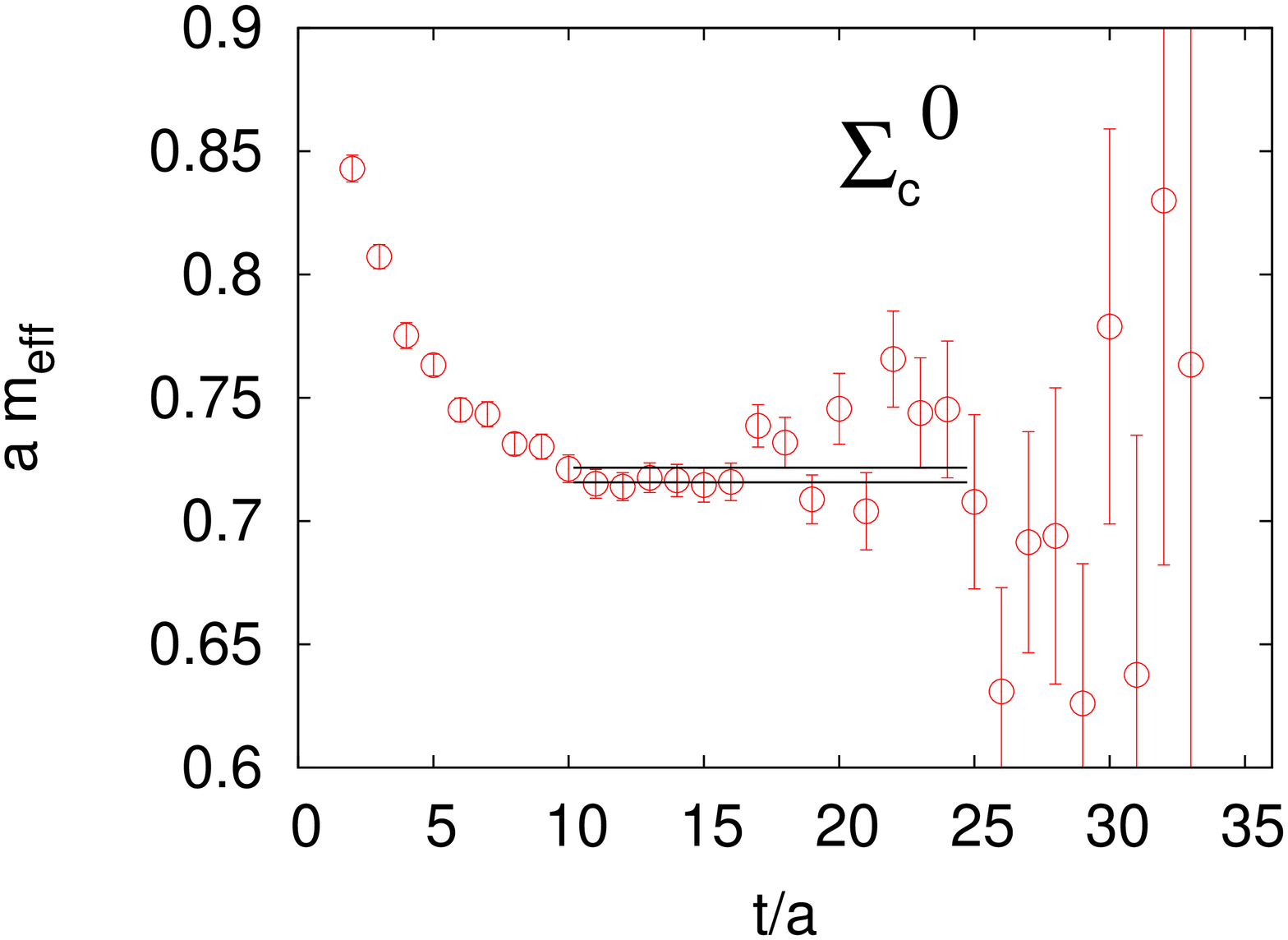} \\
\includegraphics[width=.33\textwidth, angle=270]{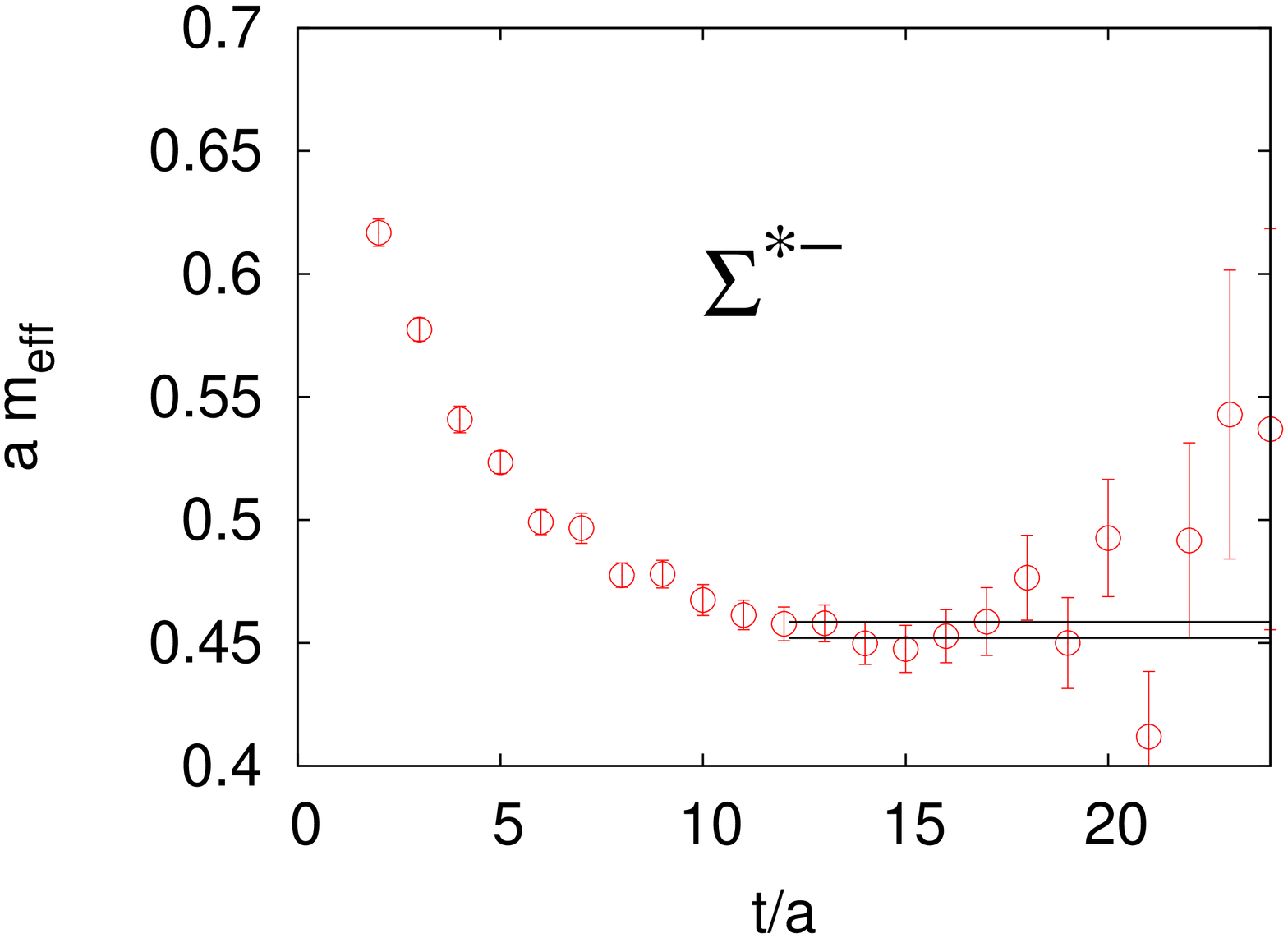} &
\includegraphics[width=.33\textwidth, angle=270]{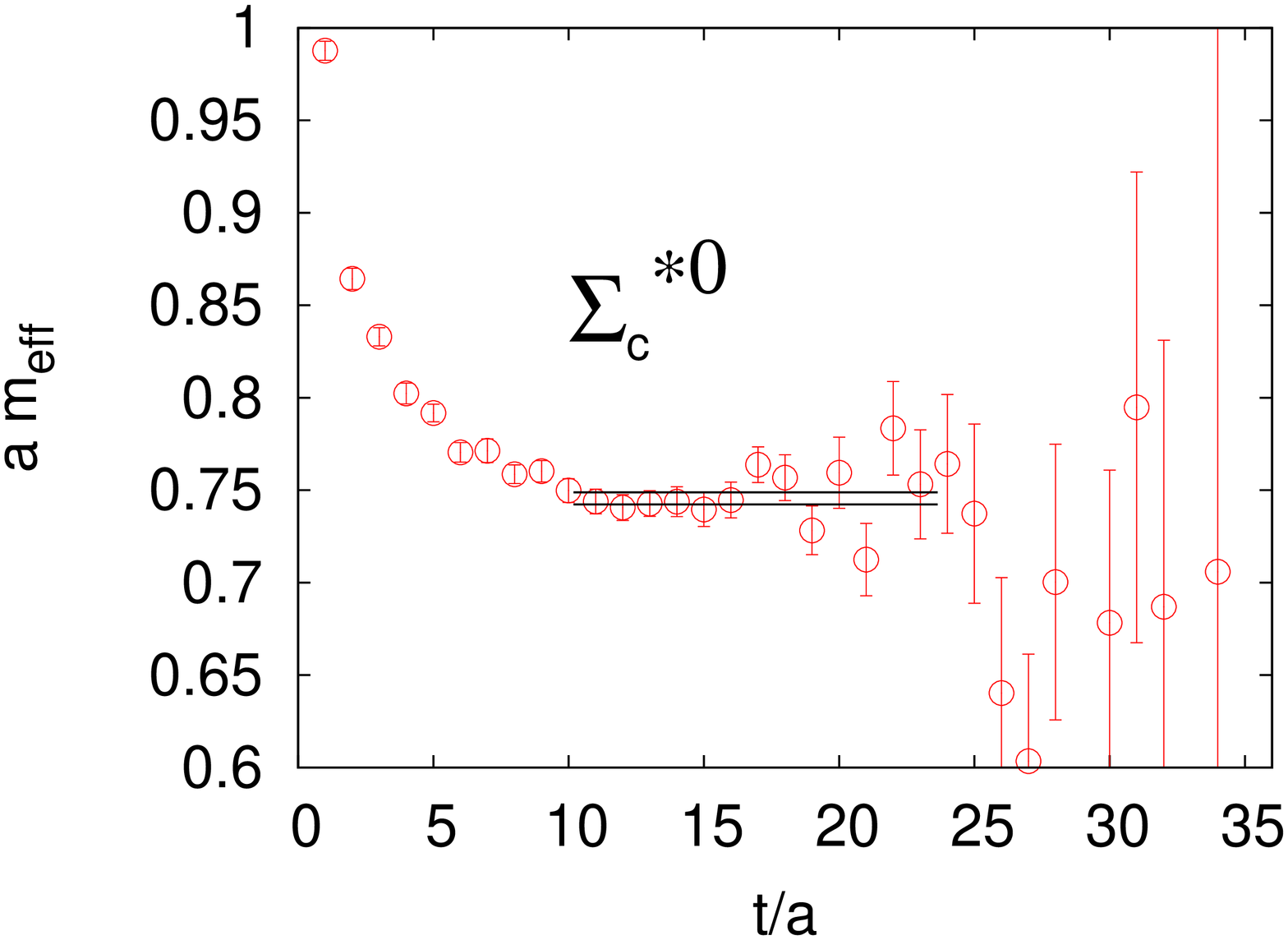} \\
\end{tabular}
\caption{Representative effective mass plots for $\beta=4.2$ and $a\mu_l=0.002$.
For the strange baryons (Left) we used $a\mu_s=0.015$, while for charm baryons (Right)
$a\mu_c=0.17$.}
\label{fig:meff}
\end{figure}

In Fig.~\ref{fig:meff}  we show representative examples of
the  effective masses of strange and charm baryons.
As can be seen, a plateau region can be clearly identified. What is
shown in these figures are effective masses extracted from correlators where
smearing is applied both at the sink and source. 
Although local correlators are expected to have the same  value in the
large time limit,
smearing suppresses excited state contributions
yielding a plateau at earlier time separations and to a  better accuracy in the mass extraction.
 We therefore extract the masses using smeared source and sink.
Our fitting procedure to extract $m_X$ is as follows:
The sum over excited states in the effective mass given in Eq.~(\ref{meff}) is 
truncated keeping only the first excited state.
Allowing a couple of time slice separation the effective mass is
fitted to the form given in Eq.~(\ref{meff}). This yields an estimate for the parameters $c_1$ and $\Delta_1$.
The lower fit range is increased until the contribution due to the first excited state 
is less than 50\% of the statistical error of $m_X$. 
This criterion is in most of the cases in agreement
 with a $\chi^2/{\rm d.o.f.}< 1$.
In the cases in which this criterion is not satisfied a careful examination of the effective mass is made to ensure that the fit range is in the
plateau region.

\section{Lattice results}

Before
we extrapolate our lattice results on the strange and charm baryon masses
 to the physical point, we need to examine their dependence
 on the heavy
quark mass as well as cut-off effects.  We collect lattice results for the masses of the strange and charm baryons in the Appendix.
The errors are evaluated using jackknife and the $\Gamma$-method~\cite{Wolff:2004} to check consistency.

\begin{figure}[h!]
%\centering
\begin{tabular}{cc}
\includegraphics[width=.33\textwidth, angle=270]{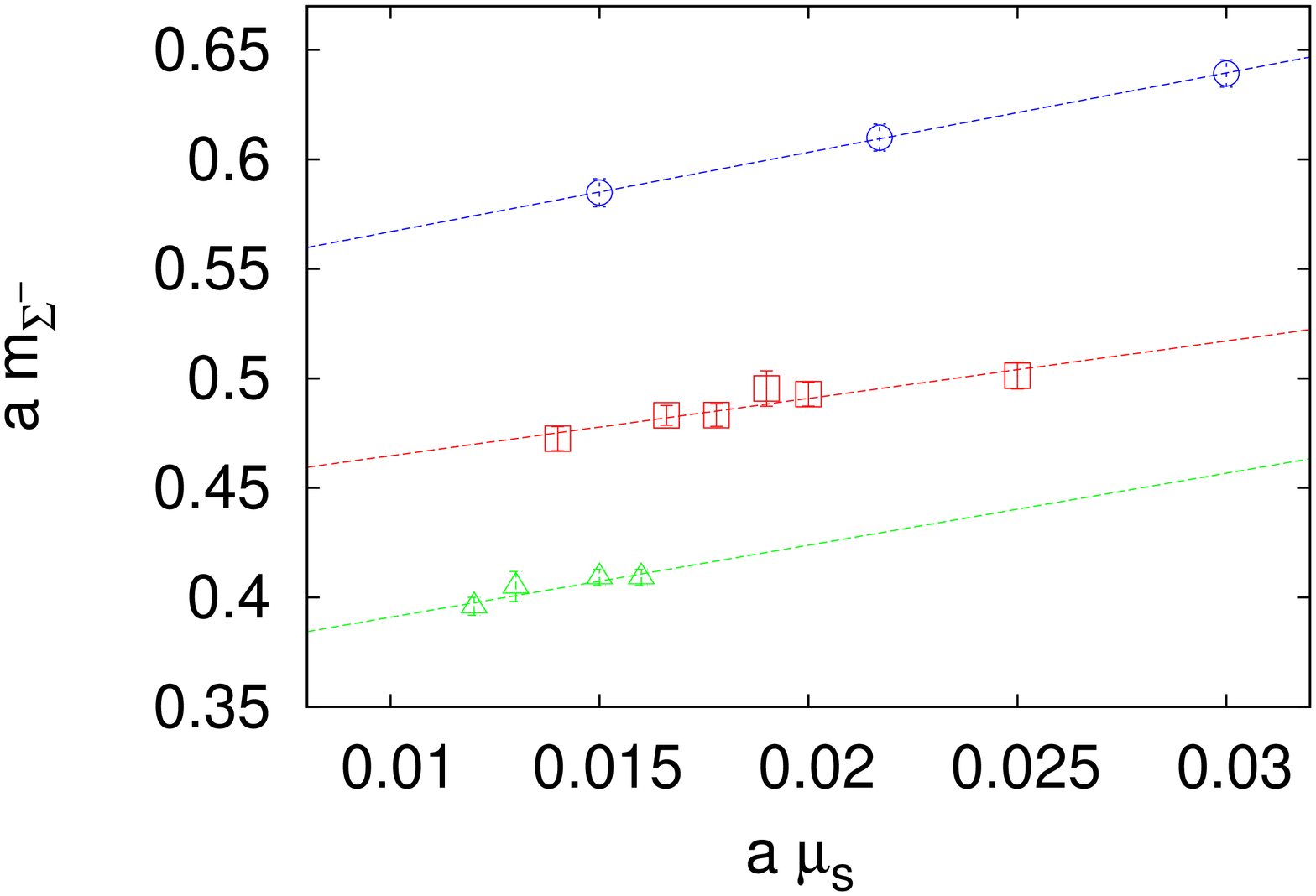} &
\includegraphics[width=.33\textwidth, angle=270]{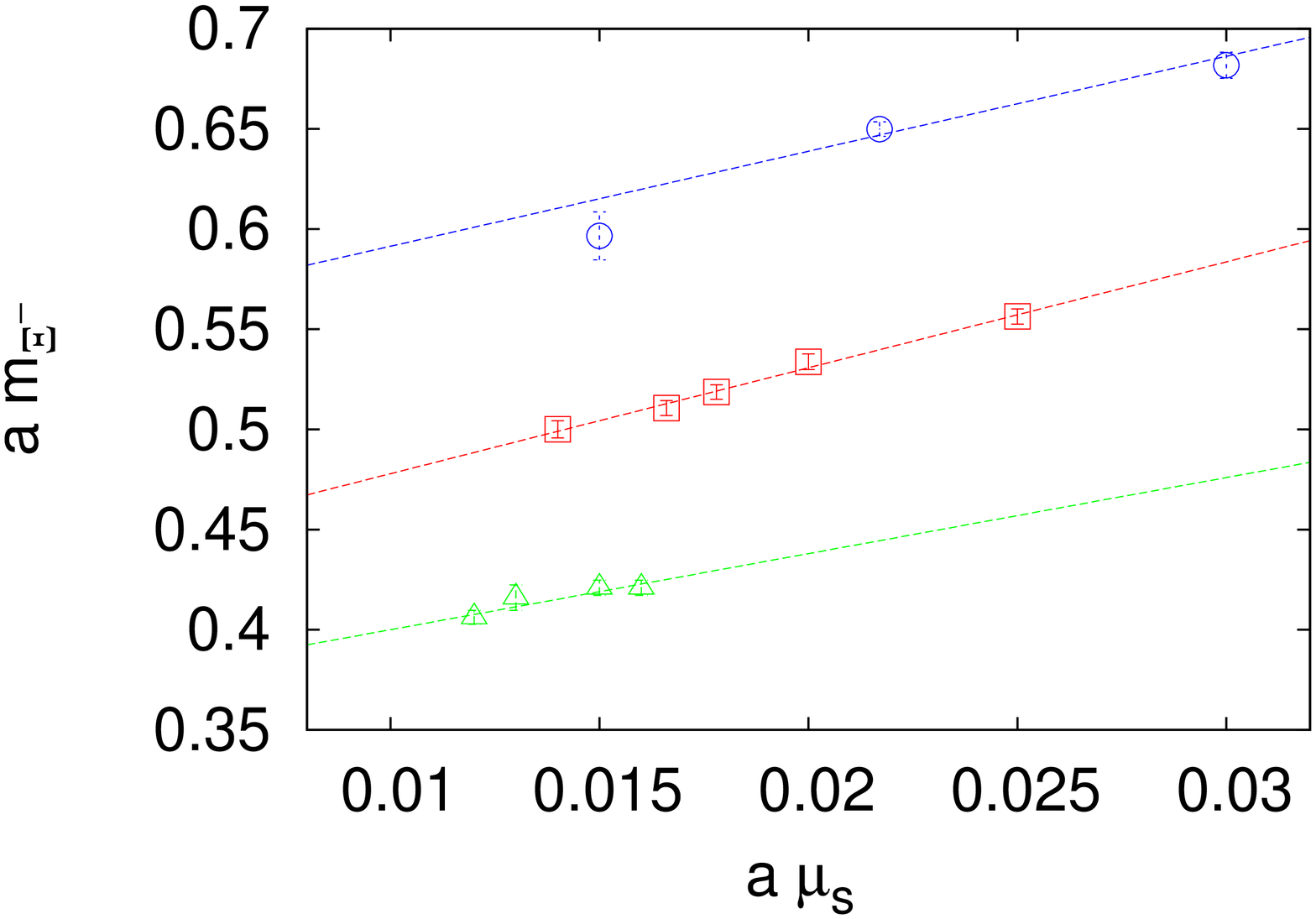} \\
\includegraphics[width=.33\textwidth, angle=270]{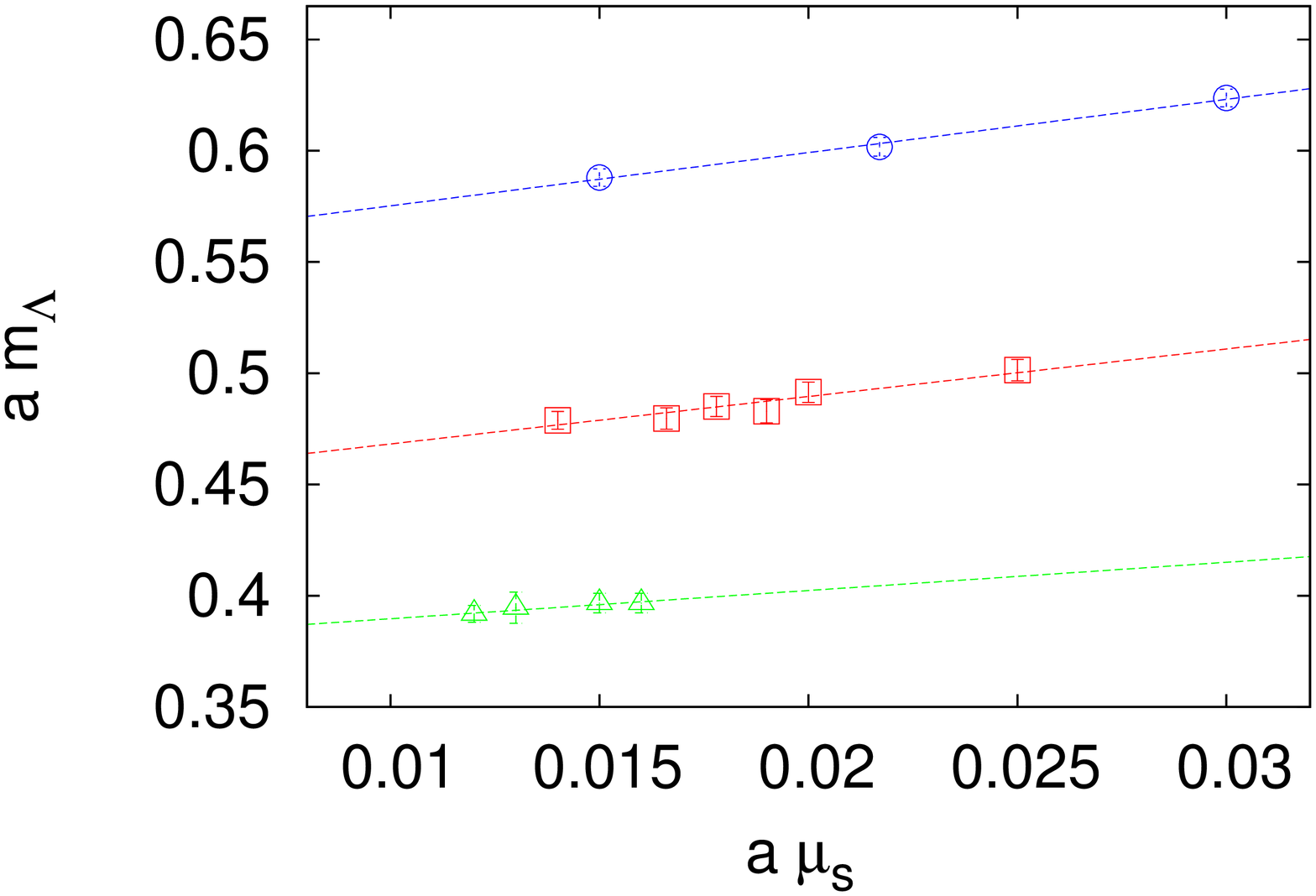} &
\includegraphics[width=.33\textwidth, angle=270]{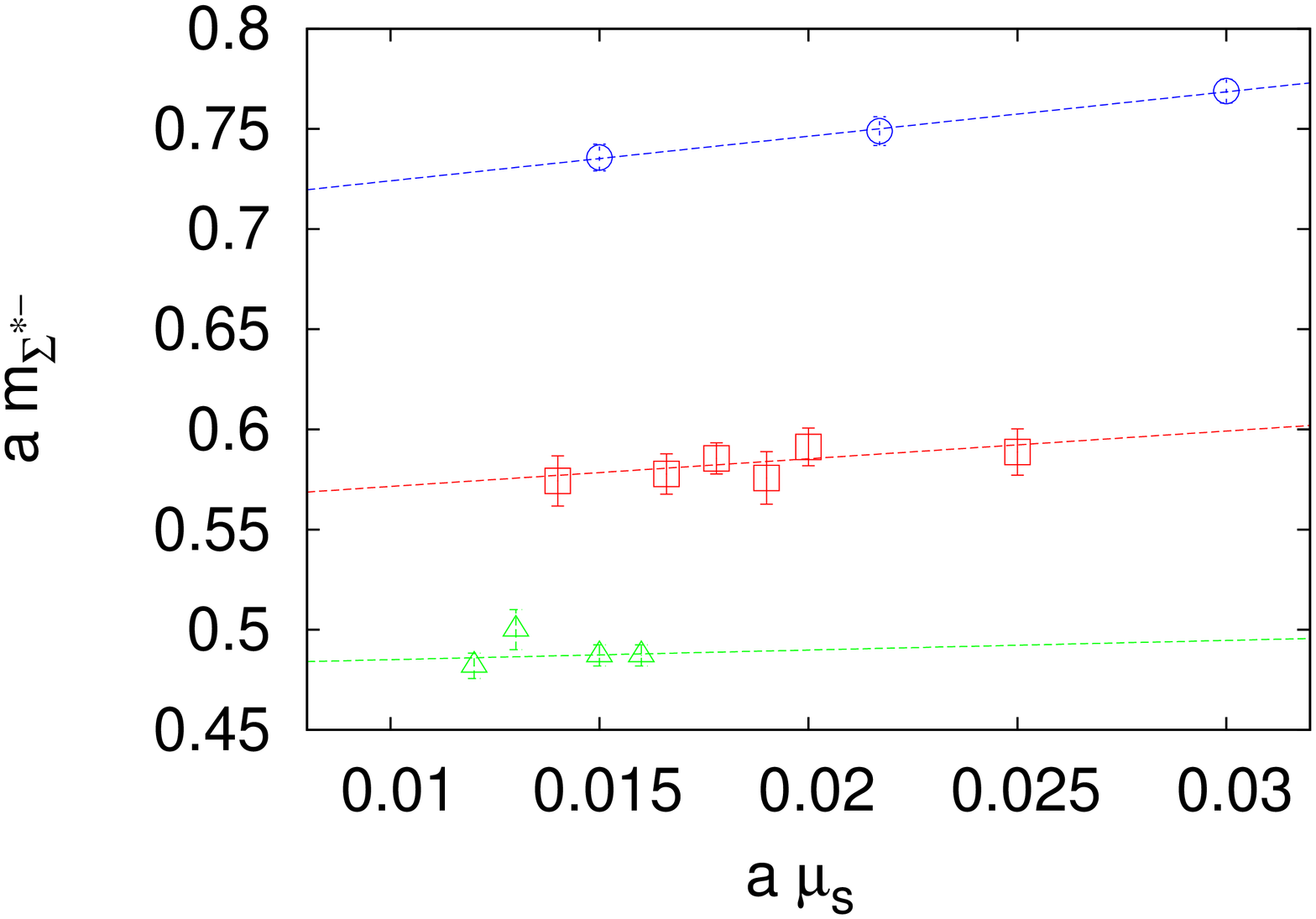} \\
\includegraphics[width=.33\textwidth, angle=270]{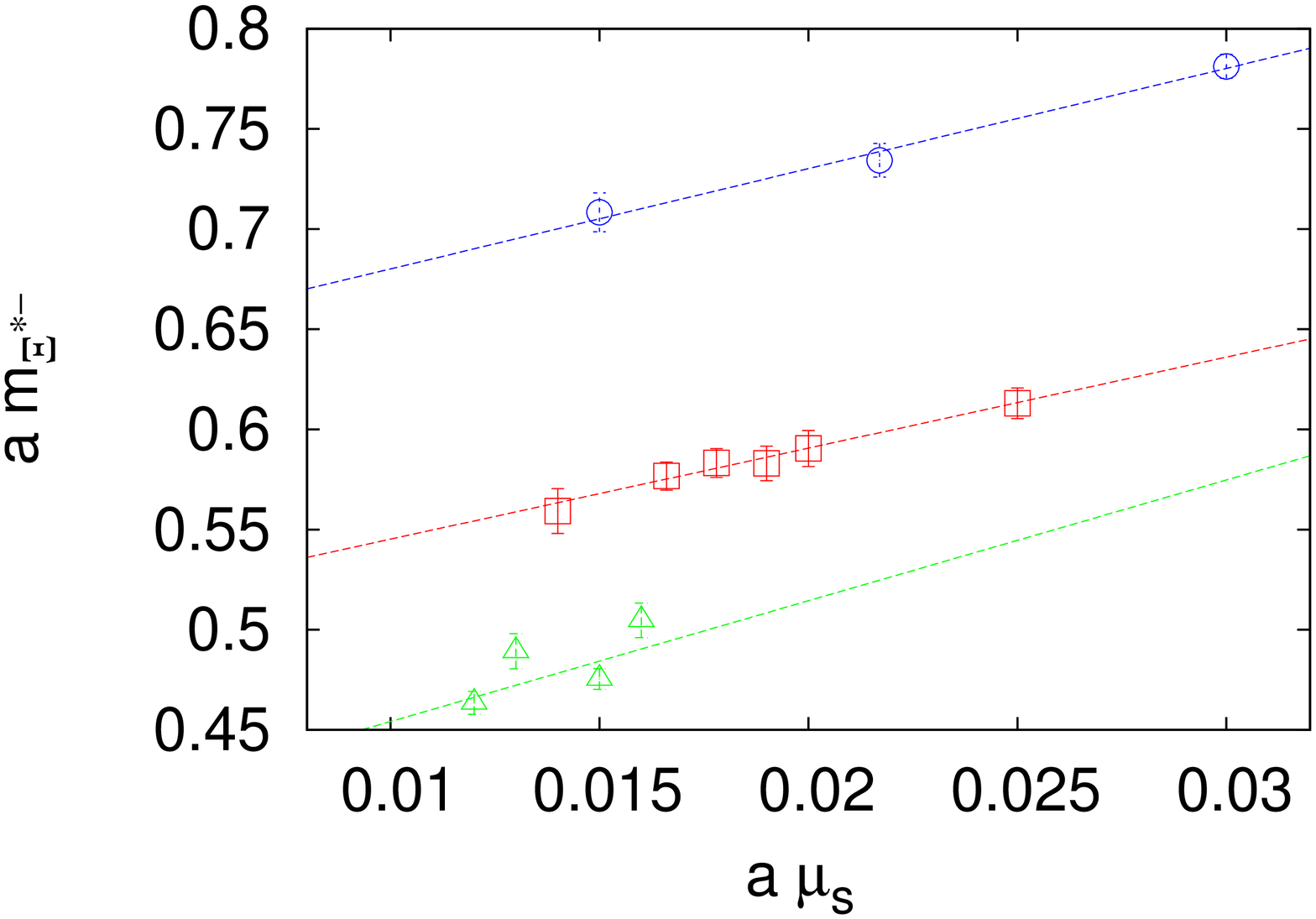} &
\includegraphics[width=.33\textwidth, angle=270]{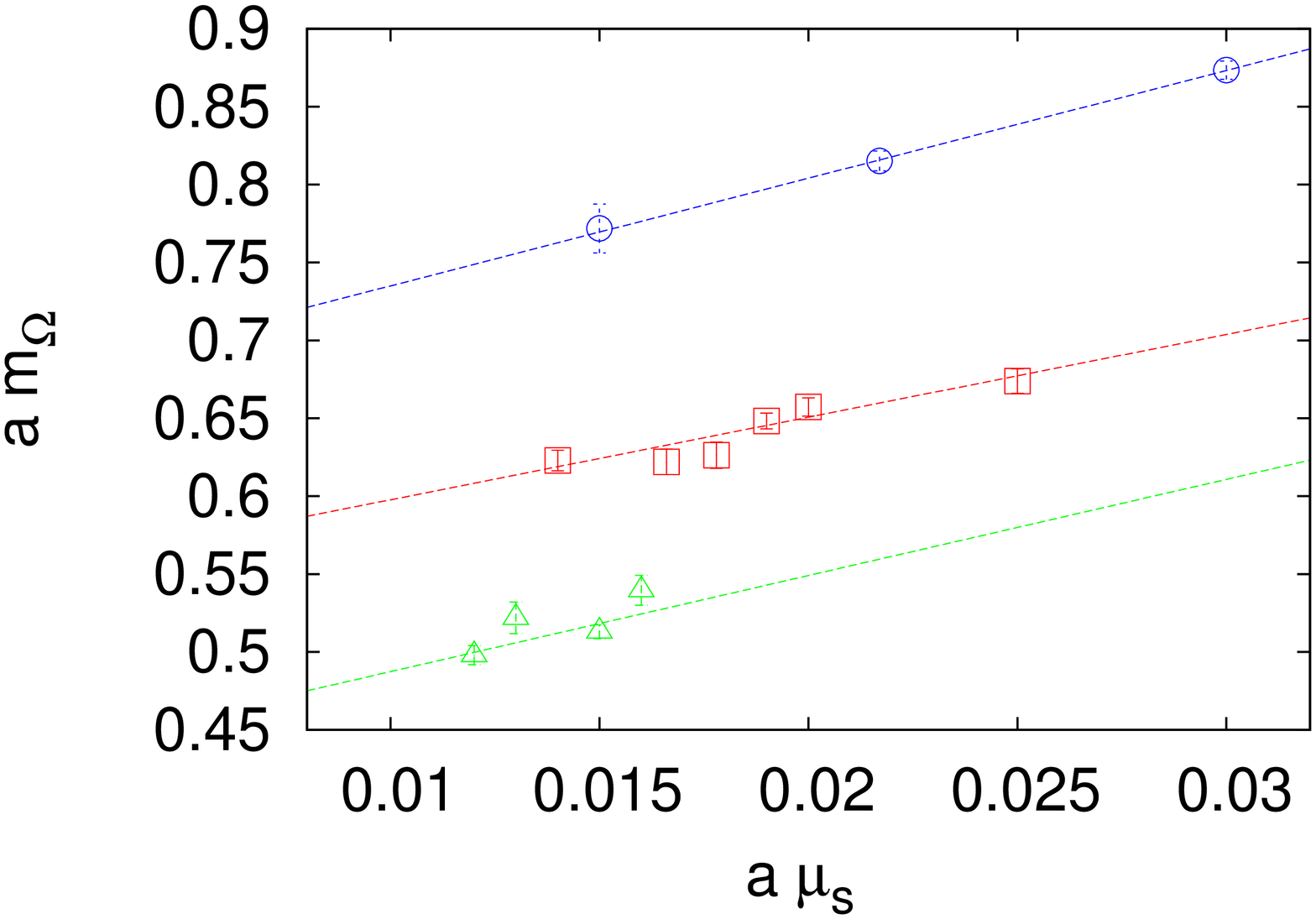} \\
\end{tabular}
\caption{Baryon masses versus  the strange quark mass for $\beta=3.9$ and $a \mu_l=0.0064$ (circles),
$\beta=4.05$ and $a \mu_l=0.006$ (squares), $\beta=4.2$ and $a \mu_l=0.0065$ (triangles).}
\label{fig:ms dependence}
\end{figure}

\begin{figure}[h!]
%\centering
\begin{tabular}{cc}
\includegraphics[width=.33\textwidth, angle=270]{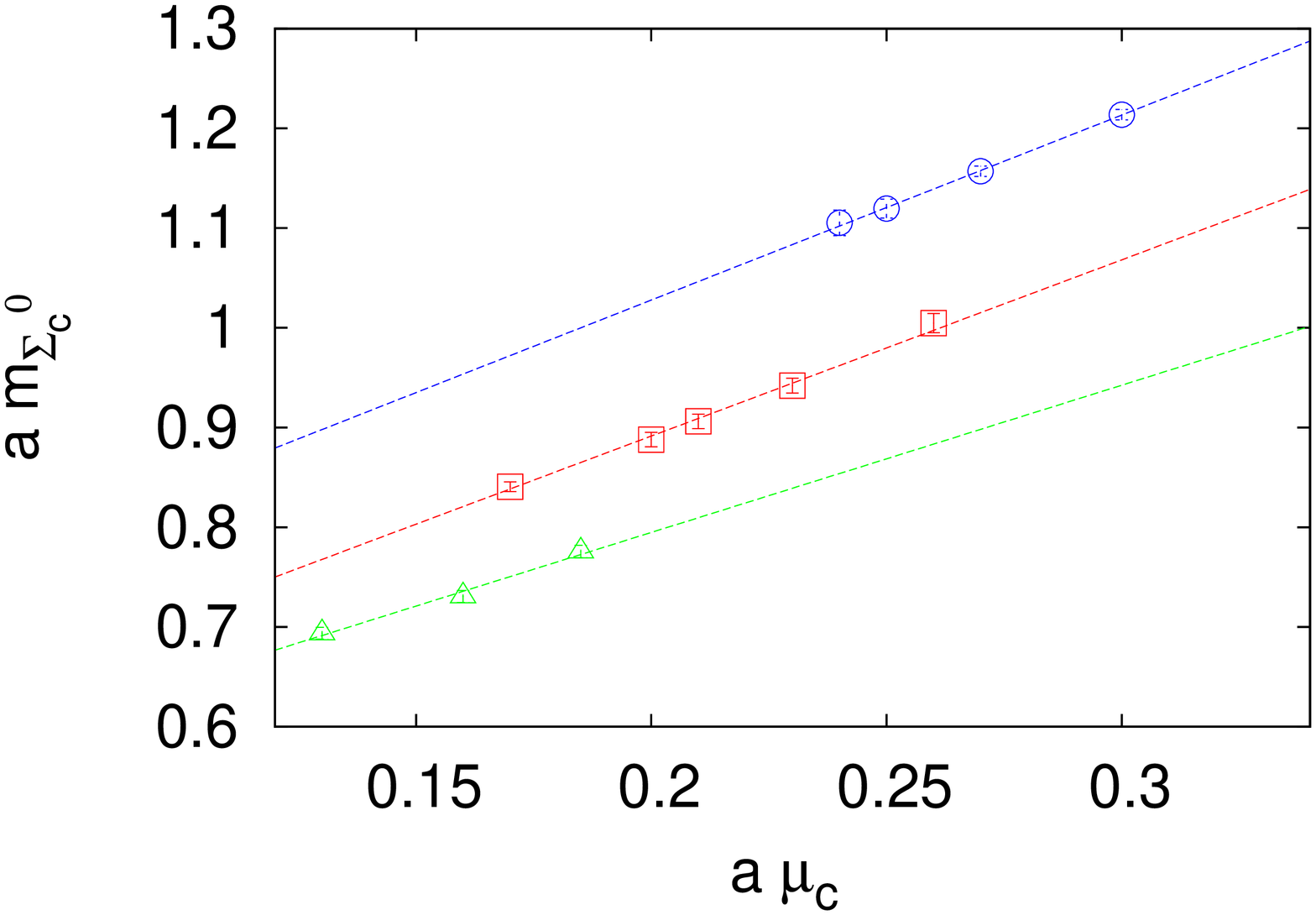} &
\includegraphics[width=.33\textwidth, angle=270]{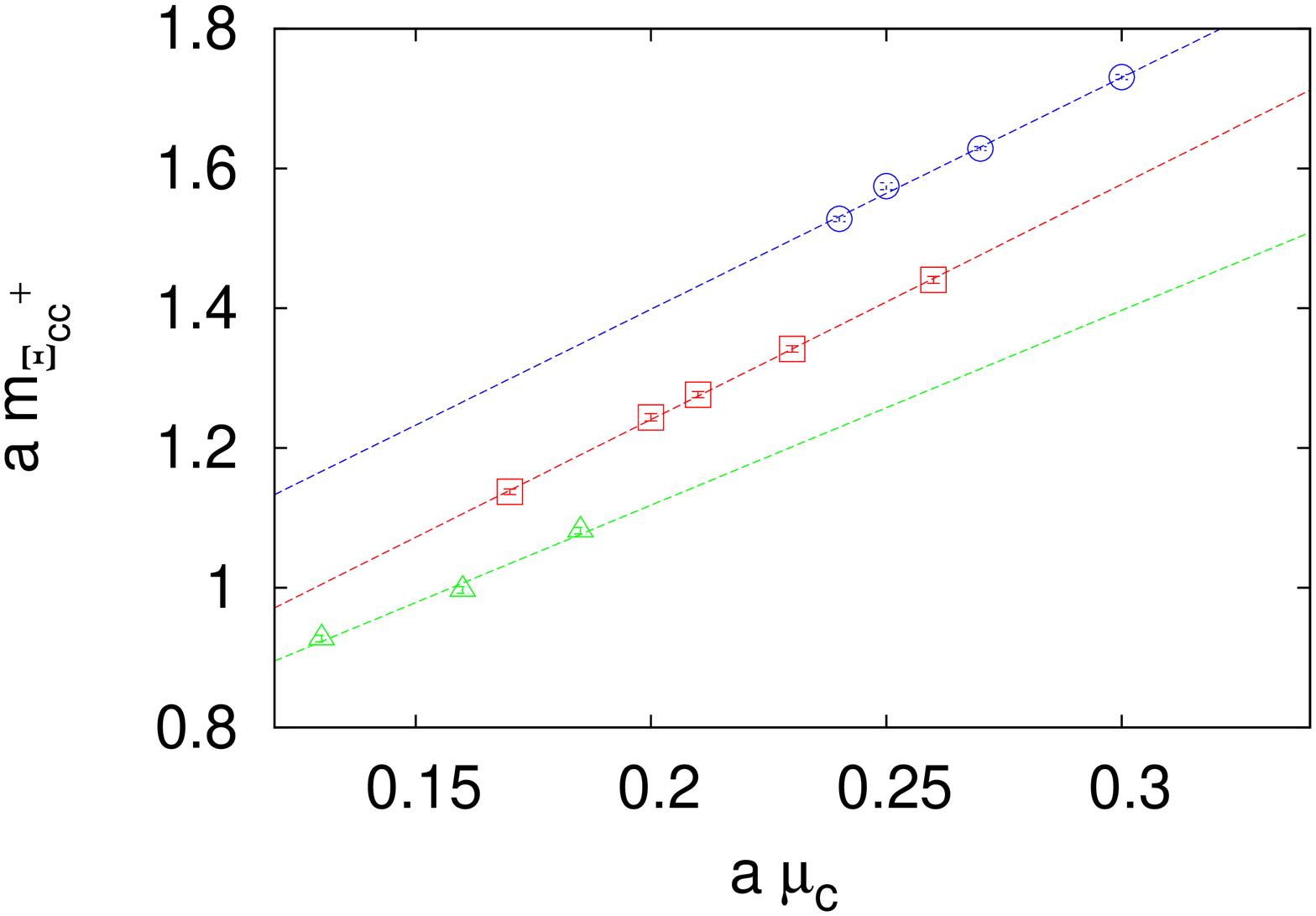} \\
\includegraphics[width=.33\textwidth, angle=270]{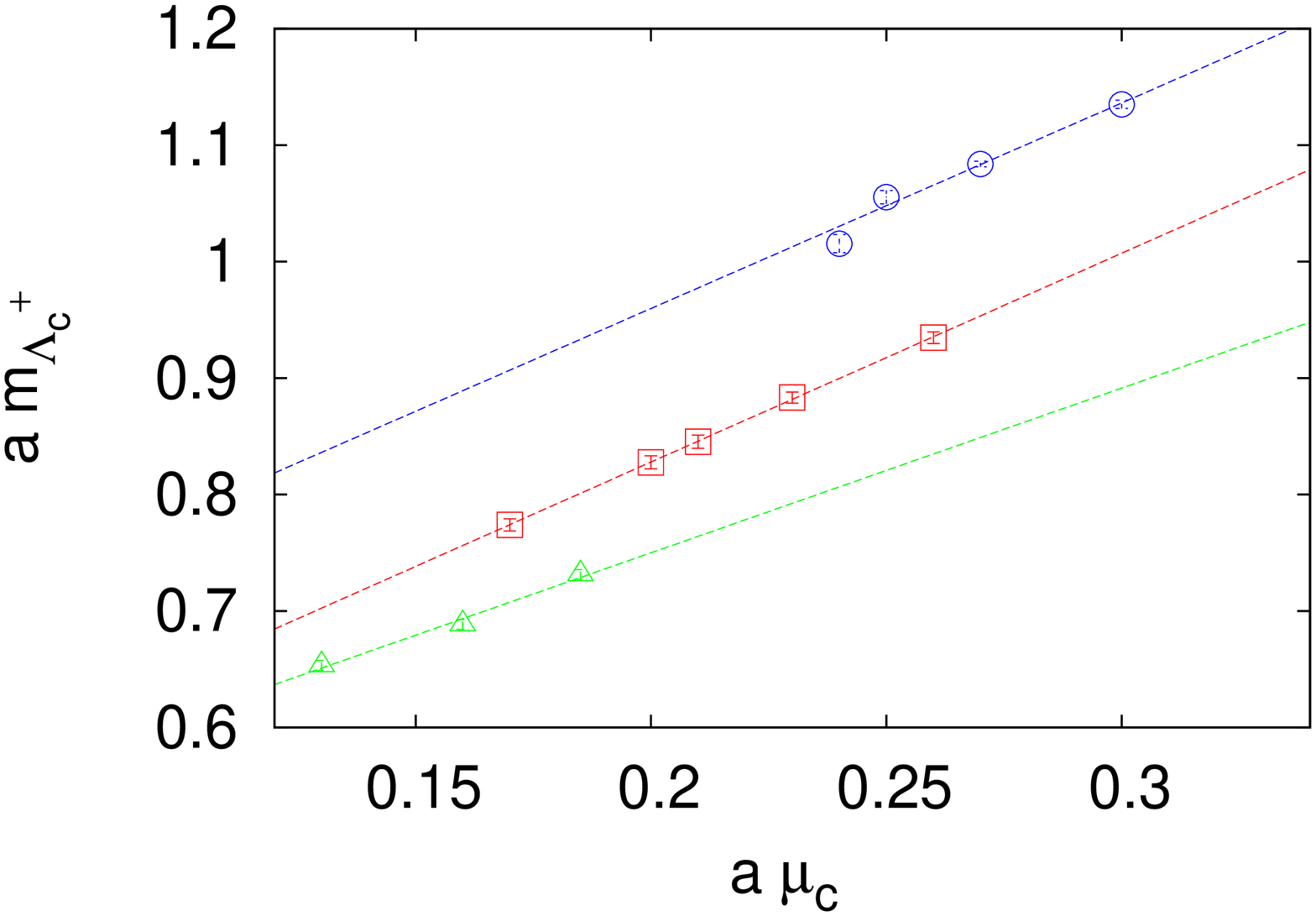} &
\includegraphics[width=.33\textwidth, angle=270]{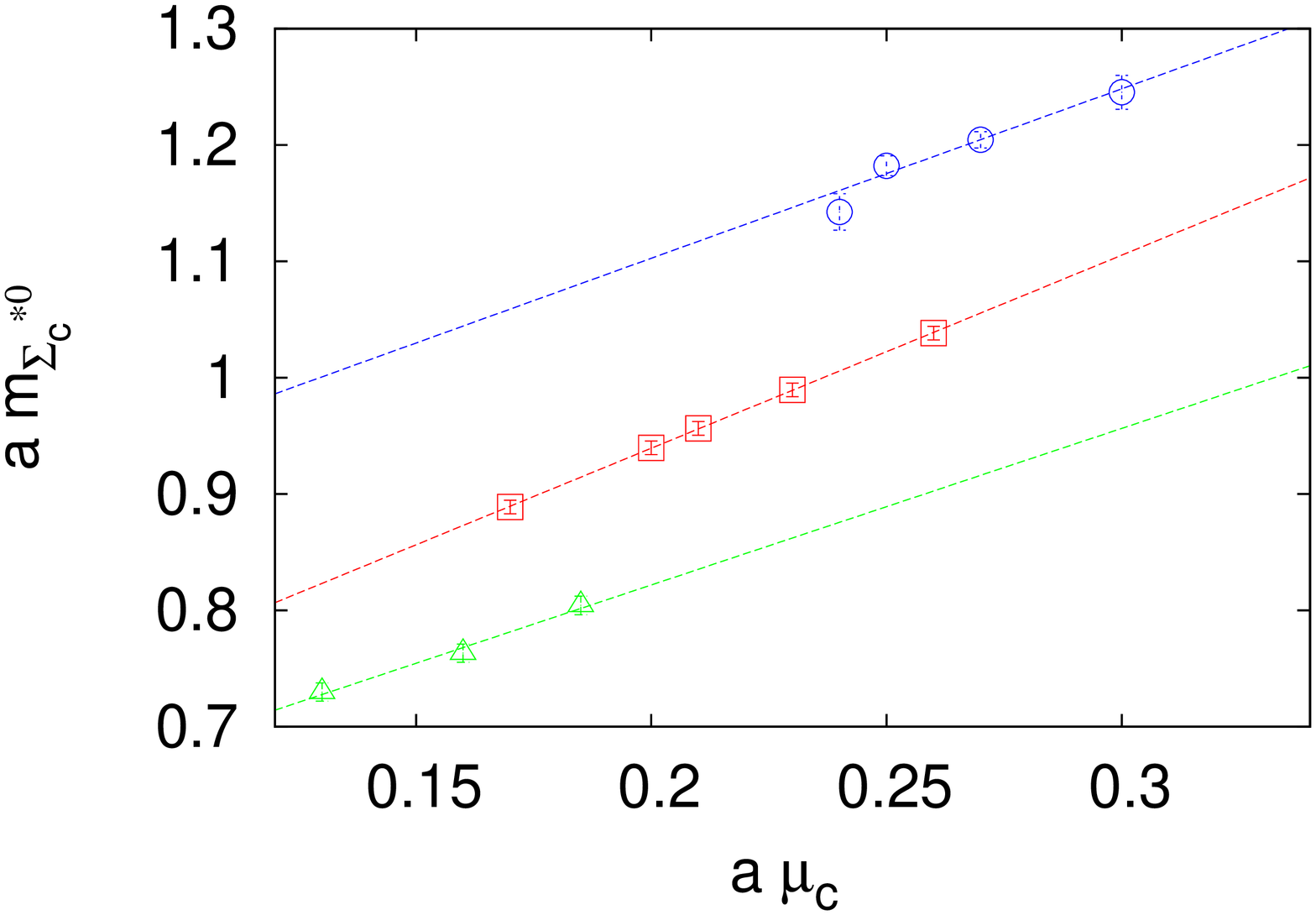} \\
\includegraphics[width=.33\textwidth, angle=270]{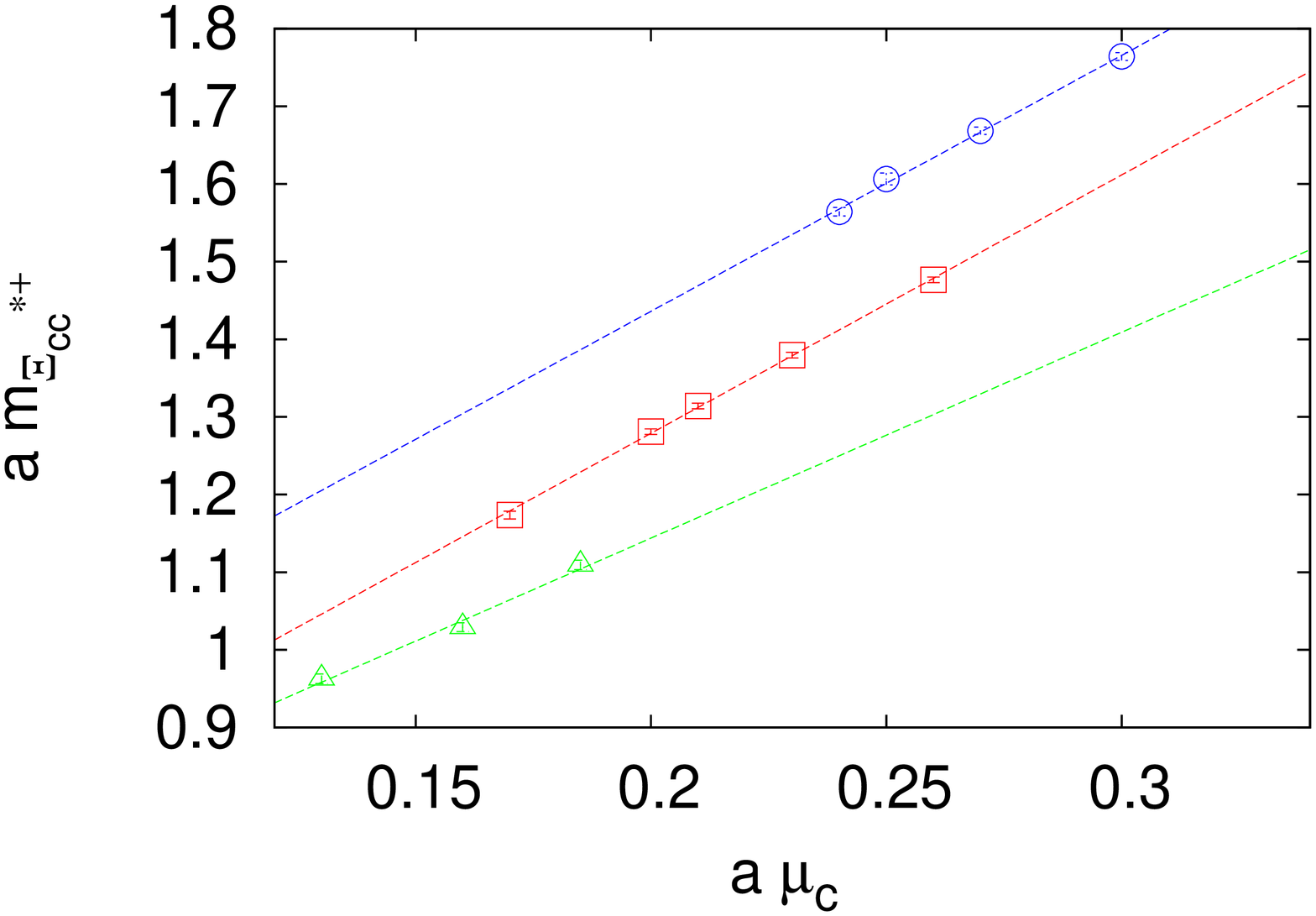} &
\includegraphics[width=.33\textwidth, angle=270]{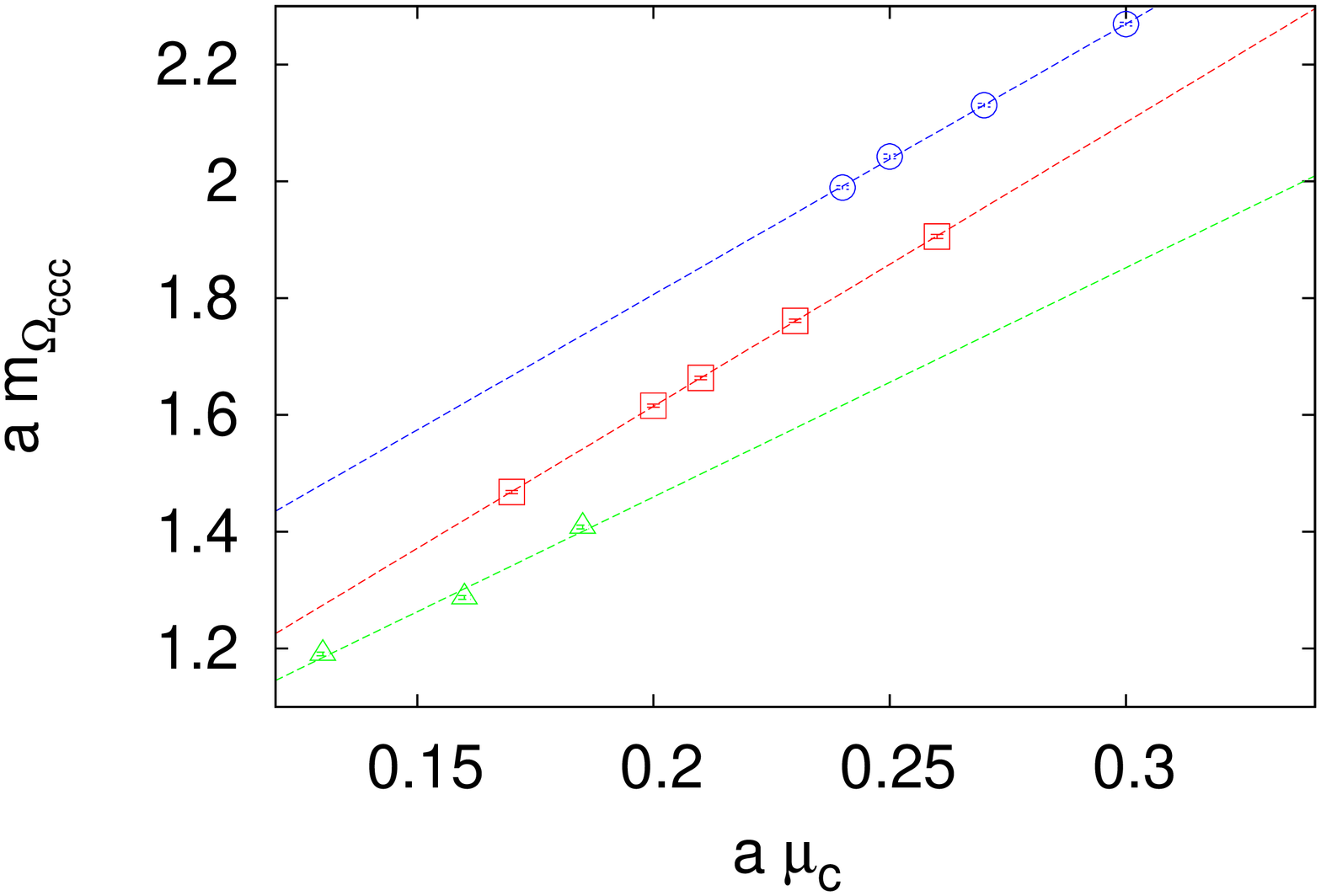} \\
\end{tabular}
\caption{Baryon masses versus the charm quark mass for $\beta=3.9$ and $a \mu_l=0.004$ (circles),$\beta=4.05$ and $a \mu_l=0.003$ (squares), $beta=4.2$ and $a \mu_l=0.0065$ (triangles).
}
\label{fig:mc dependence}
\end{figure}

In Figs.~\ref{fig:ms dependence} and \ref{fig:mc dependence} we show the dependence of the strange and charm baryon masses on the the strange and charm quark mass, respectively.
 Overall, the data display a linear dependence on  both the
strange  and charm quark mass. One can therefore  interpolate between different
values of quark masses, if needed.
 
\subsection{Strange baryon mass with strange quark mass tuned to its physical value}

In this section we restrict our analysis only to the subset of data
obtained at the tuned values of
the strange quark mass. 
Namely,  for $\beta=3.9$ and $\beta=4.05$, we use the
tuned value given in Table~\ref{tab:tuned quarks}, whereas
for $\beta=4.2$ we use $a \mu_s=0.015$
which agrees with the tuned strange quark mass within error bars.

\begin{figure}[h!]
\includegraphics[width=.4\textwidth, angle=-90]{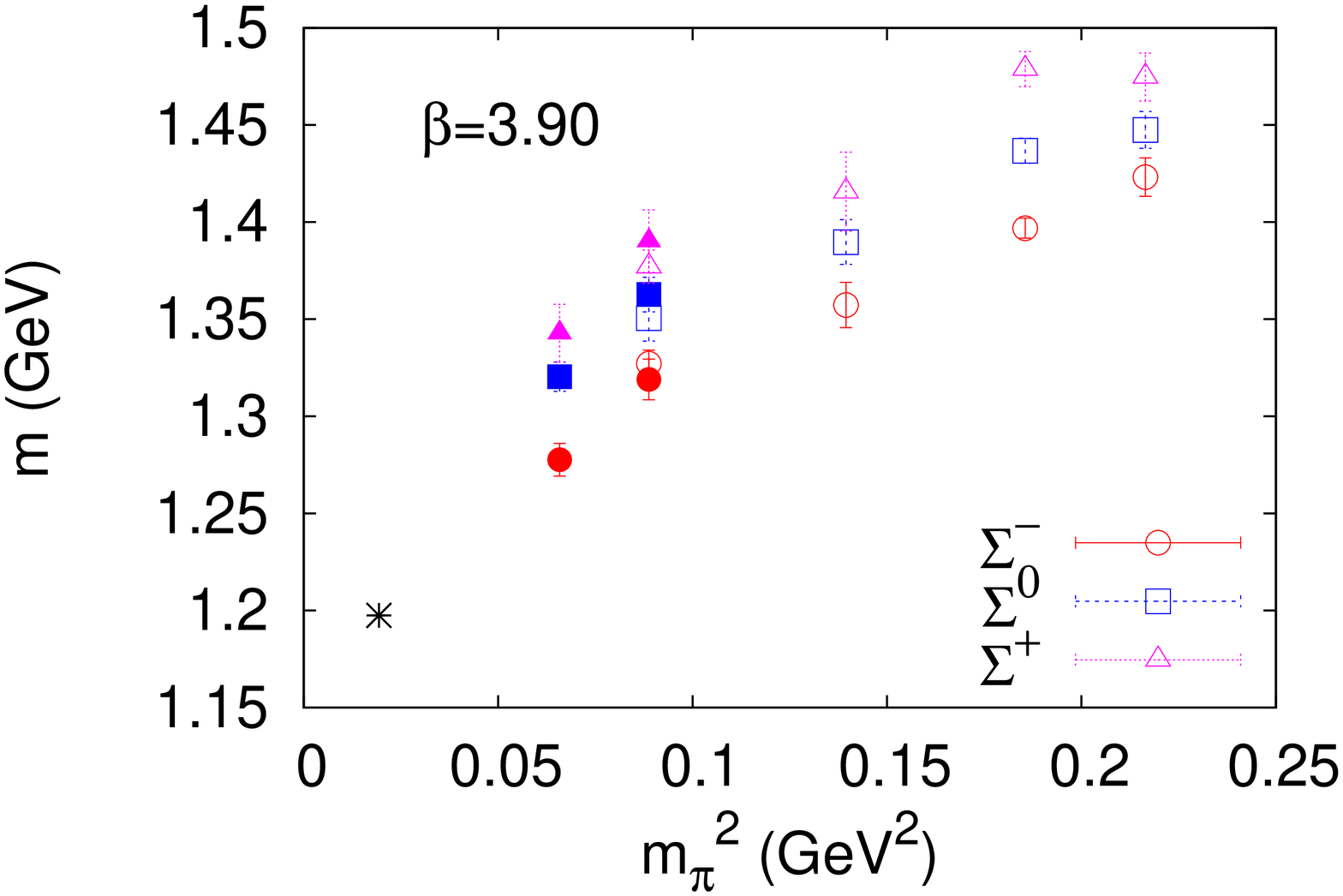} \\
\includegraphics[width=.4\textwidth, angle=-90]{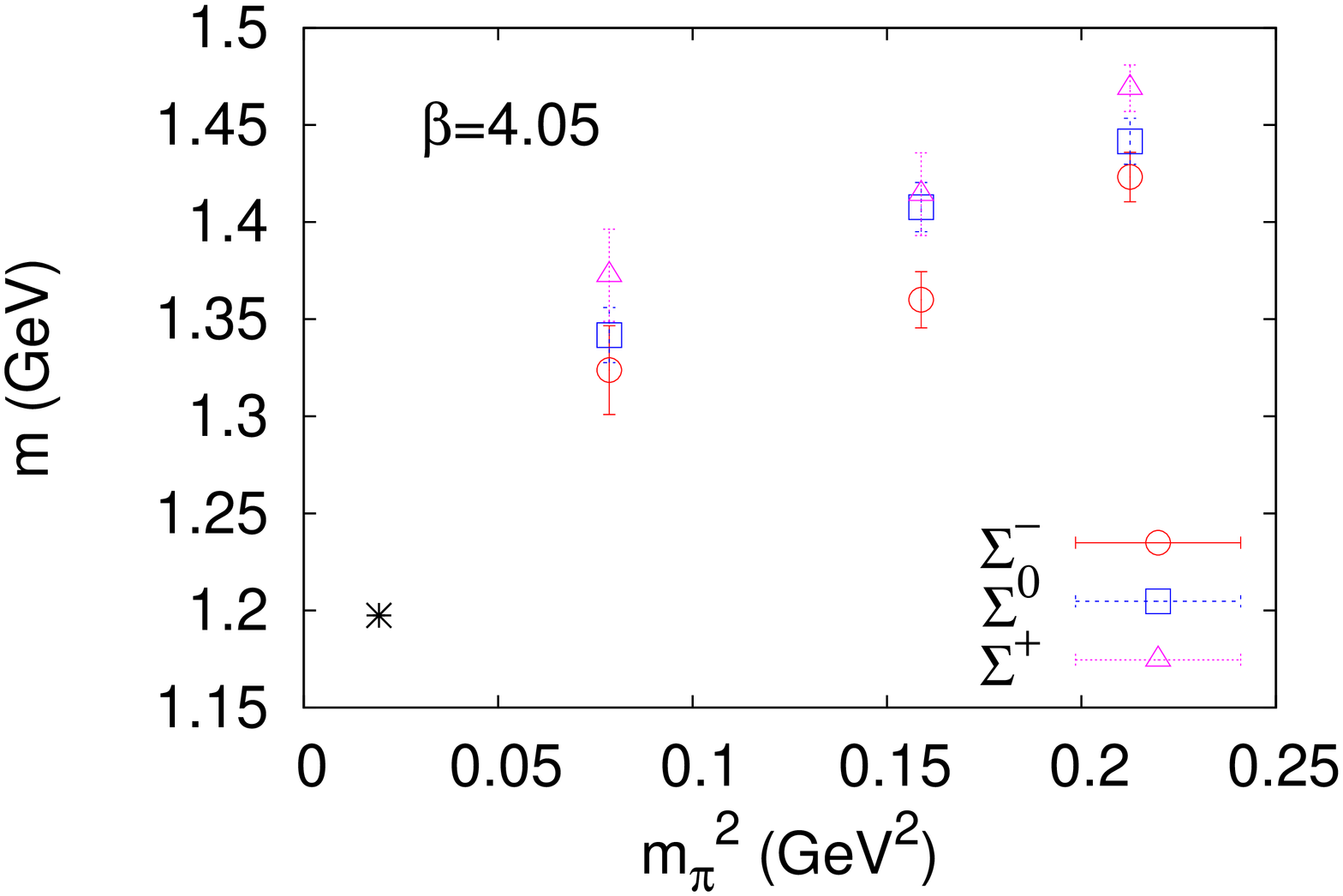} \\
\includegraphics[width=.4\textwidth, angle=-90]{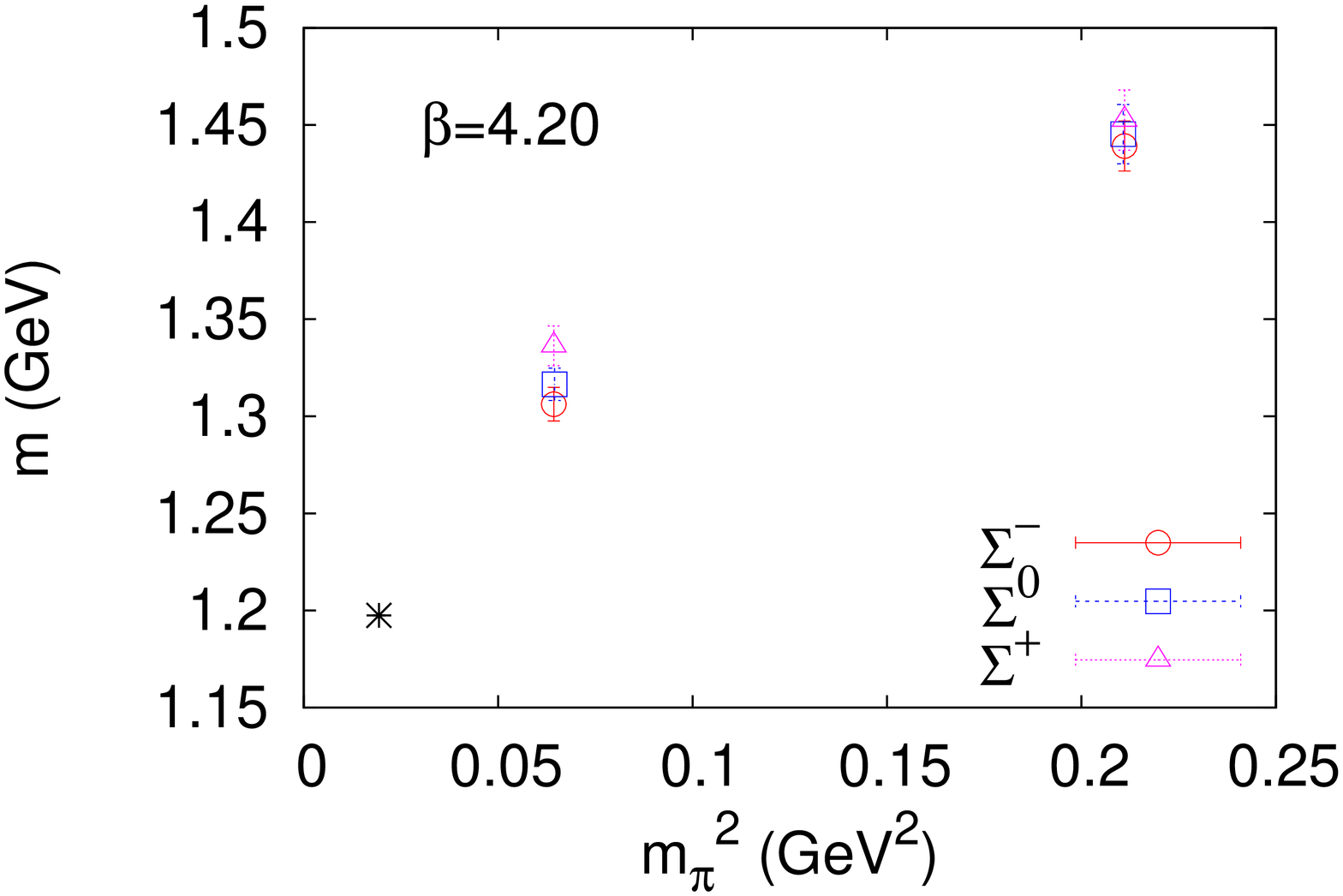}
\caption{The masses of
$\Sigma^+$, $\Sigma^0$ and $\Sigma^-$ at  $\beta=3.9,\, 4.05$ and 4.2. Filled symbols at $\beta=3.9$
refer to the larger volume.}
\label{fig:isospin}
\end{figure}

\begin{figure}[!h]
\centering
\begin{tabular}{cc}
\includegraphics[width=.3\textwidth, angle=270]{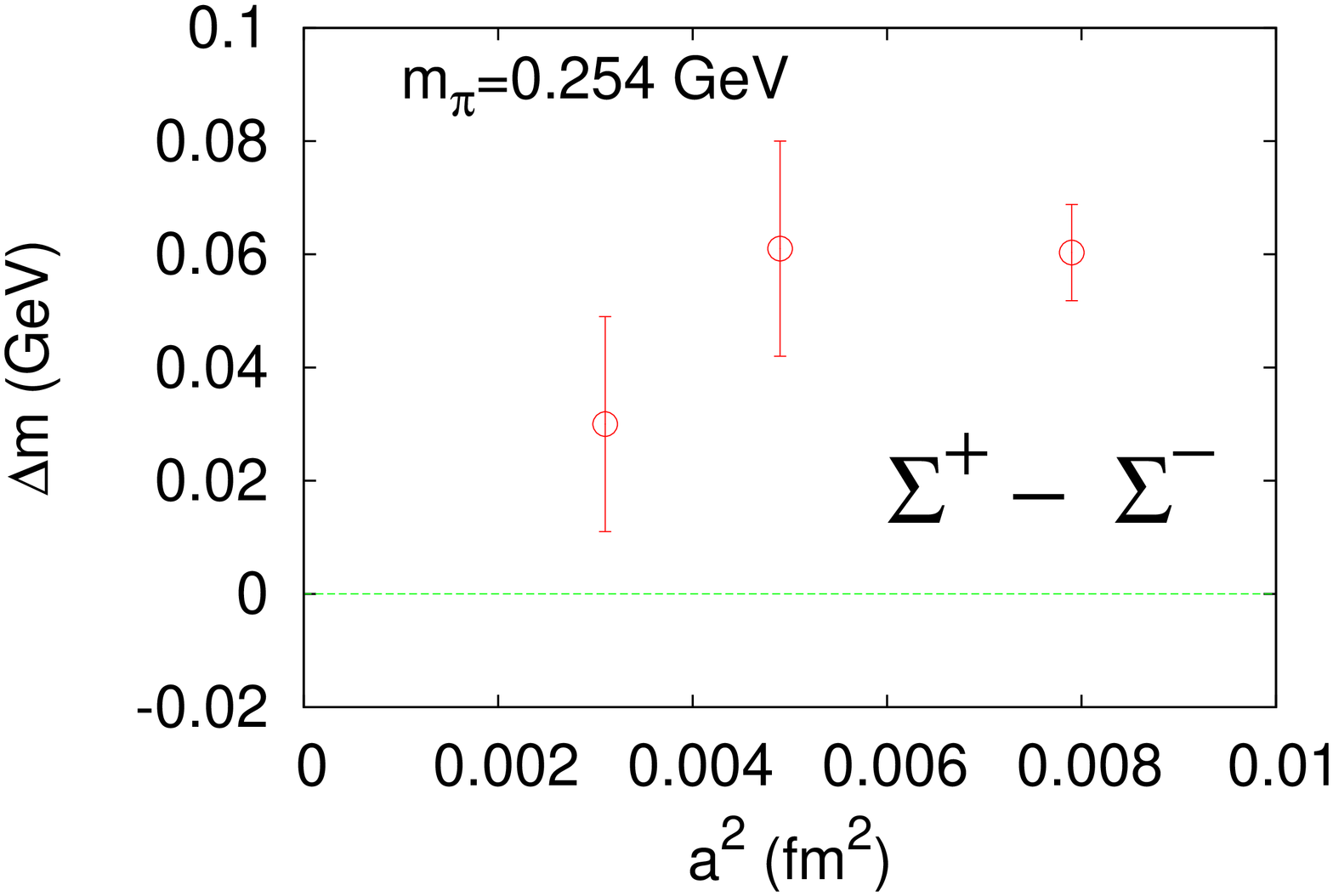} &
\includegraphics[width=.3\textwidth, angle=270]{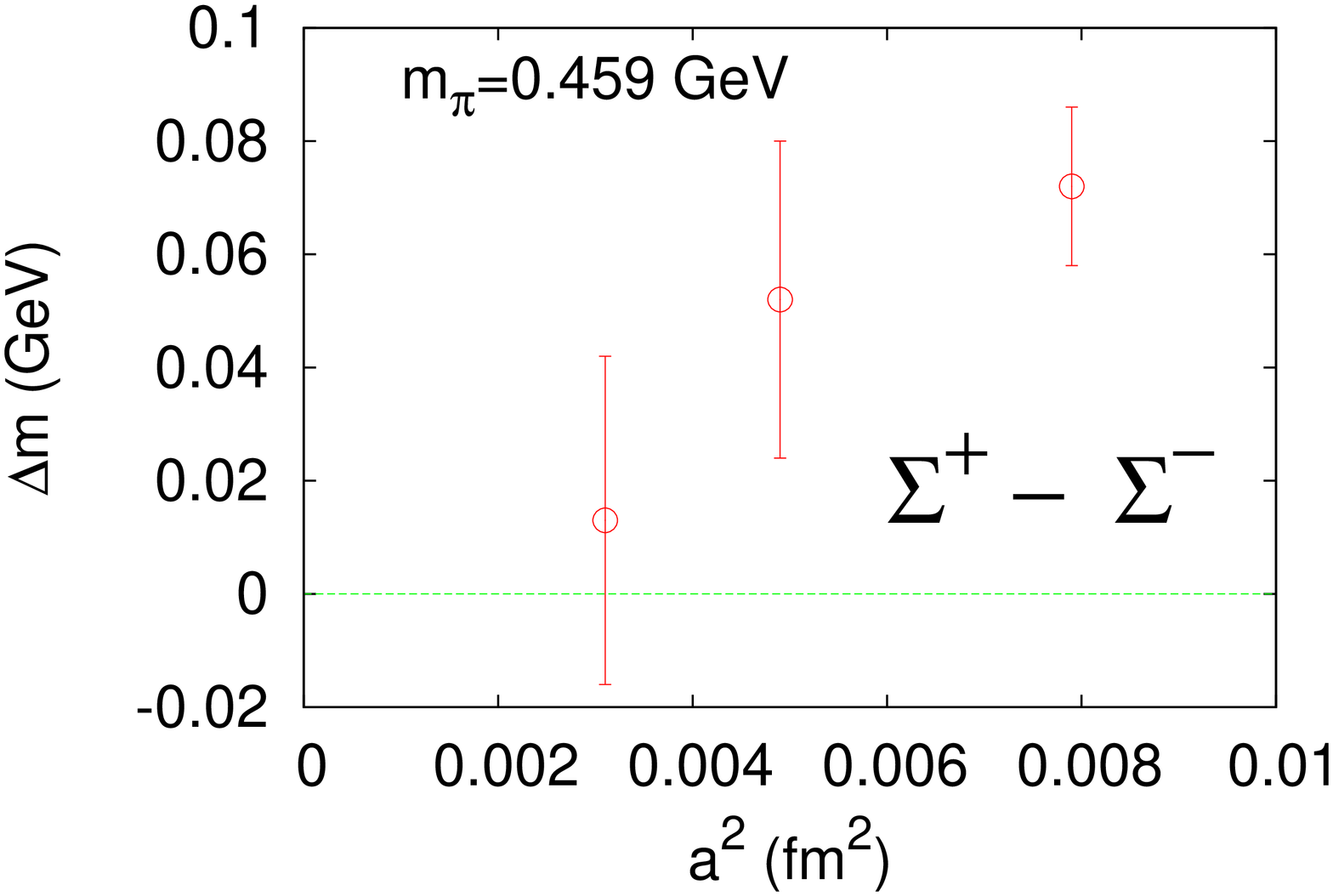} \\
\includegraphics[width=.3\textwidth, angle=270]{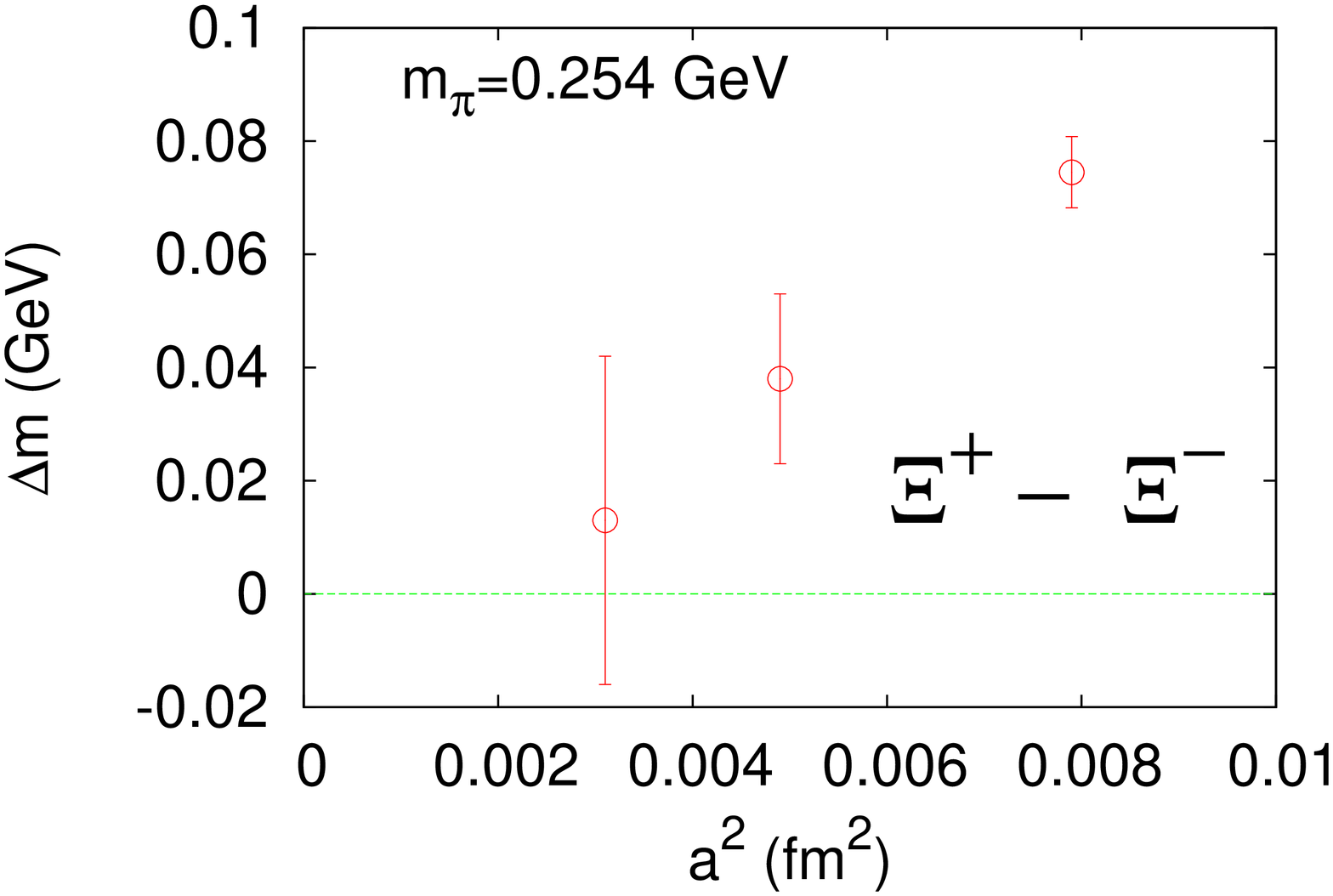} &
\includegraphics[width=.3\textwidth, angle=270]{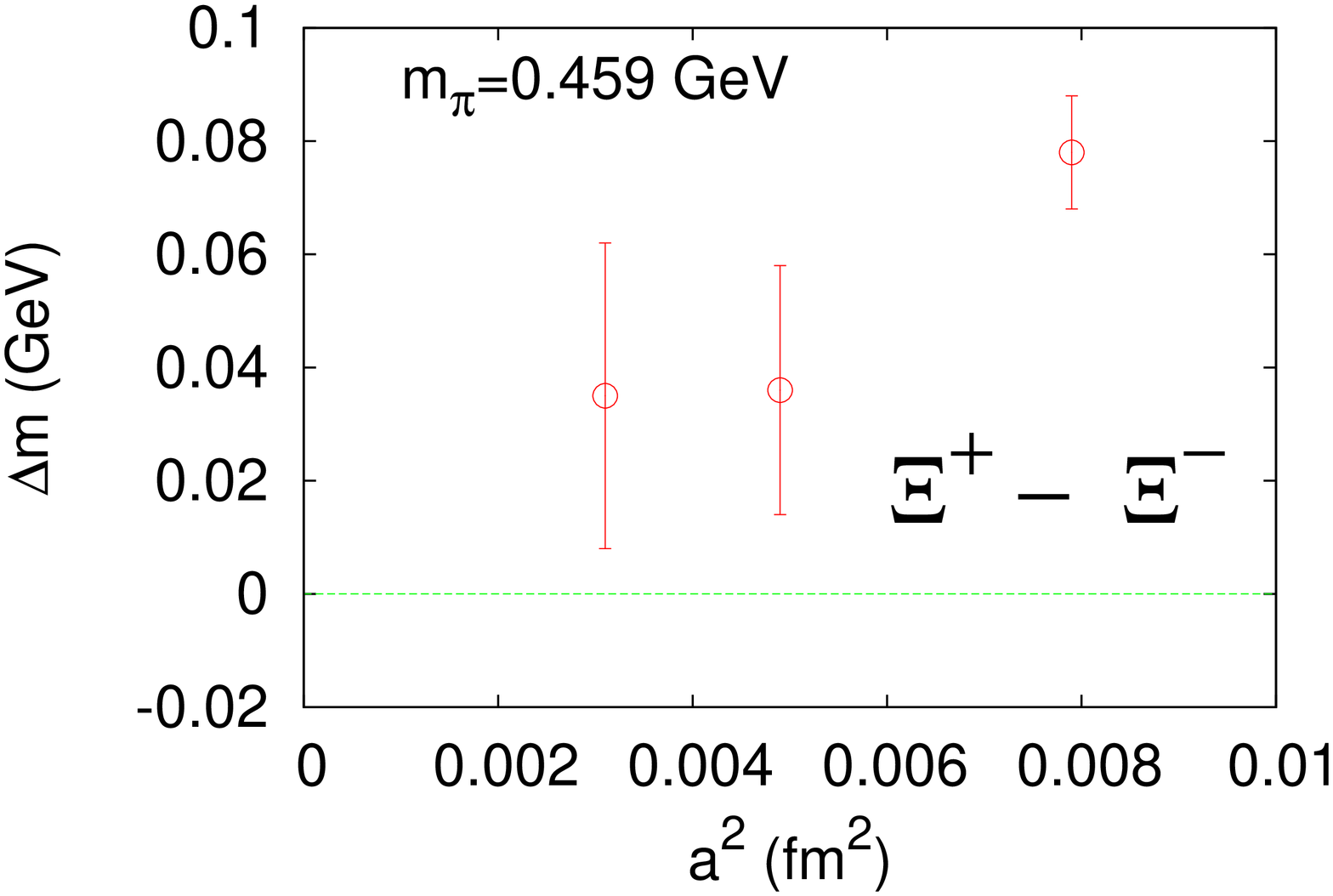} \\
\includegraphics[width=.3\textwidth, angle=270]{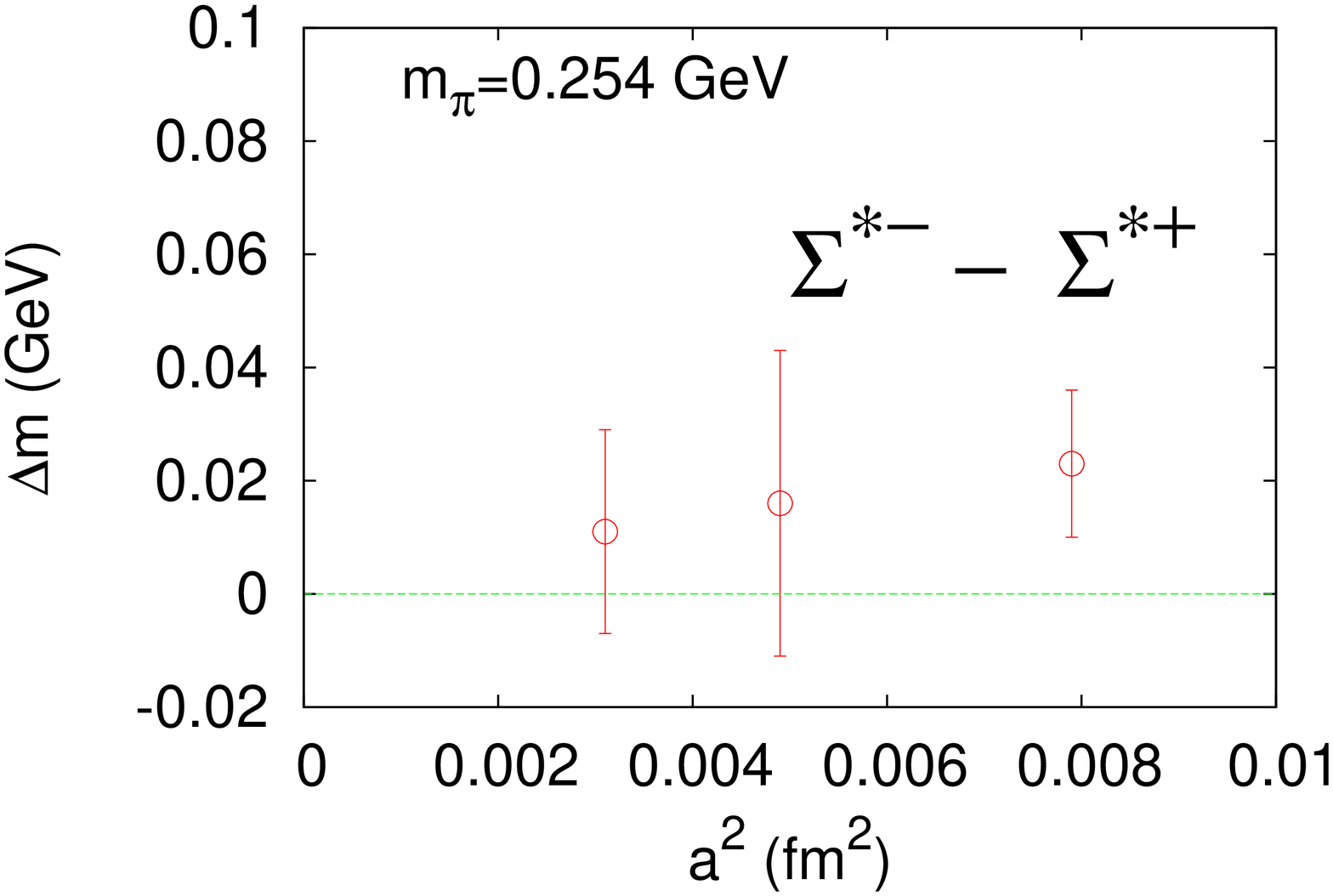} &
\includegraphics[width=.3\textwidth, angle=270]{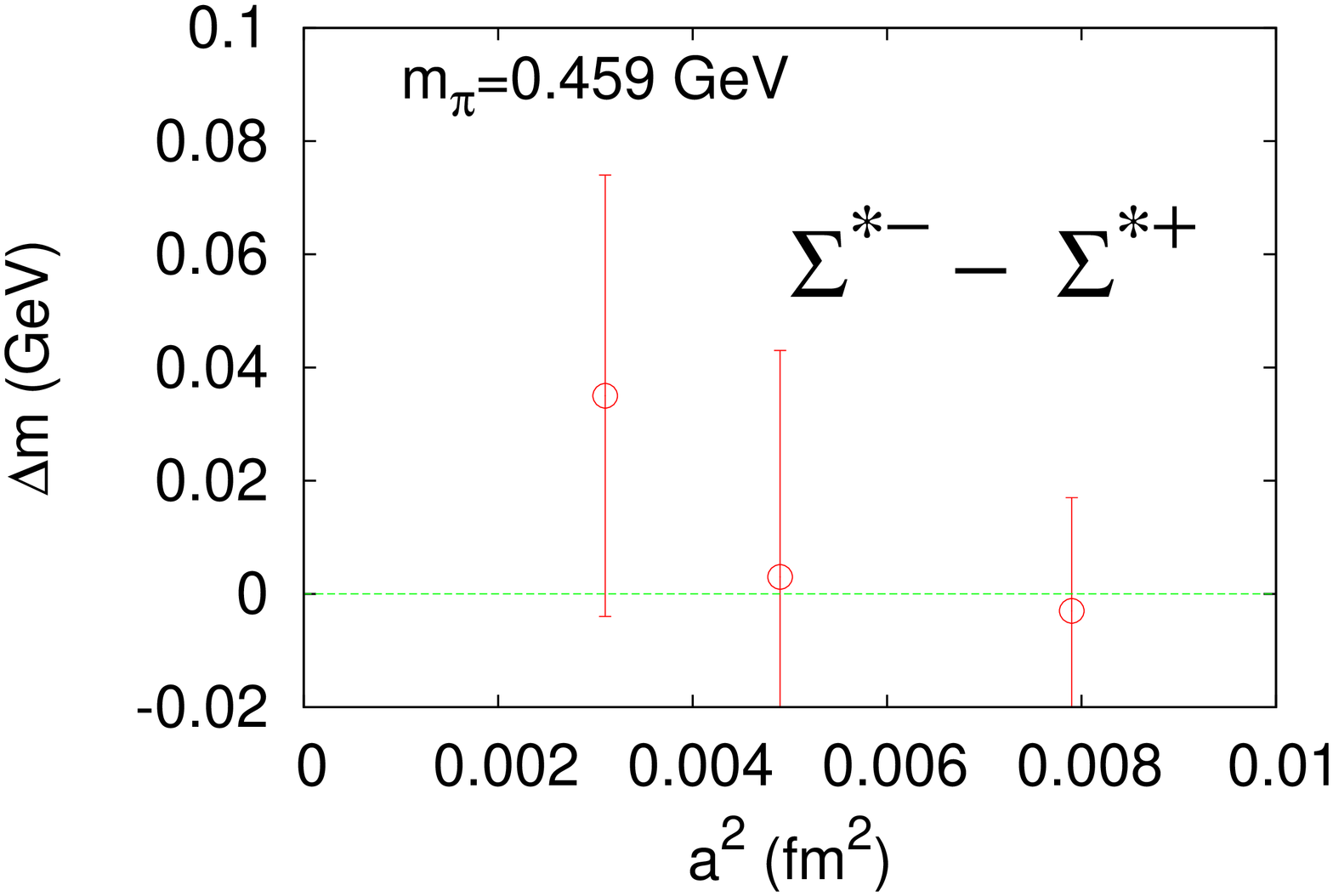} \\
\includegraphics[width=.3\textwidth, angle=270]{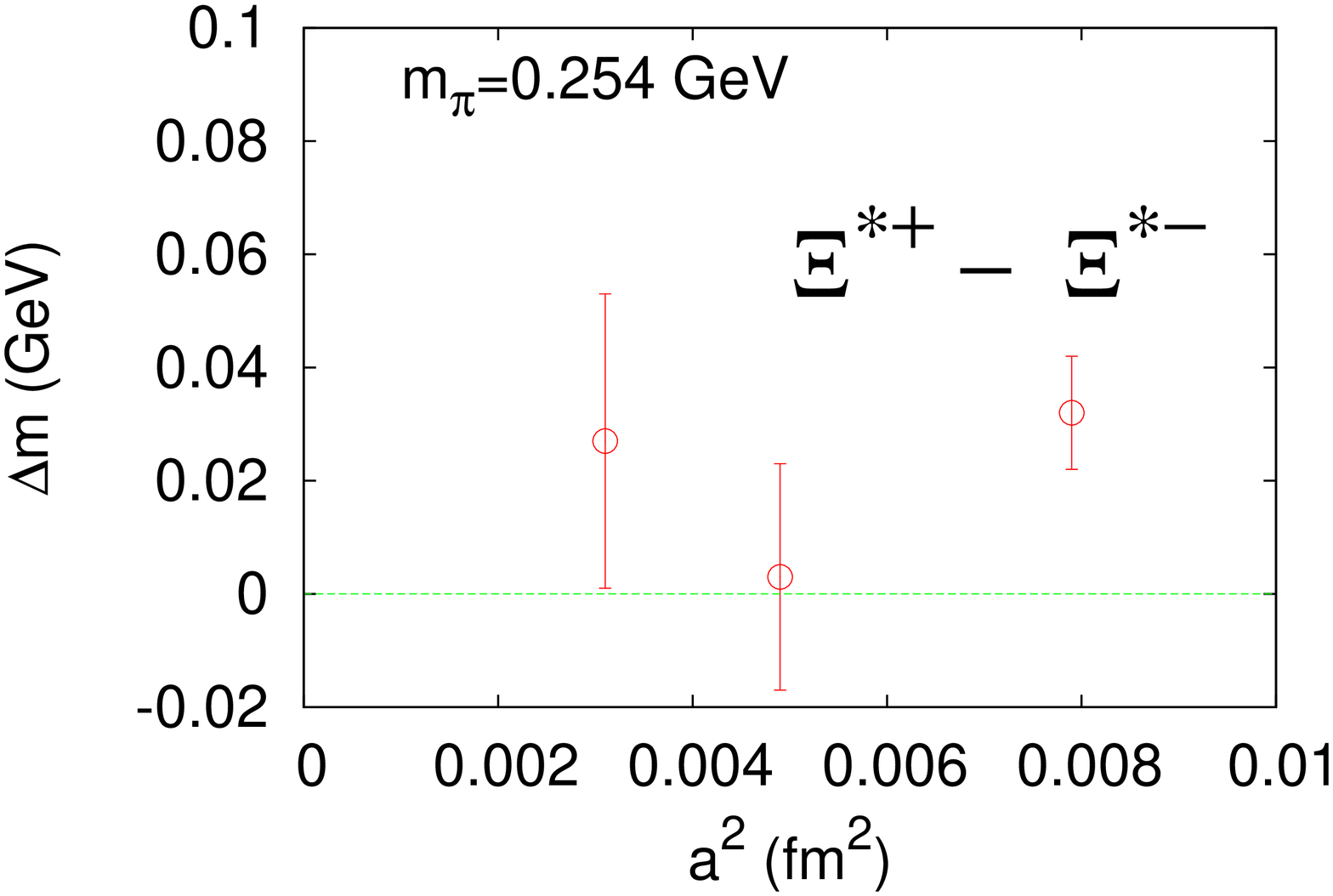} &
\includegraphics[width=.3\textwidth, angle=270]{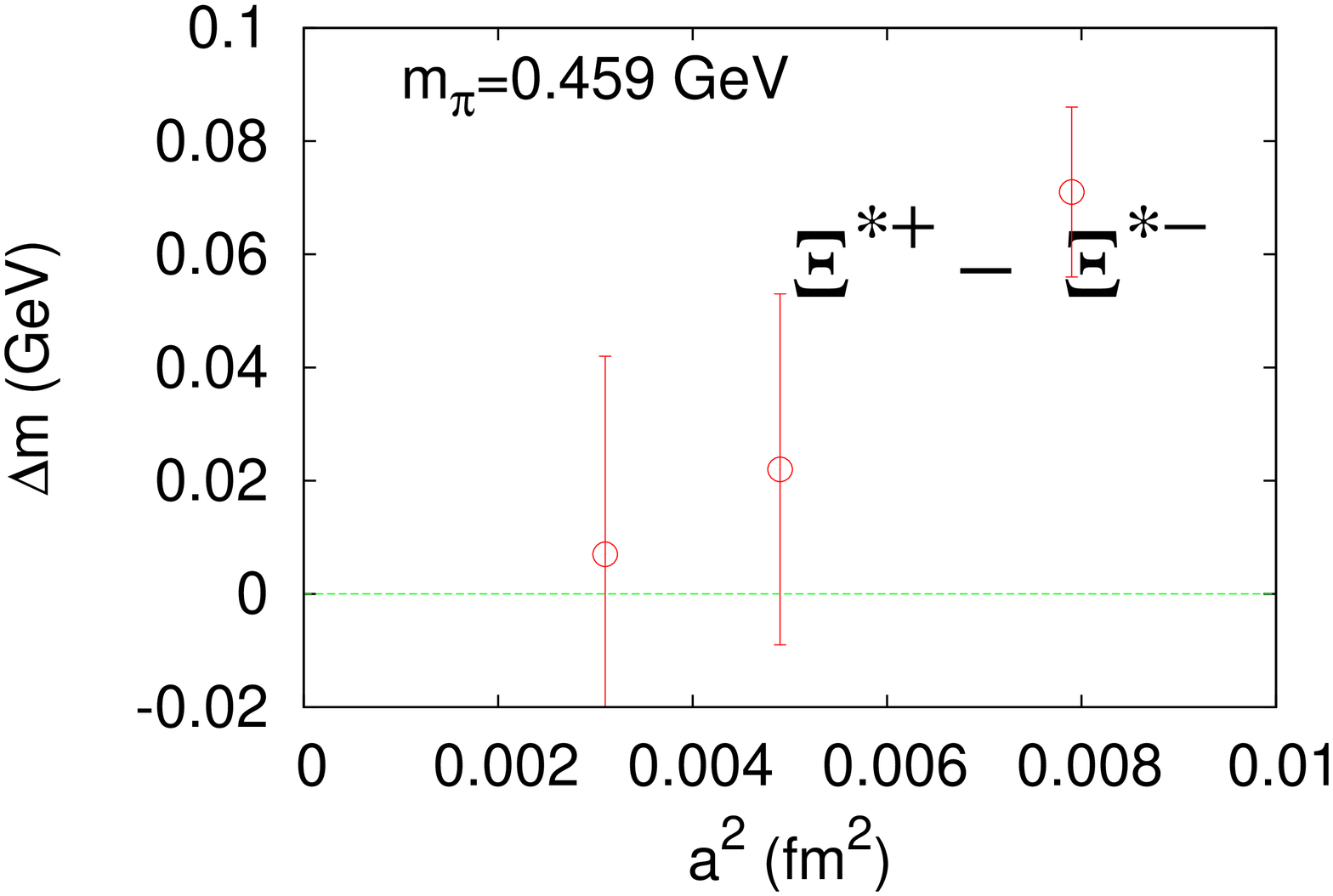} \\
\end{tabular}
\caption{Isospin mass splittings in the strange-quark sector  at the smallest (left)  and largest (right) pion masses used in our computation.}
\label{fig:isospin2}
\end{figure}

It is interesting to examine the degree of isospin splitting as a function
of the lattice spacing. 
The splitting is expected to be 
zero in the continuum limit.
In Fig.~\ref{fig:isospin} we show the mass of $\Sigma^+$, $\Sigma^0$ and $\Sigma^-$ at $\beta=3.9, 4.05$ and 4.2. As expected the mass splitting among the three
charge states of the $\Sigma$ decreases with the lattice spacing. The same behavior is observed for the other strange particles studied in this work. This
is shown in Fig.~\ref{fig:isospin2} where we plot the mass difference
as a function of $a^2$ at our smallest and heaviest pion mass. As can be seen, the mass difference is consistent with zero for all particles at the smallest lattice spacing. The small non-zero 
values seen for the $\Sigma$ and $\Xi$ particles are just outside one standard deviation.
Therefore, the general conclusion is that indeed isospin splitting is small
at these values of the lattice spacing and it vanishes at the continuum limit.
Since for finite $a$ there are small differnces, for the chiral extrapolation
where we use all lattice data we do not
average the masses for the different charge states of the $\Sigma$, $\Xi$ and $\Xi^*$.

\begin{figure}[h!]
%\centering
\begin{tabular}{cc}
\includegraphics[width=.33\textwidth, angle=270]{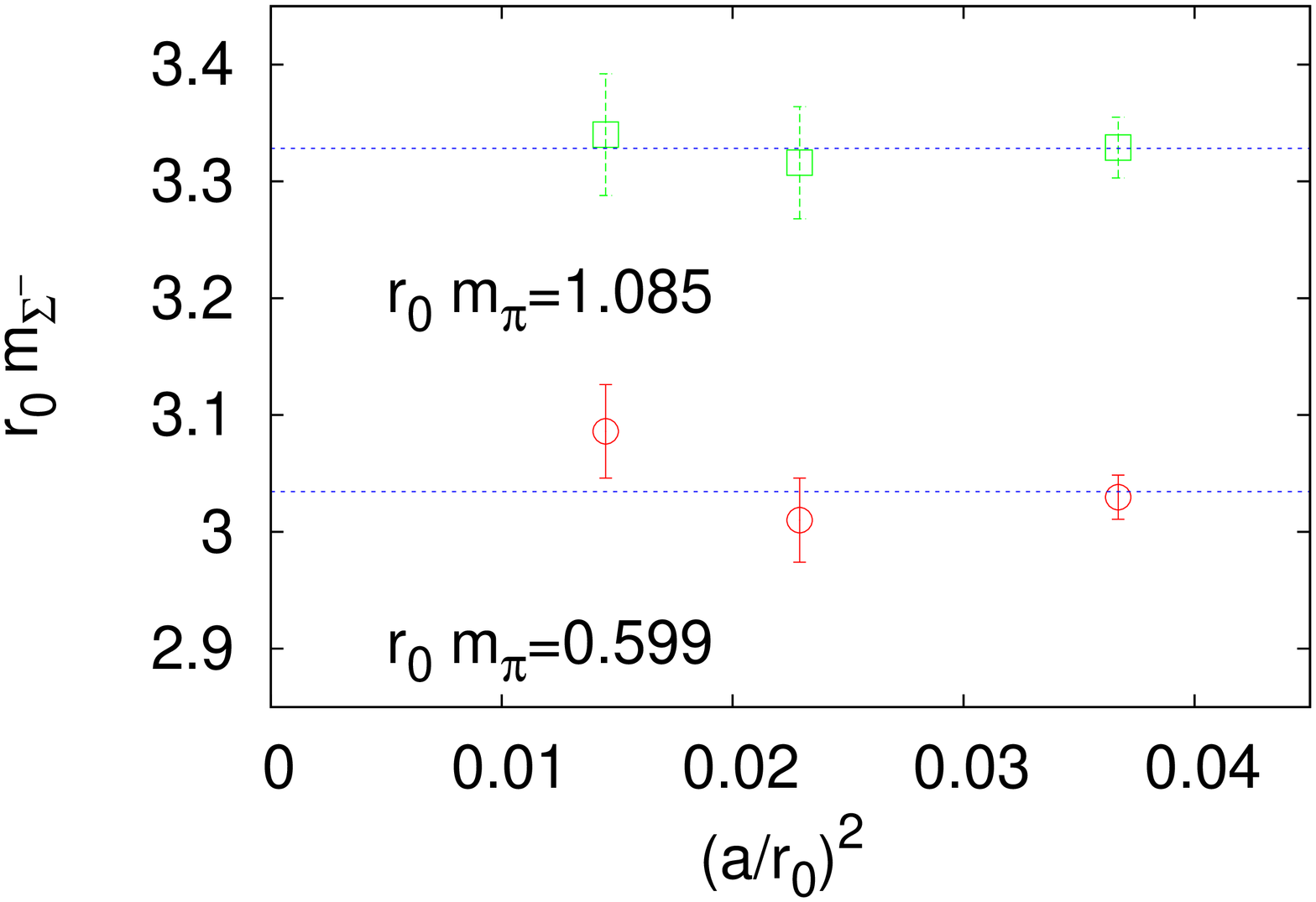} &
\includegraphics[width=.33\textwidth, angle=270]{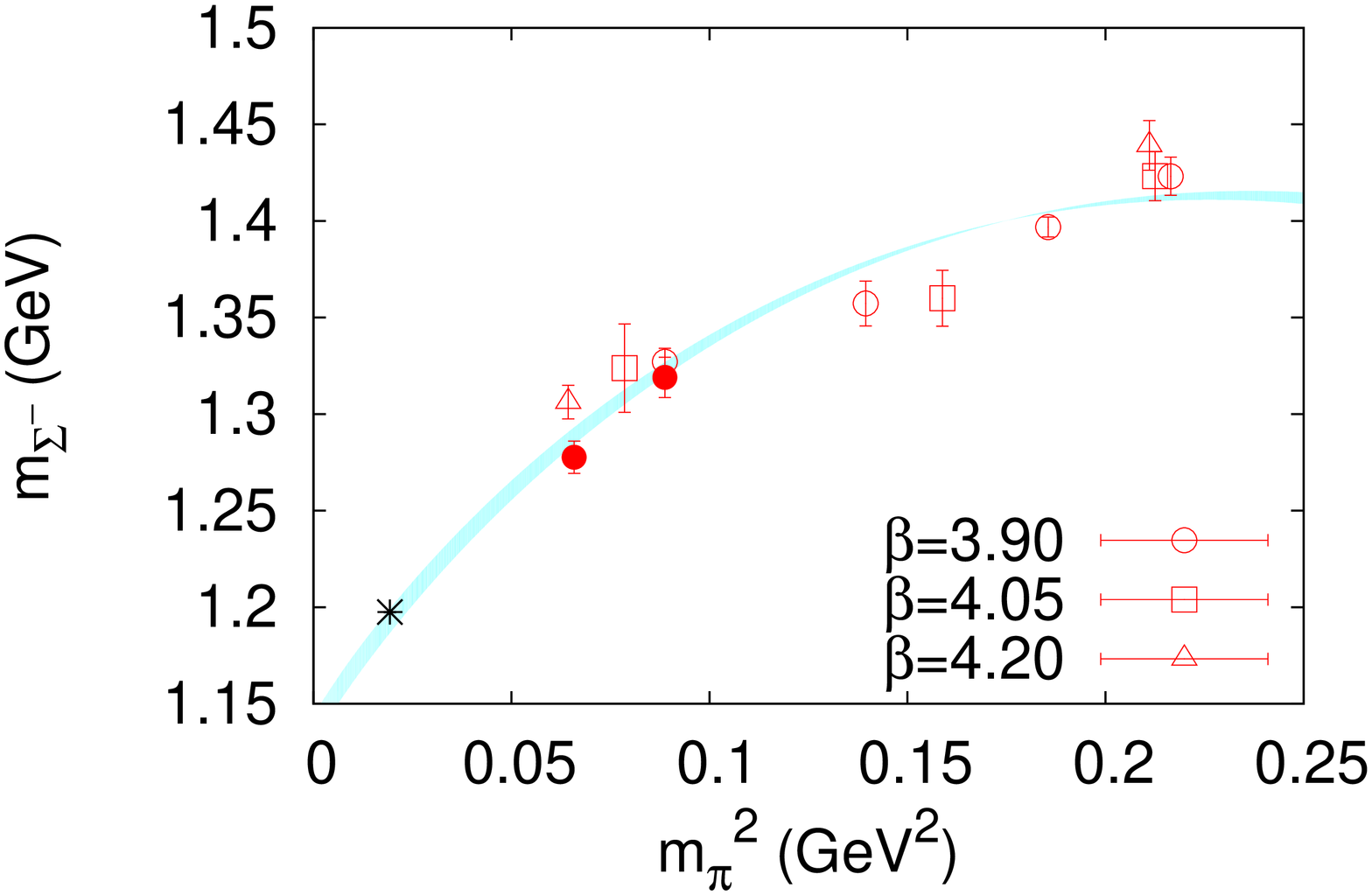} \\
\includegraphics[width=.33\textwidth, angle=270]{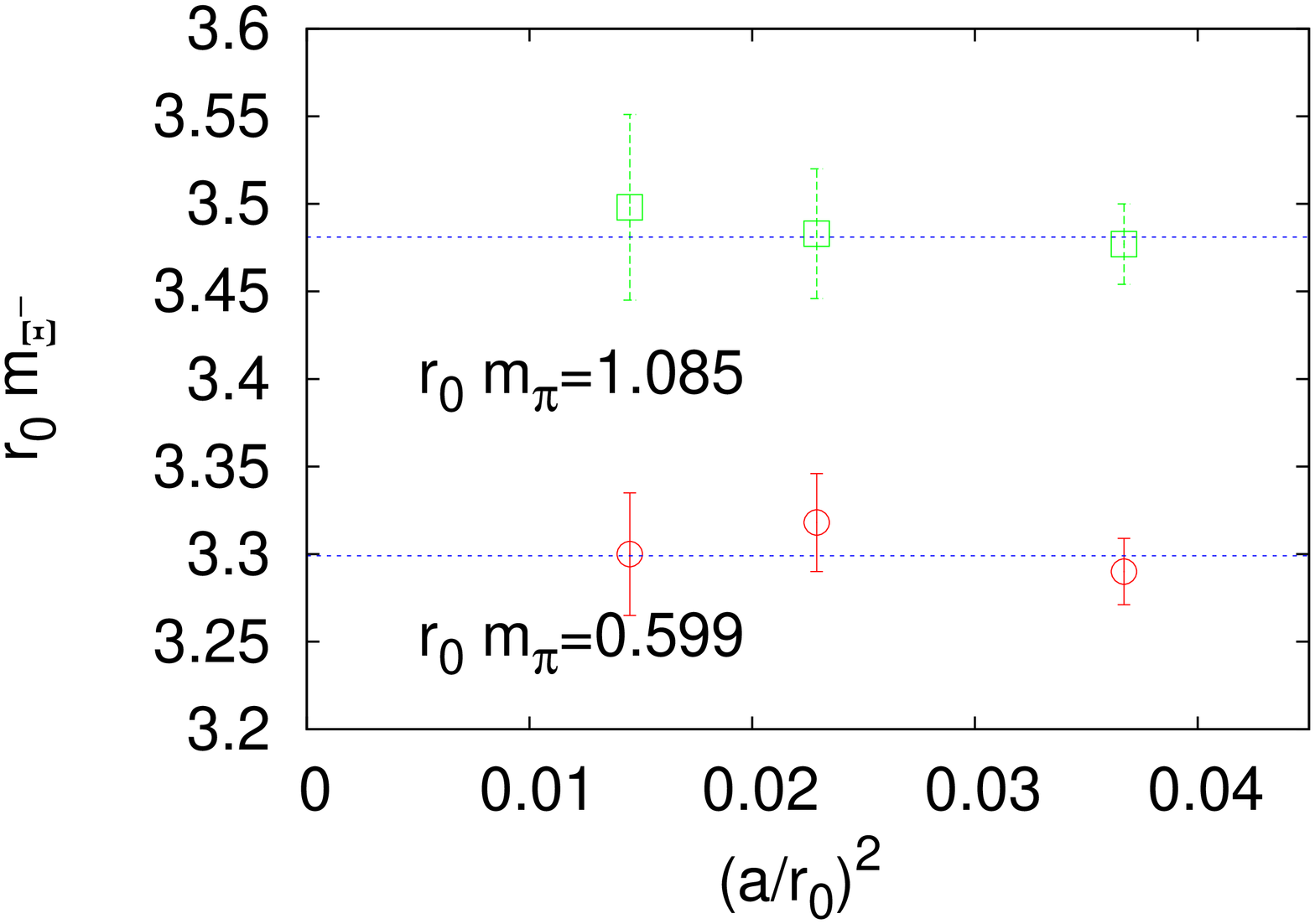} &
\includegraphics[width=.33\textwidth, angle=270]{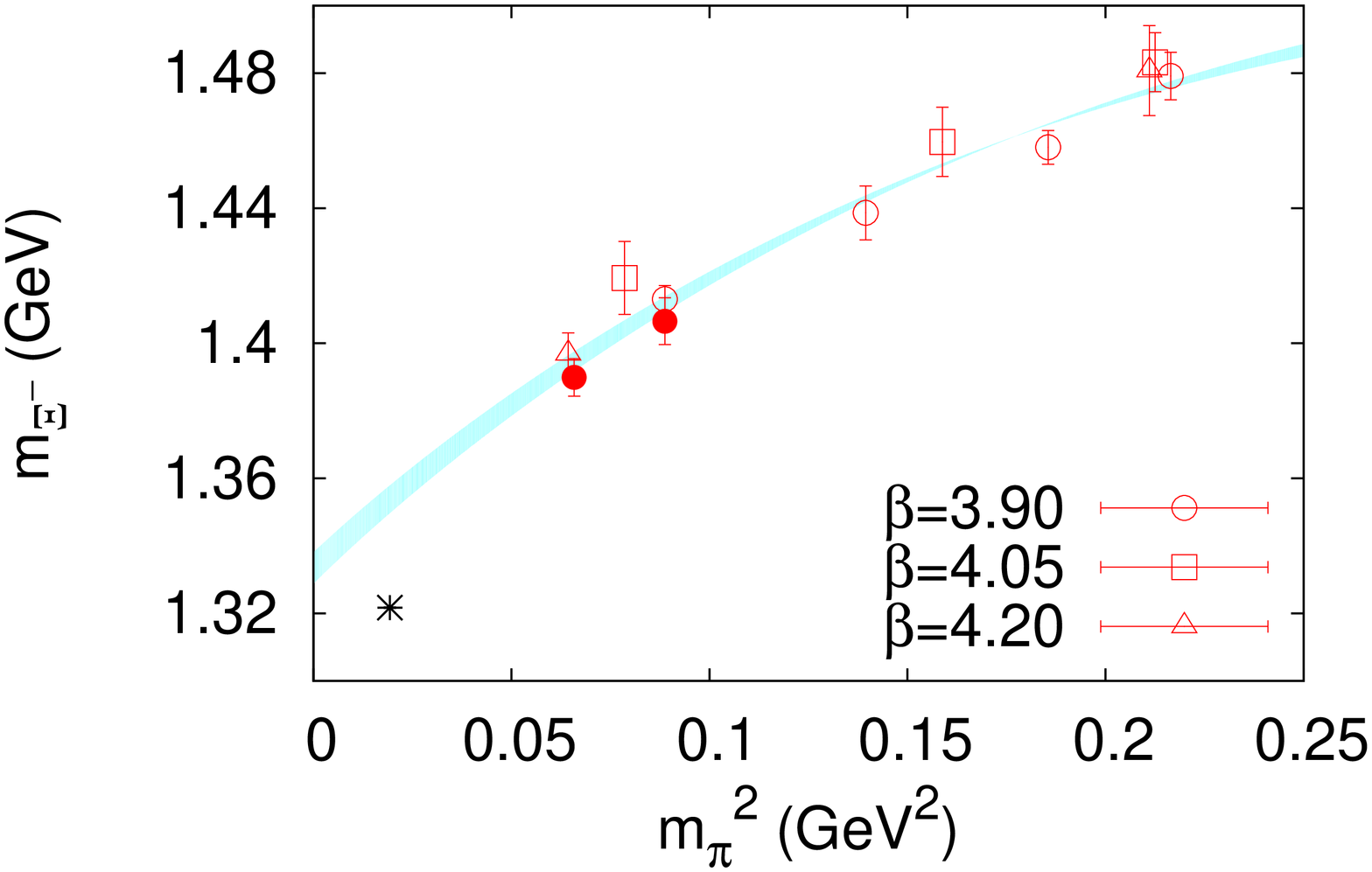} \\
\includegraphics[width=.33\textwidth, angle=270]{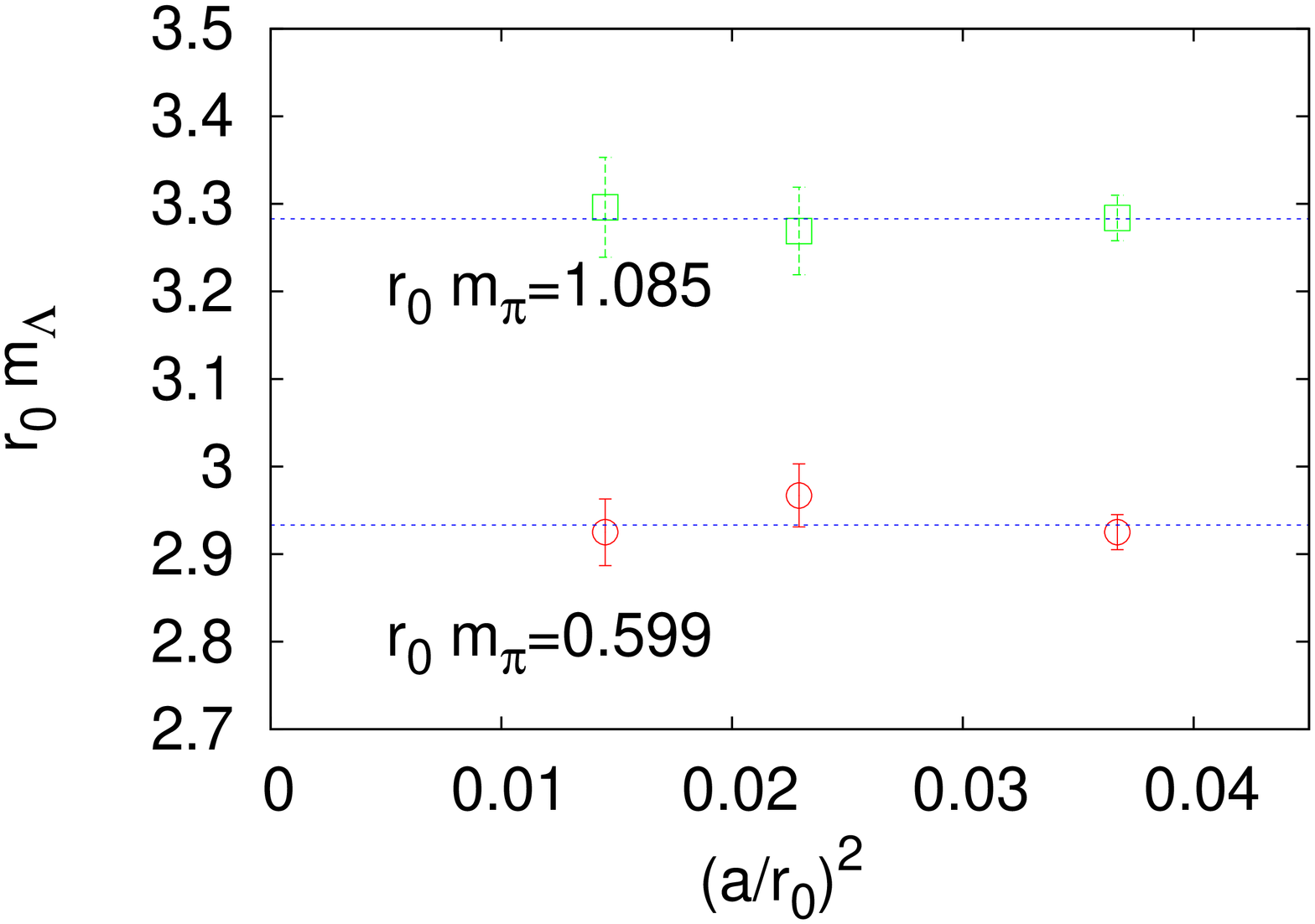} &
\includegraphics[width=.33\textwidth, angle=270]{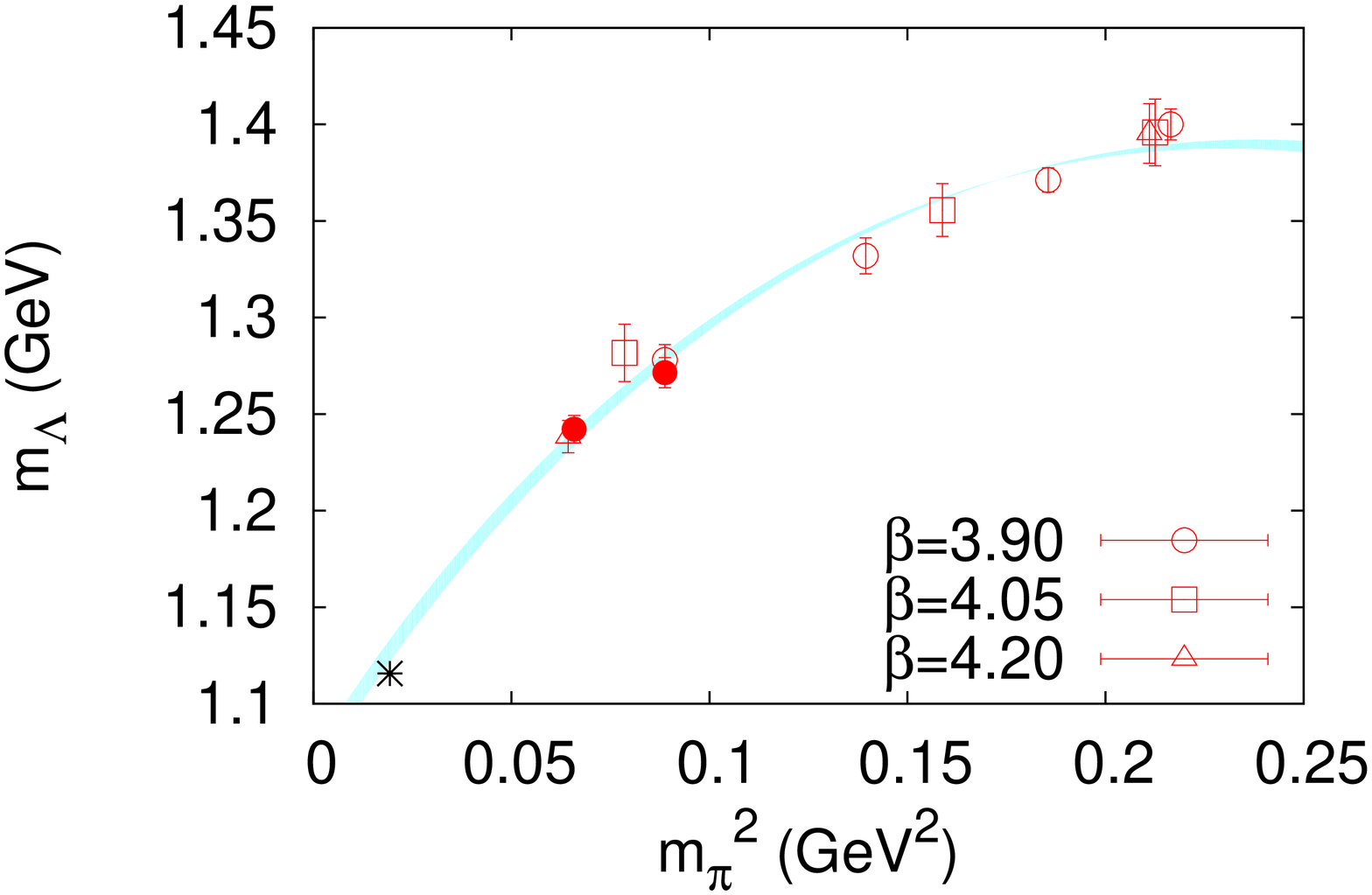} \\
\end{tabular}
\caption{Left: The strange octet baryon mass in units of $r_0$ versus $(a/r_0)^2$ for several particles in the octet and decuplet at two
fixed values of $m_\pi r_0$. The solid line is a linear fit in $(a/r_0)^2$
and the dashed line to a constant.
 Right: Chiral extrapolation at fixed
strange quark mass. Results at
 $\beta=3.9$ are shown with the open circles for $L=2.1$~fm and with the filled
circles for $L=2.7$~fm, at $\beta=4.05$ with the open squares and at $\beta=4.2$ with the open triangles.}
\label{fig:continuum octet}
\end{figure}

\begin{figure}[h!]
%\centering
\begin{tabular}{cc}
\includegraphics[width=.33\textwidth, angle=270]{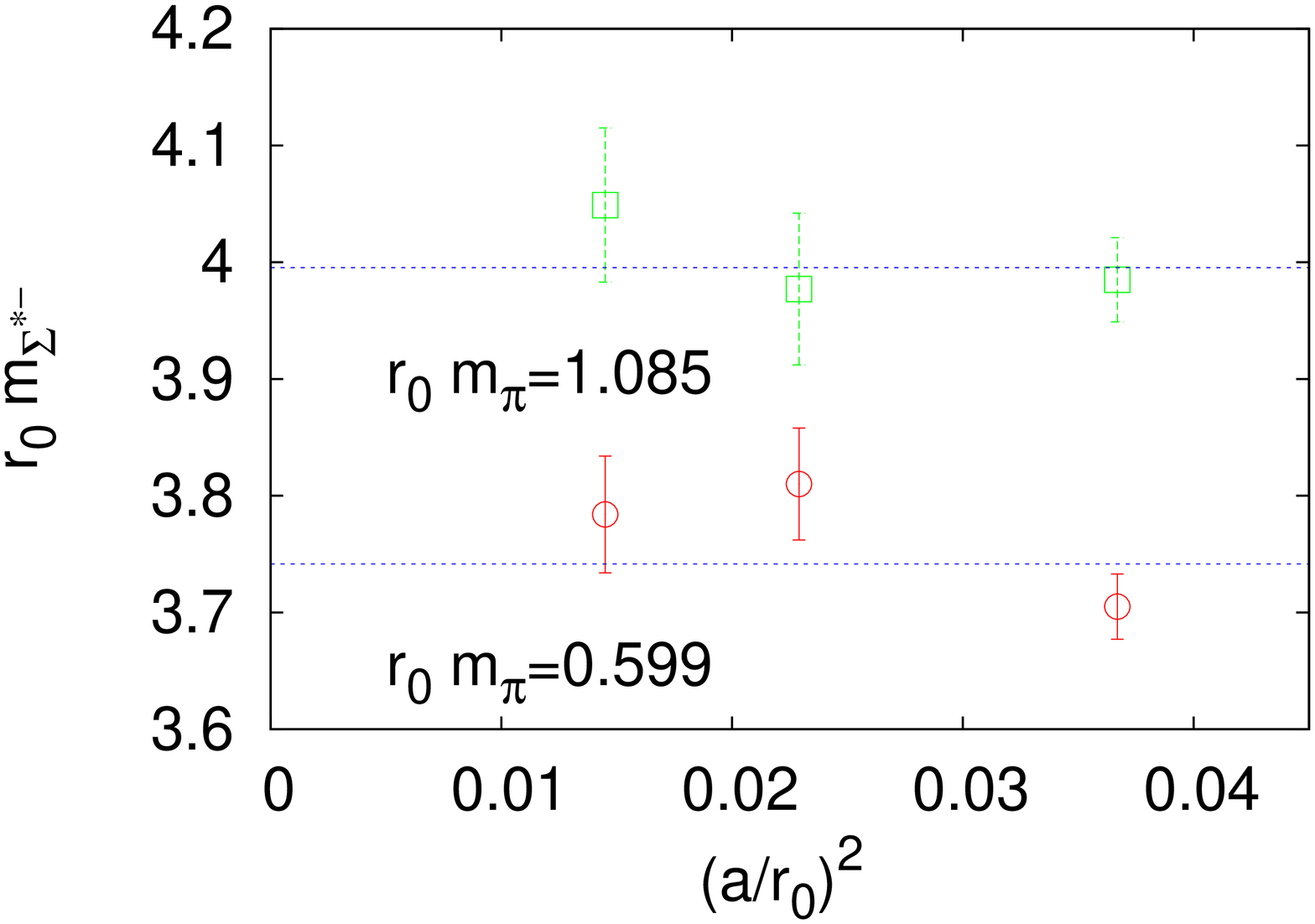} &
\includegraphics[width=.33\textwidth, angle=270]{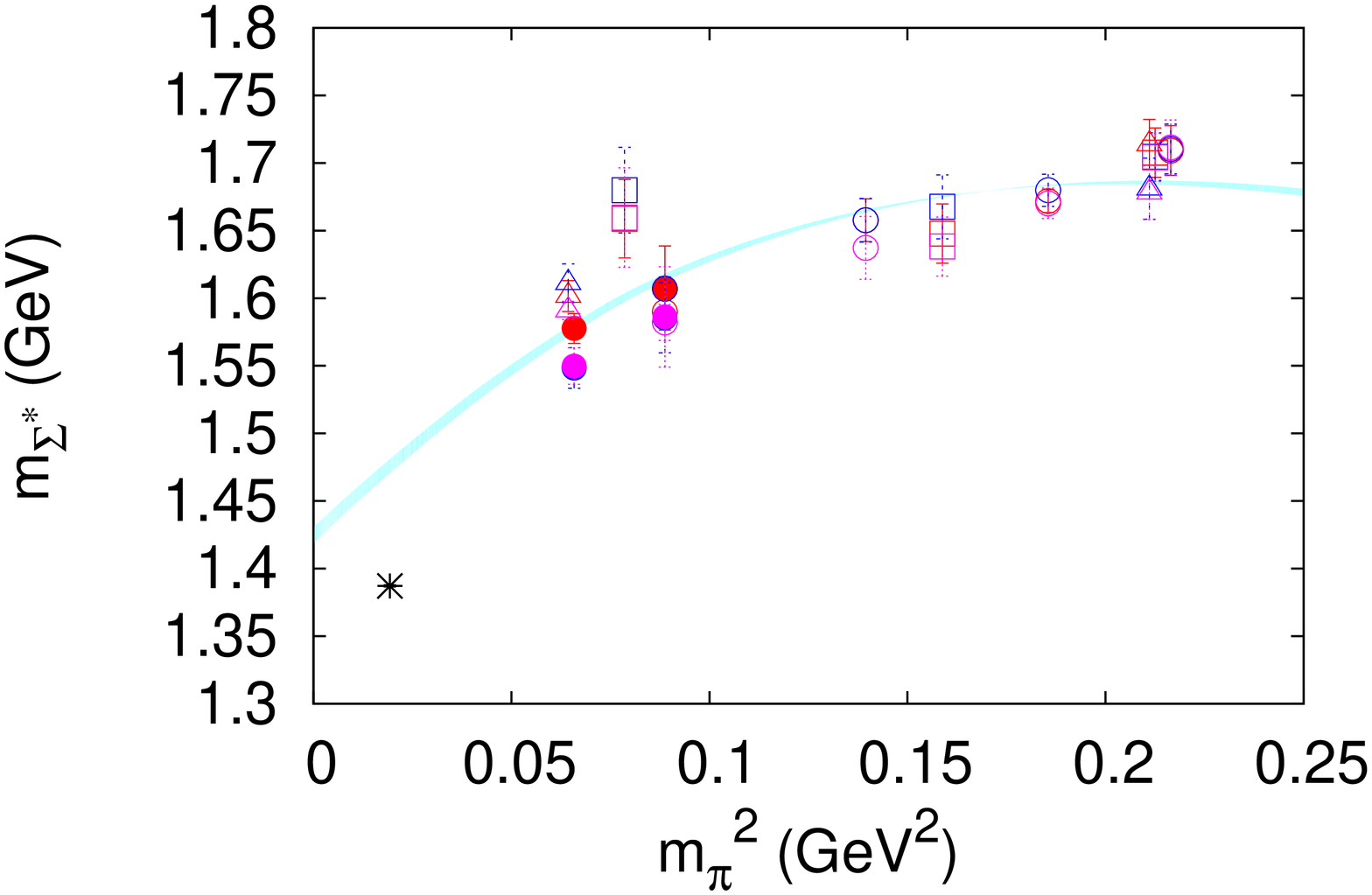} \\
\includegraphics[width=.33\textwidth, angle=270]{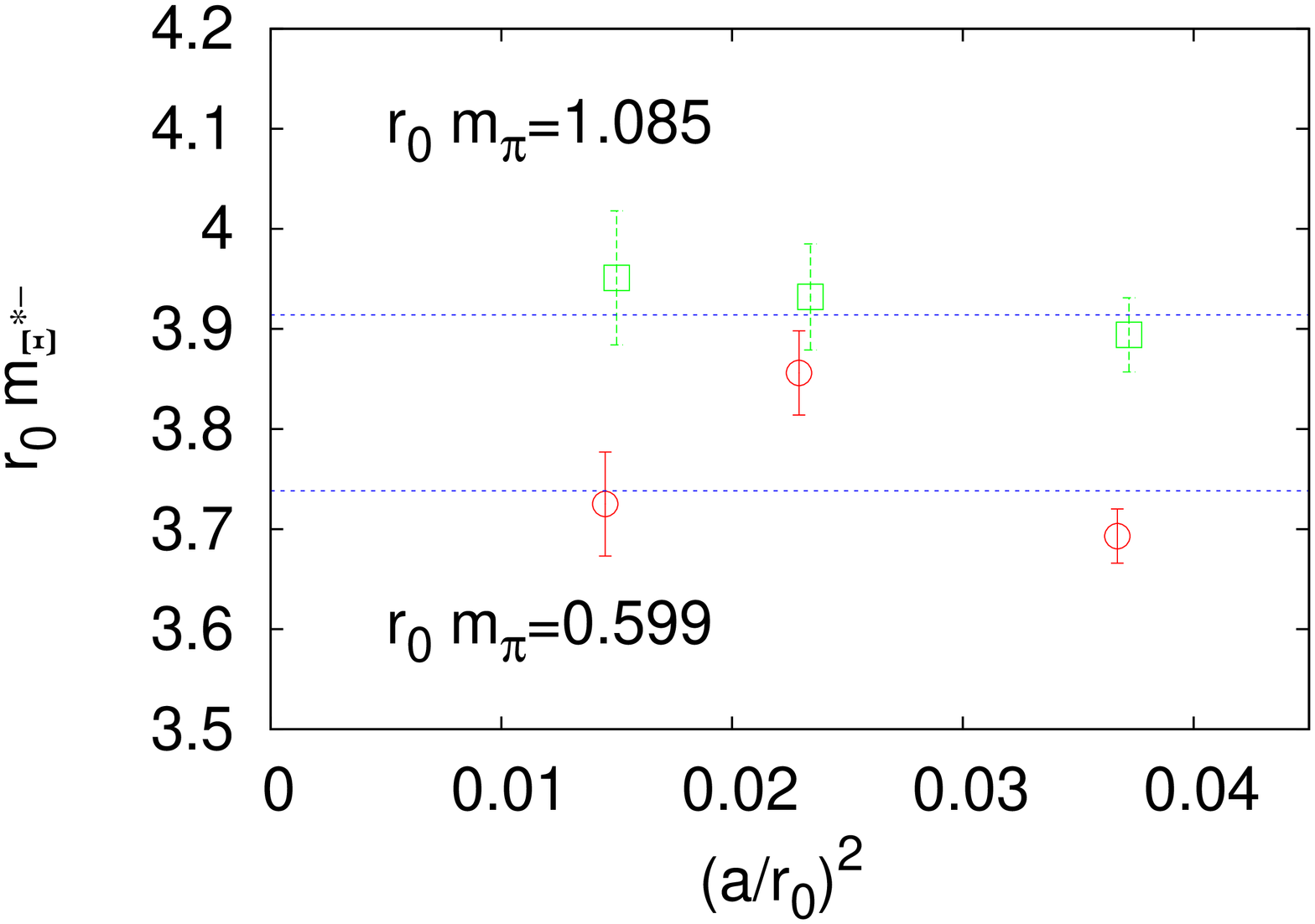} &
\includegraphics[width=.33\textwidth, angle=270]{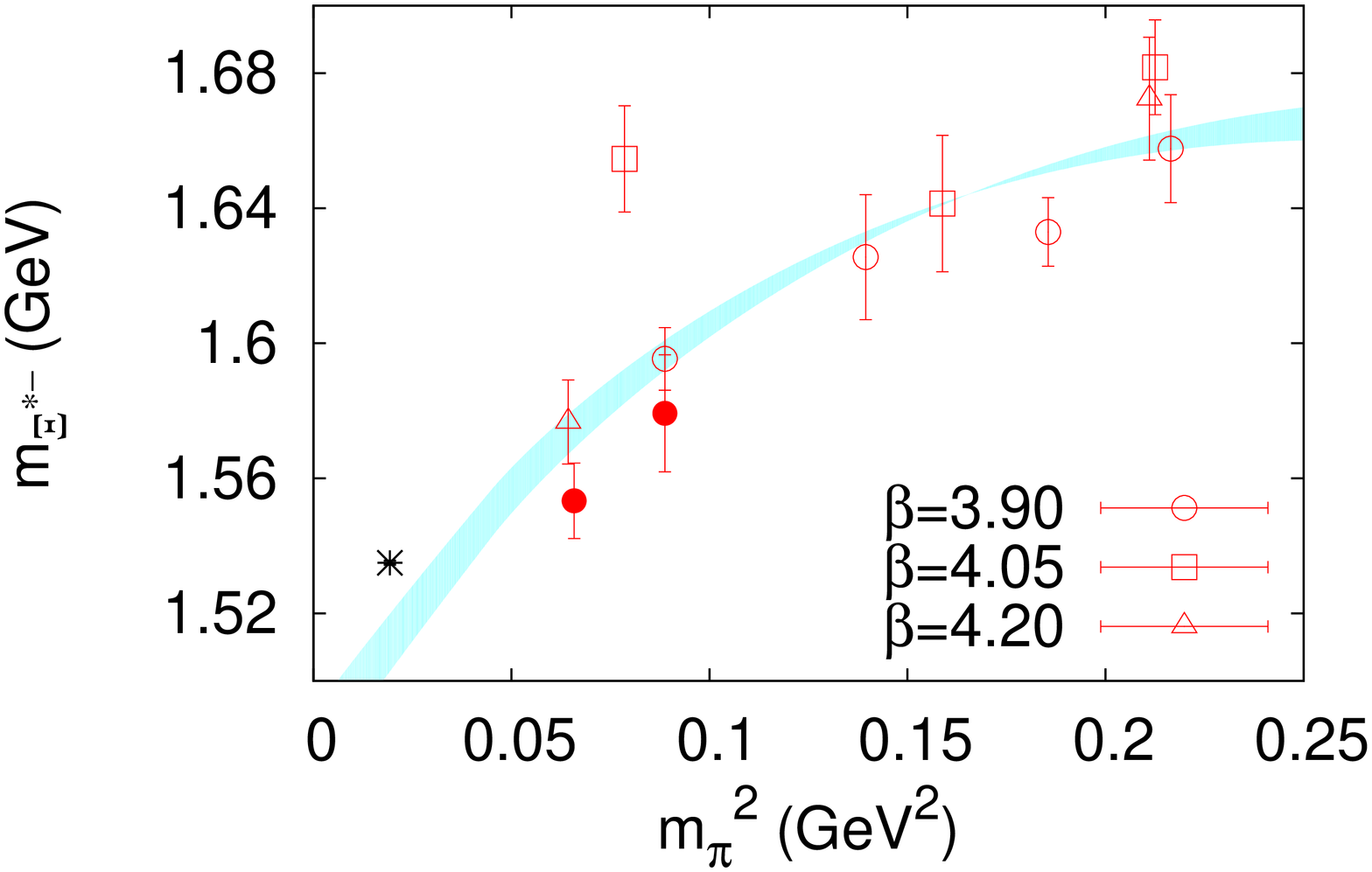} \\
\includegraphics[width=.33\textwidth, angle=270]{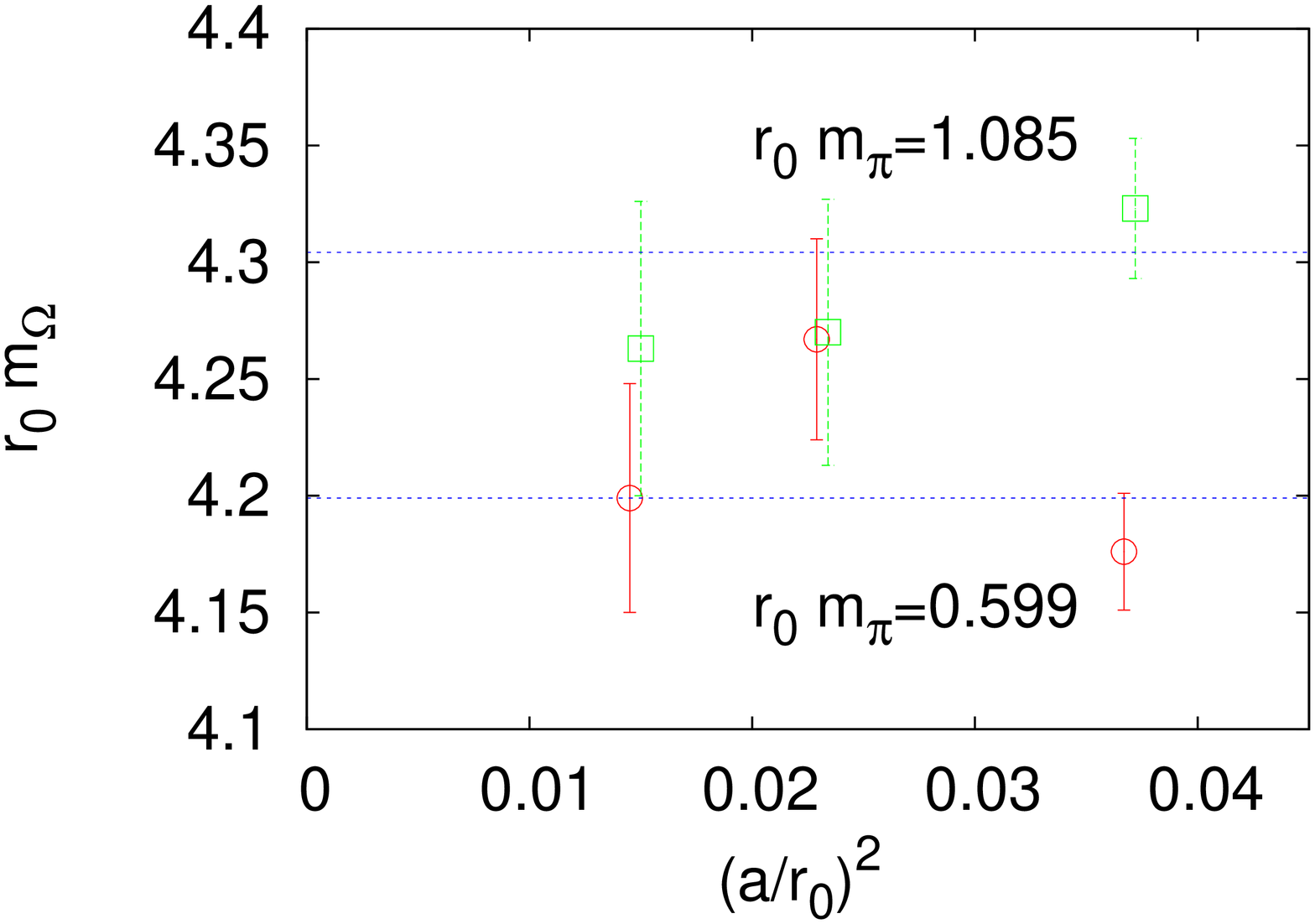} &
\includegraphics[width=.33\textwidth, angle=270]{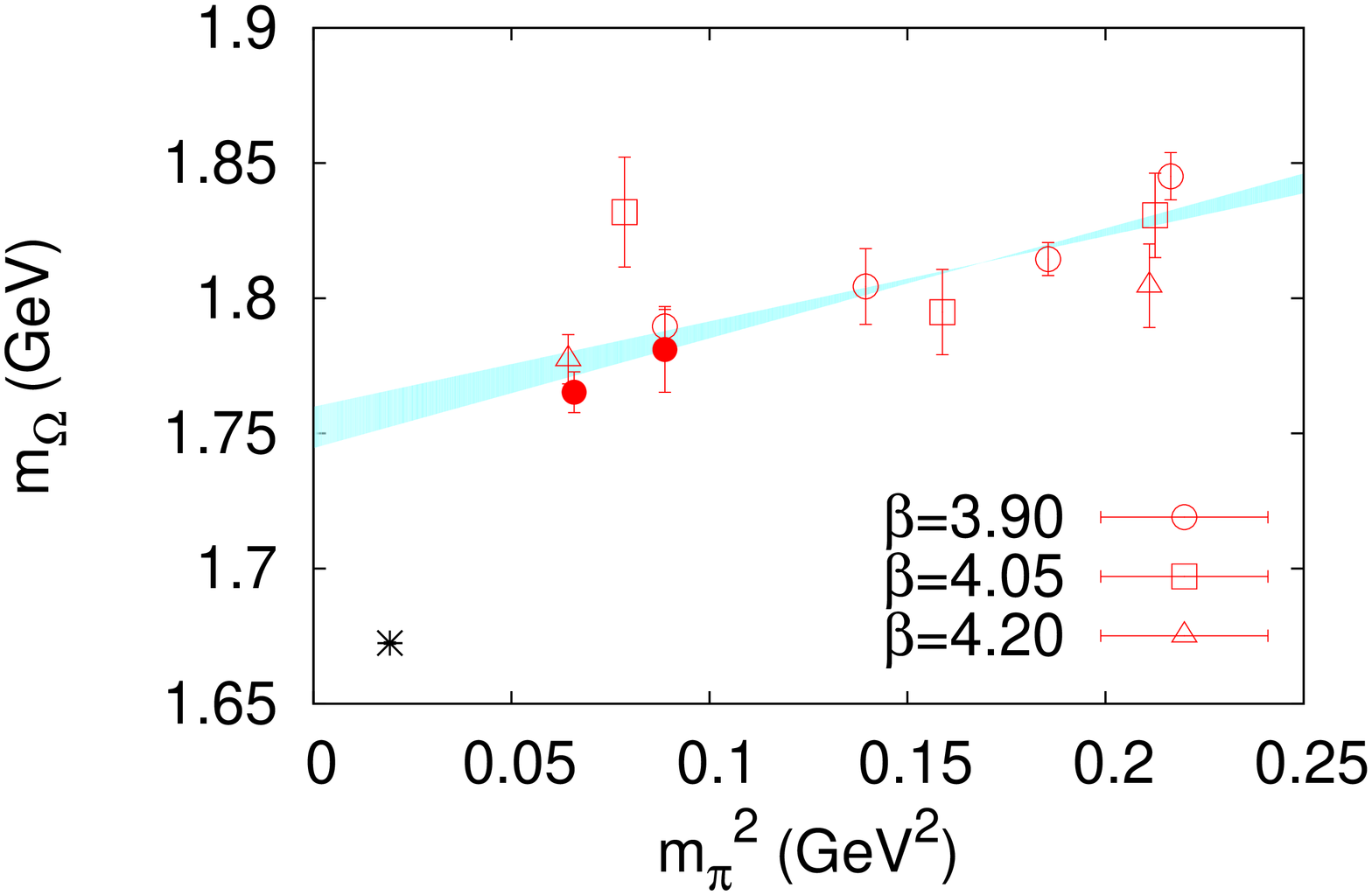} \\
\end{tabular}
\caption{Left: The strange decuplet baryon mass in units of $r_0$ versus $(a/r_0)^2$  at two
fixed values of $m_\pi r_0$. 
The solid line is a linear fit in $(a/r_0)^2$
and the dashed line to a constant.
 Right: Chiral extrapolation at fixed strange quark mass.
In the case
of the $\Sigma^*$ we show the three charge states
${\Sigma^*}^+$ (purple), ${\Sigma^*}^0$ (blue), and ${\Sigma^*}^-$ (red).
The spread of data at a given pion mass with the same symbol indicates the level of isospin breaking for the  $\Sigma^*$ particle.
Results at
 $\beta=3.9$ are shown with the open circles for $L=2.1$~fm and with the filled
circles for $L=2.7$~fm, at $\beta=4.05$ with the open squares and at $\beta=4.2$ with the open triangles.}
\label{fig:continuum decuplet}
\end{figure}

Volume effects can be studied at $\beta=3.9$ where we have simulations
at two volumes at  pion mass of about 300~MeV. As can be seen in Figs.~\ref{fig:isospin}, \ref{fig:continuum octet} and \ref{fig:continuum decuplet} 
results at different volumes are consistent.
Therefore, we conclude any volume effects are smaller than our statistical 
accuracy.

%{\bf Mario: I wouldn't put this table. For me the corresponding plots are enough}

%\begin{table}
%\begin{tabular}{|l|l|l|l|}
%\hline
%$m_\pi=0.254$ GeV & & & \\ \hline
%Particles & $\Delta m(3.9)$ & $\Delta m(4.05)$ & $\Delta m(4.2) $\\ \hline 
%$\Sigma^+ - \Sigma^-$ & 0.0603(85)  & 0.061(19) & 0.030(19) \\ \hline  
%$ \Xi^0 - \Xi^-$ & 0.0745(63) & 0.038(15) & 0.013(29) \\ \hline
%$\Sigma^{*-} - \Sigma^{*+}$ & 0.023(13)  & 0.016(27) & 0.011(18) \\ \hline
%$\Xi^{*0} - \Xi^{*-}$ & 0.032(10)  & 0.003(20) & 0.027(26) \\ \hline
%\hline
%$m_\pi=0.459$ GeV & & & \\ \hline
%Particles & $\Delta m(3.9)$ & $\Delta m(4.05)$ & $\Delta m(4.2) $\\ \hline
%$\Sigma^+ - \Sigma^-$ & 0.072(14)  & 0.052(28) & 0.013(29) \\ \hline
%$ \Xi^0 - \Xi^-$ & 0.078(10) & 0.036(22) & 0.035(27) \\ \hline
%$\Sigma^{*-} - \Sigma^{*+}$ & -0.003(20)  & 0.003(40) & 0.035(39) \\ \hline
%$\Xi^{*0} - \Xi^{*-}$ & 0.071(15)  & 0.022(31) & 0.007(35) \\ \hline
%\end{tabular}
%\caption{Difference of mass in GeV for particles with different isospin in 
%the strange baryon case.}
%\label{tab:mass strange}
%\end{table}

In order to examine the continuum limit we interpolate our lattice result
at a given pion mass in units of $r_0$. For $\beta=4.2$ we have simulations 
at only two
values of the pion mass at the upper and lower
range of pion masses considered in this
work, namely for $m_\pi=0.254$~GeV and $m_\pi=0.459$~GeV. 
Therefore we interpolate the results for the other two values of $\beta$
to these two pion masses.
In Figs.~\ref{fig:continuum octet} and \ref{fig:continuum decuplet},  we show results for the octet and
decuplet strange baryon masses, respectively,
 at our three values of the lattice spacing.
We perform a continuum extrapolation by perfroming a linear fit in $(a/r_0)^2$
as well as to constant. As can be seen from Fig.~\ref{fig:continuum octet}, the values obtained in the
continuum limit agree for all octet baryons. In the case of the decuplet the
statistical errors are larger and the value obtained at $a=0$ with the linear
fit carries a large error. The value obtained using a constant fit has
a smaller error and it is  compatible with one the one obtained using
a linear fit. Therefore,  for a given charge state and within
the current statistical accuracy, the $(a/r_0)^2$ term
can be taken as negligible.
%%
%{\bf MARIO: here explain the new plots I produced, where it is shown the 
%continuum limit carried out with a linear function (filled triangle) 
%and with a constant (empty triangle). The two limits allways agree. The 
%only small deviation is observed in the small pion mass region for $\Sigma^*$ 
%and $\Xi^*$. Also the captions of Fig. 10 and 11 shuold be adapted, commenting 
%the comparison between the two contimuun limits.}
%%
Therefore, we can use results at all
$\beta$-values to extrapolate to the
physical point since cut-off are small for a given charge state. 
 There are two exceptions in the case of the decuplet.
 At $\beta=4.05$ the mass of the 
$\Xi^*$ and $\Omega$  at the lowest pion mass are systematically higher 
than that at the other two $\beta$-values. 
Since the results at $\beta=3.9$, with
larger lattice space are consistent with those at $\beta=4.2$ 
we conclude that this is not a cut-off effect.

\subsection{Charm baryon mass with charm quark mass tuned to its physical value}

As in the previous subsection
we consider results  obtained at the tuned charm mass given in Table~\ref{tab:tuned quarks}. The only exception is 
at $\beta=4.2$ at the heavy pion mass, where  we have results
close to the tuned value, namely at $a\mu_c=0.16,0.185,0.21$. 
%{\bf Dina give the values used)}. 
As we have seen, the dependence on the heavy quark is linear and 
therefore the charm baryon
masses at the tuned value can be easily  determined by a linear interpolation.

\begin{figure}[h!]
\begin{tabular}{cc}
\includegraphics[width=.3\textwidth, angle=270]{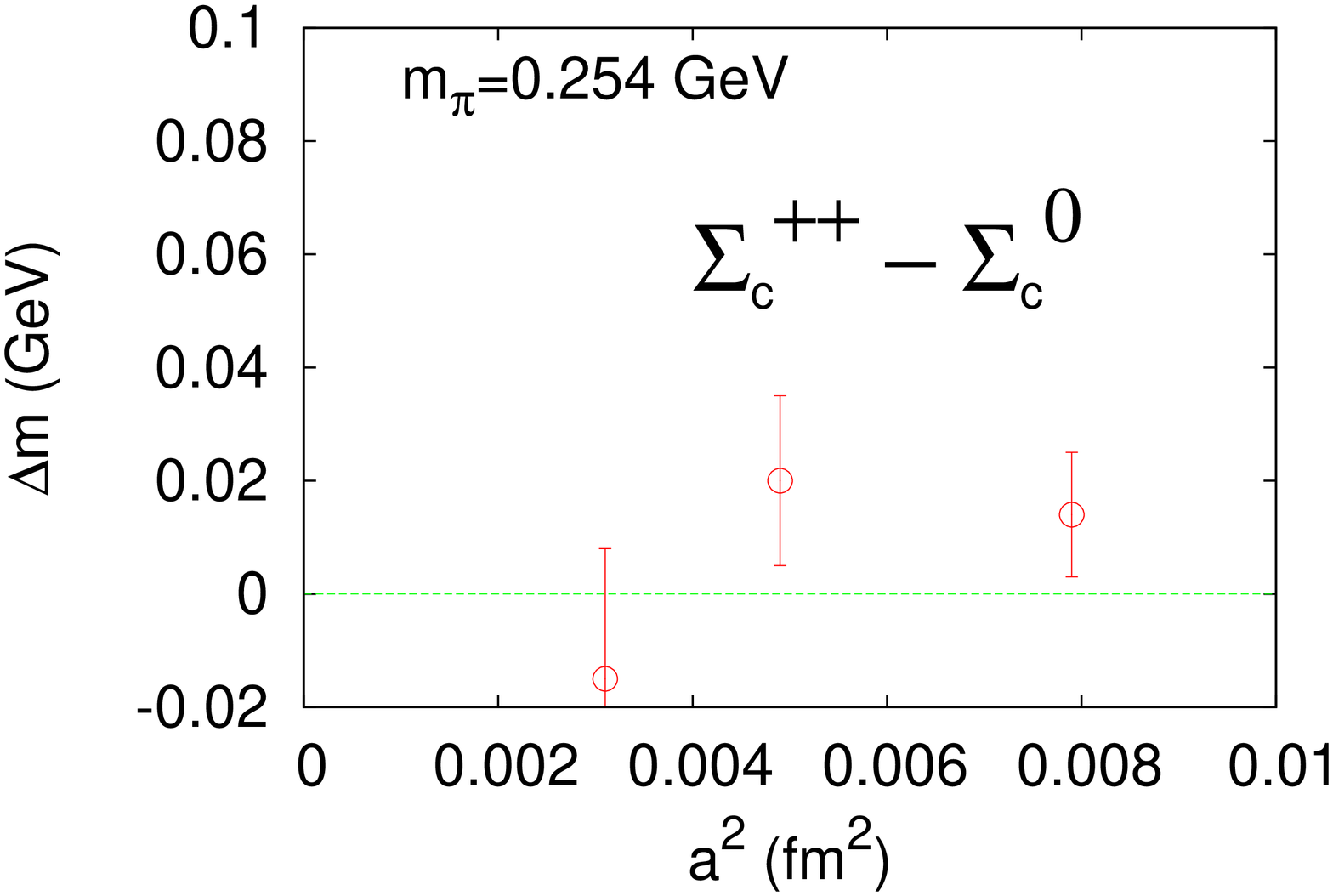} &
\includegraphics[width=.3\textwidth, angle=270]{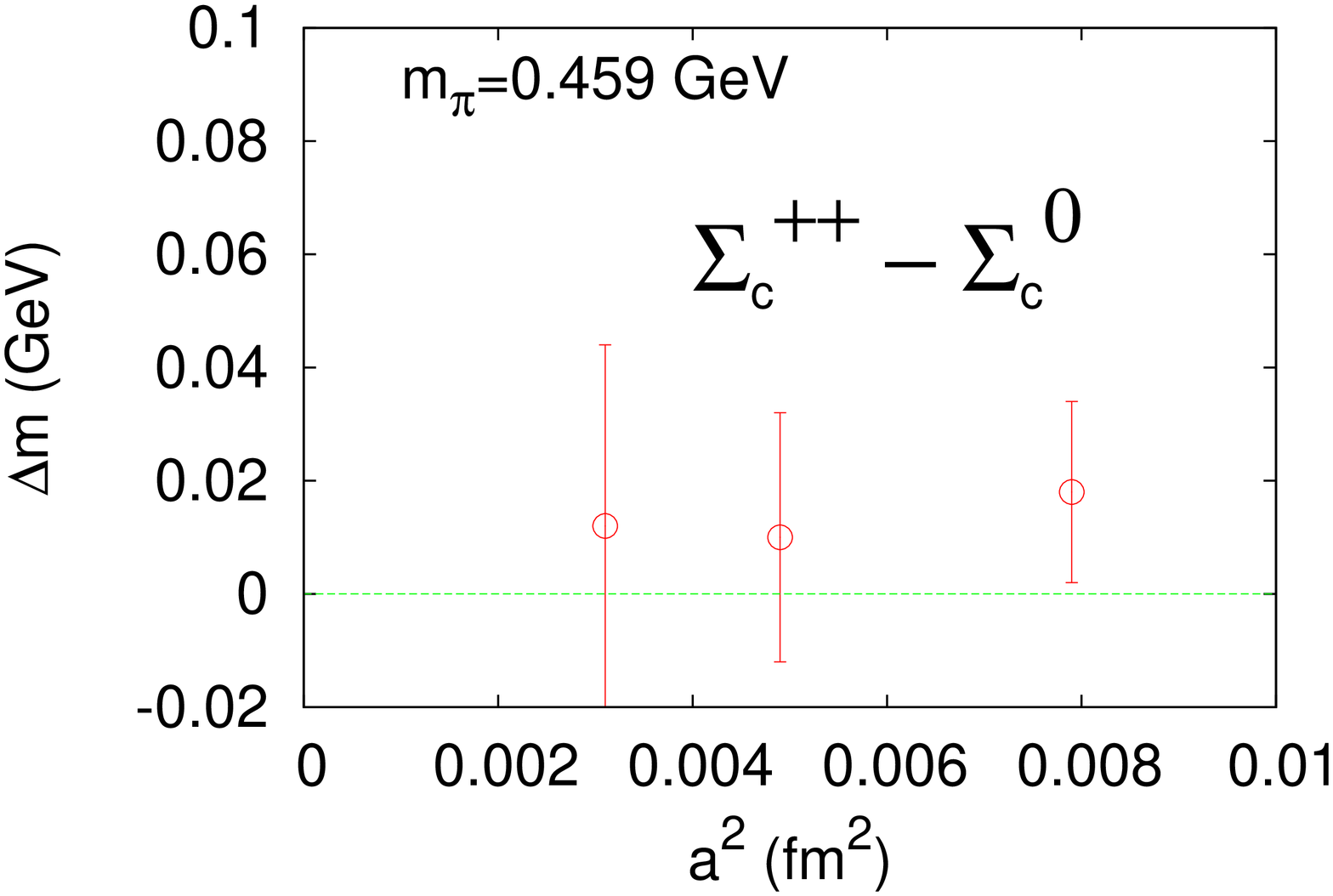} \\
\includegraphics[width=.3\textwidth, angle=270]{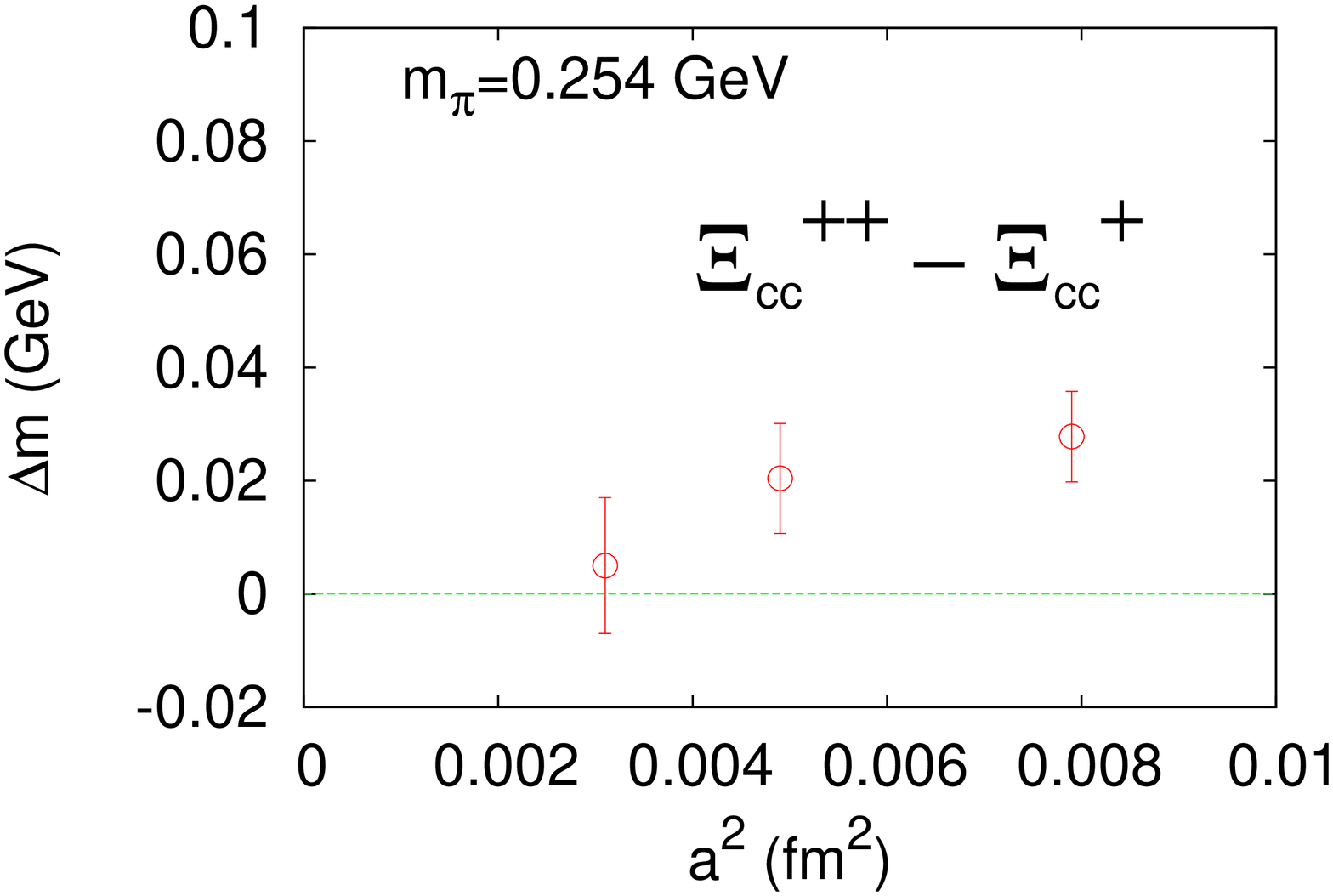} &
\includegraphics[width=.3\textwidth, angle=270]{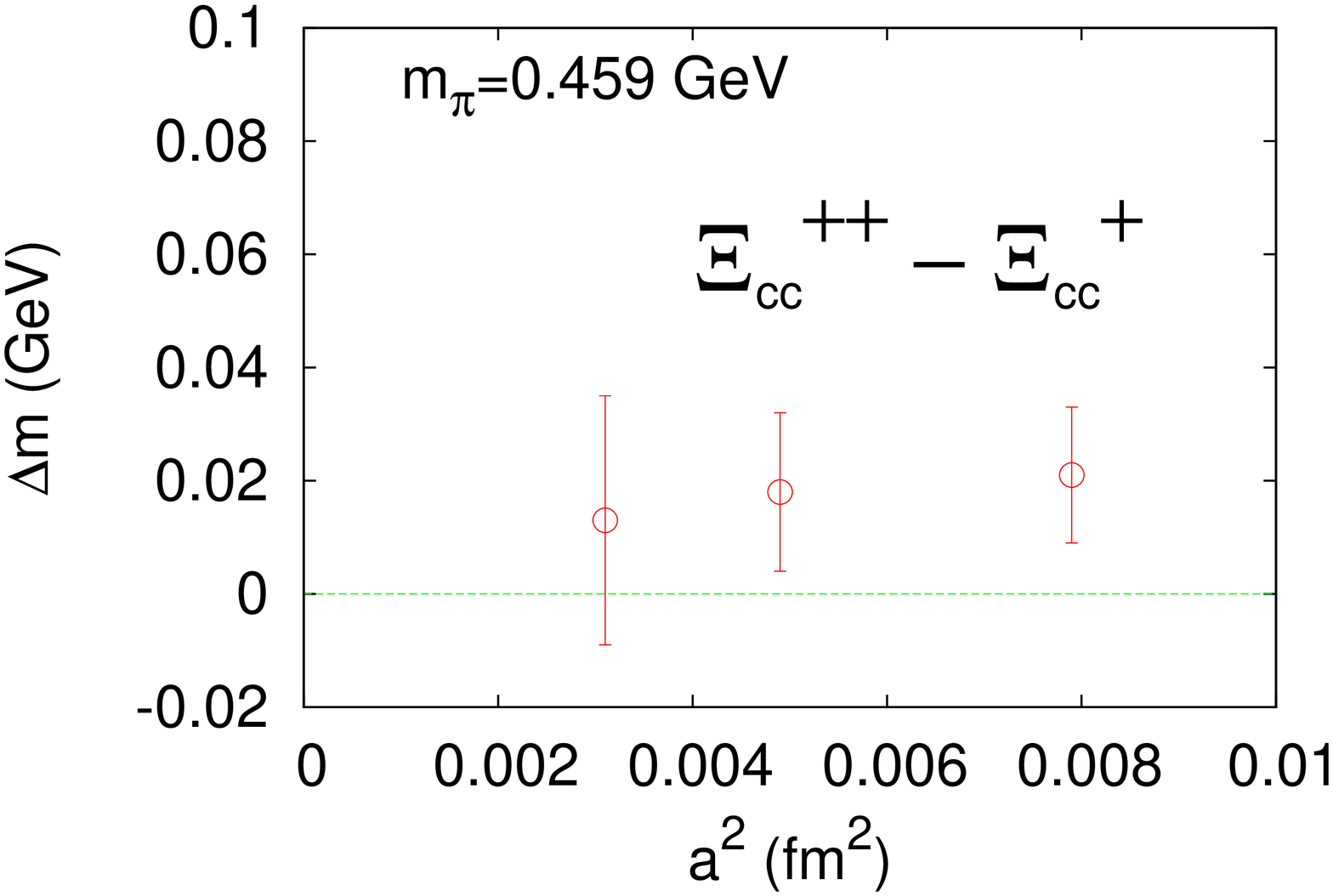} \\
\includegraphics[width=.3\textwidth, angle=270]{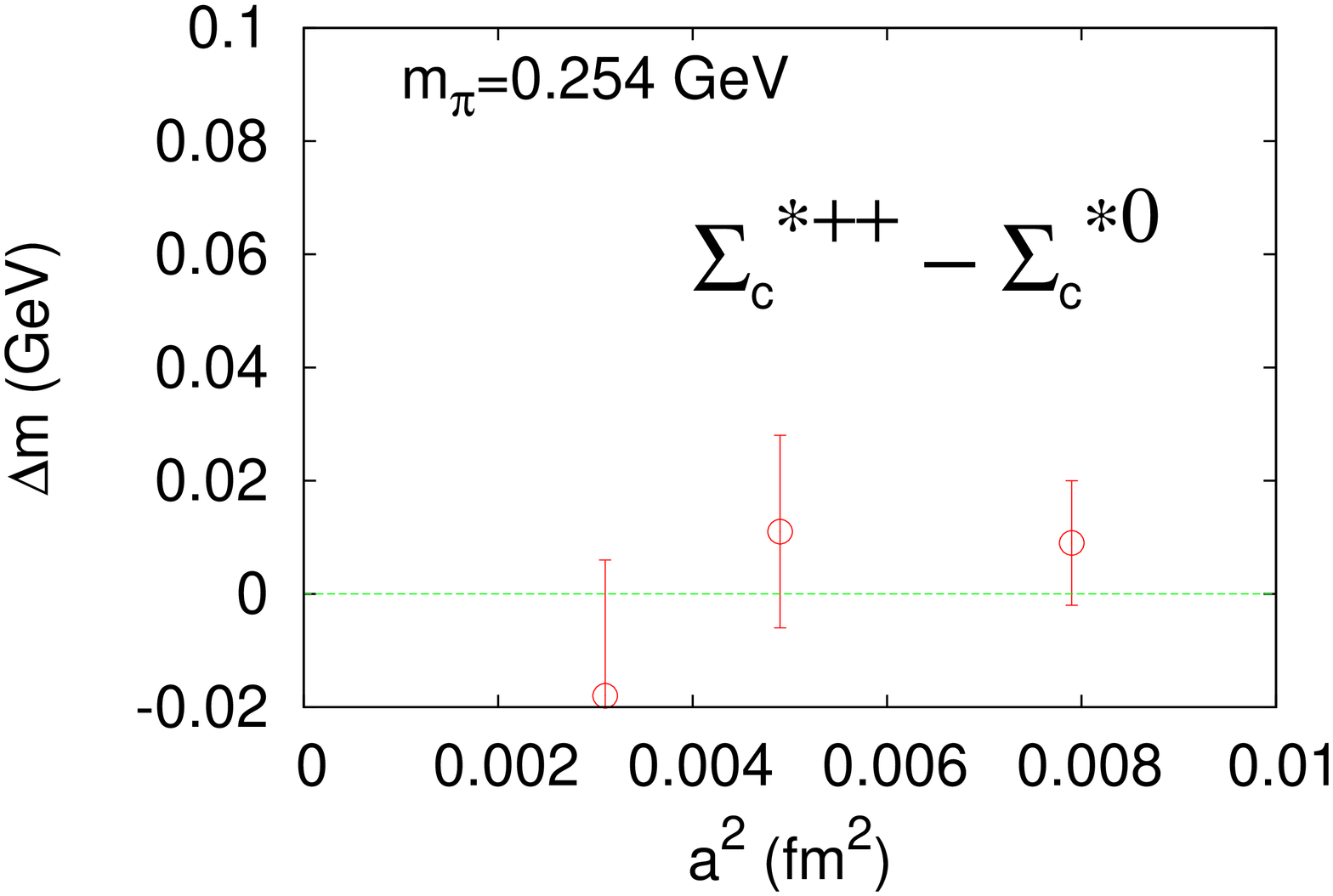} &
\includegraphics[width=.3\textwidth, angle=270]{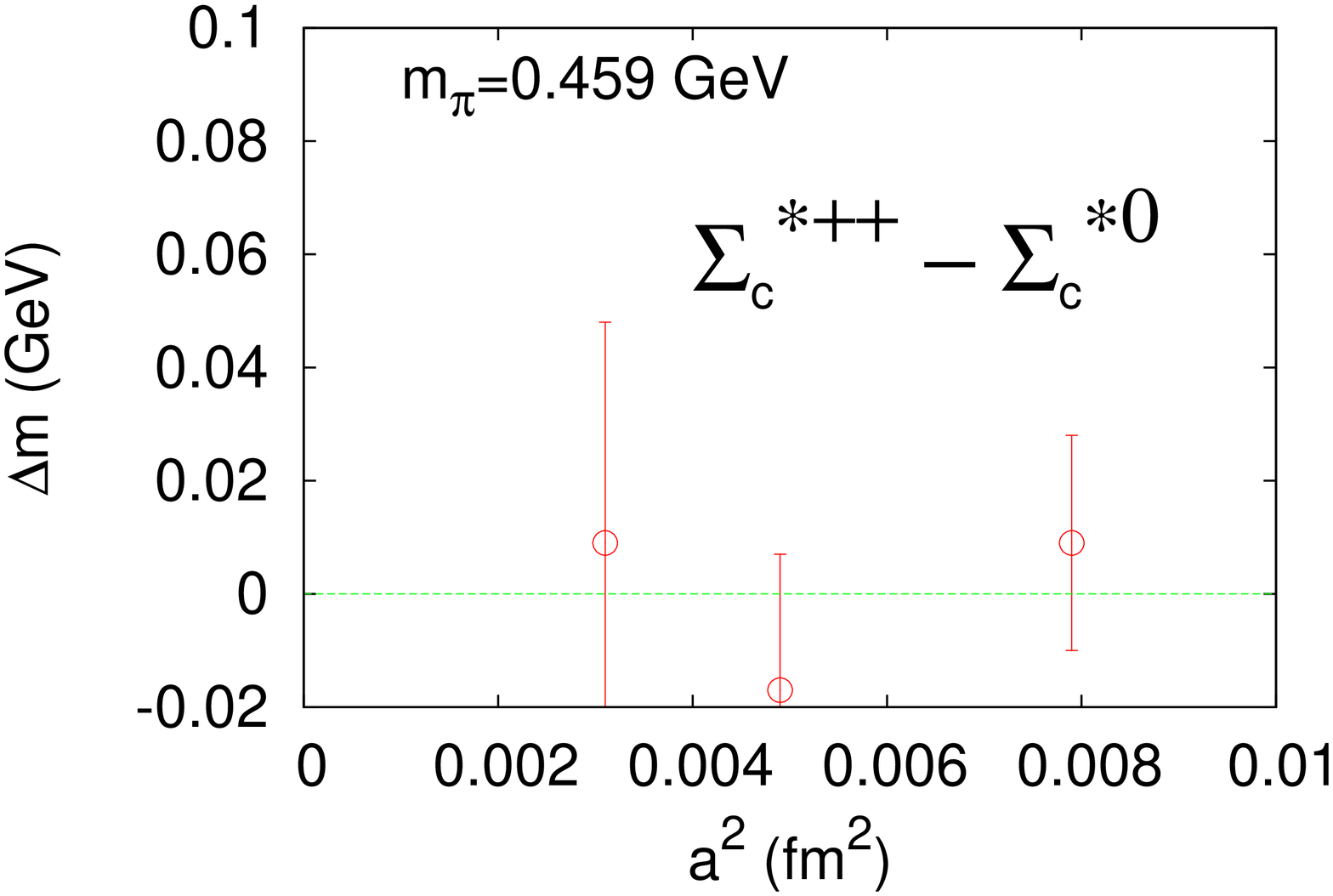} \\
\includegraphics[width=.3\textwidth, angle=270]{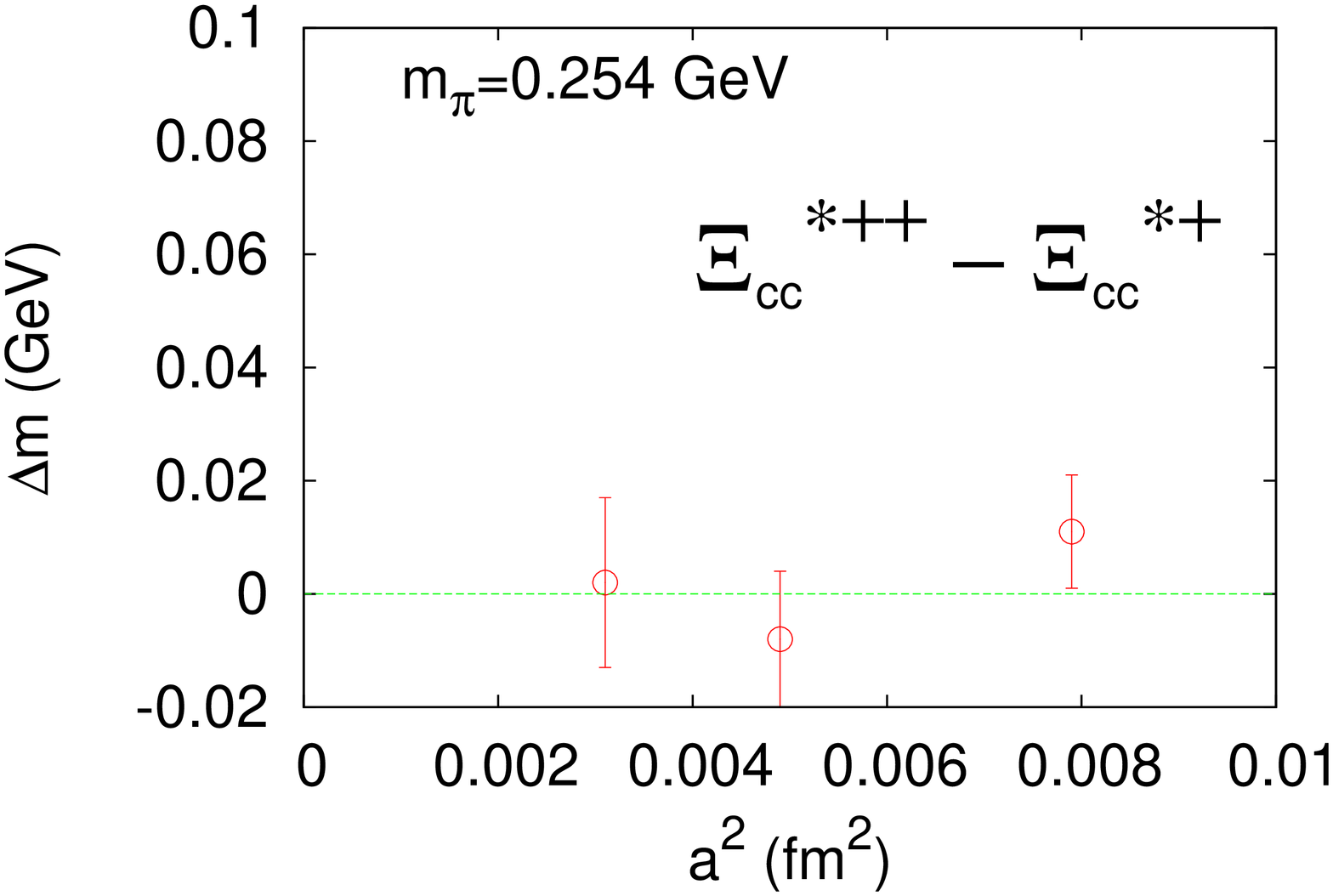} &
\includegraphics[width=.3\textwidth, angle=270]{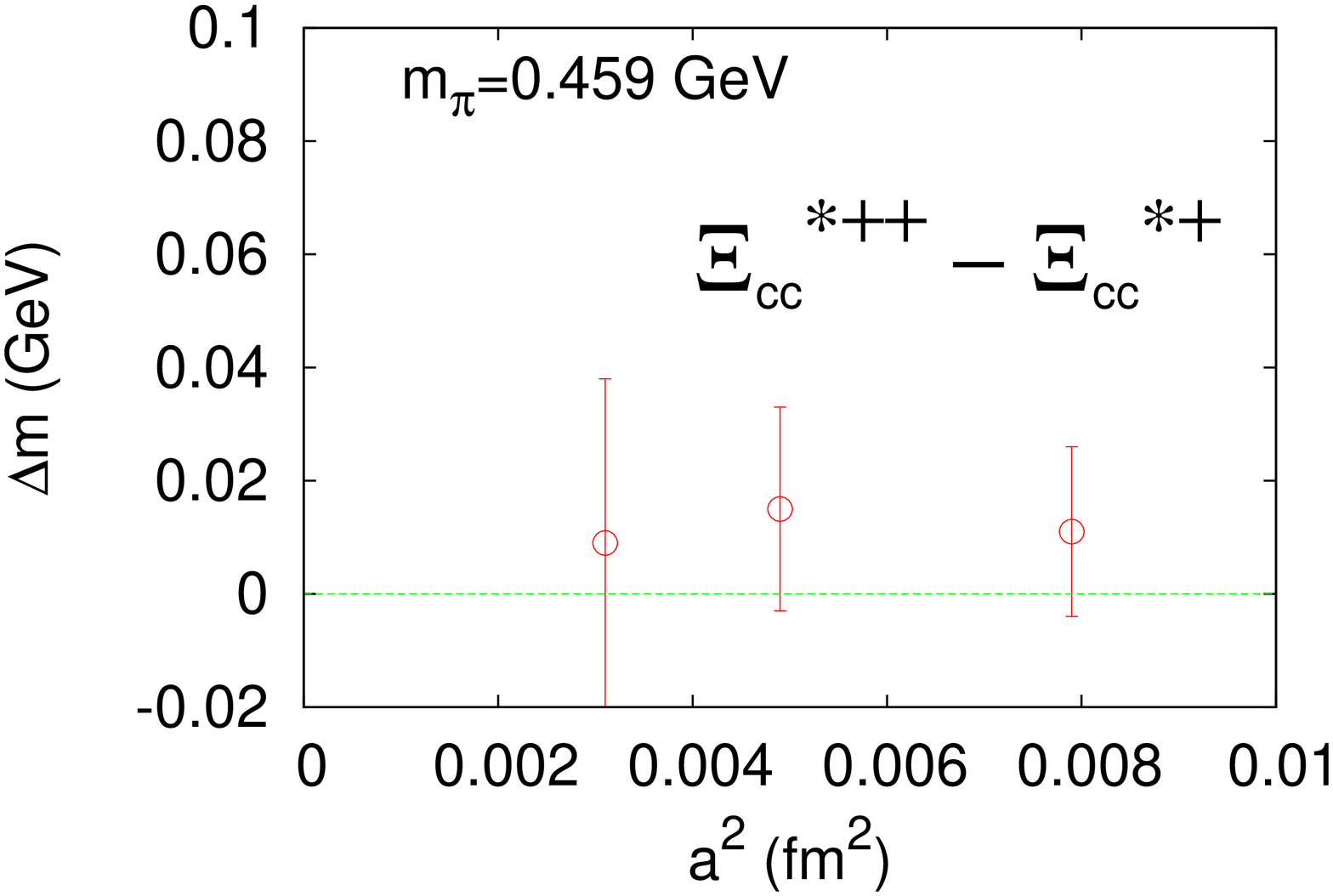} \\
\end{tabular}
\caption{ Isospin mass splitting in the charm sector 
 at the smallest and largest values of the pion mass used in this work. }
\label{fig:isospin charm}
\end{figure}

\begin{figure}[h!]
%\centering
\begin{tabular}{cc}
\includegraphics[width=.33\textwidth, angle=270]{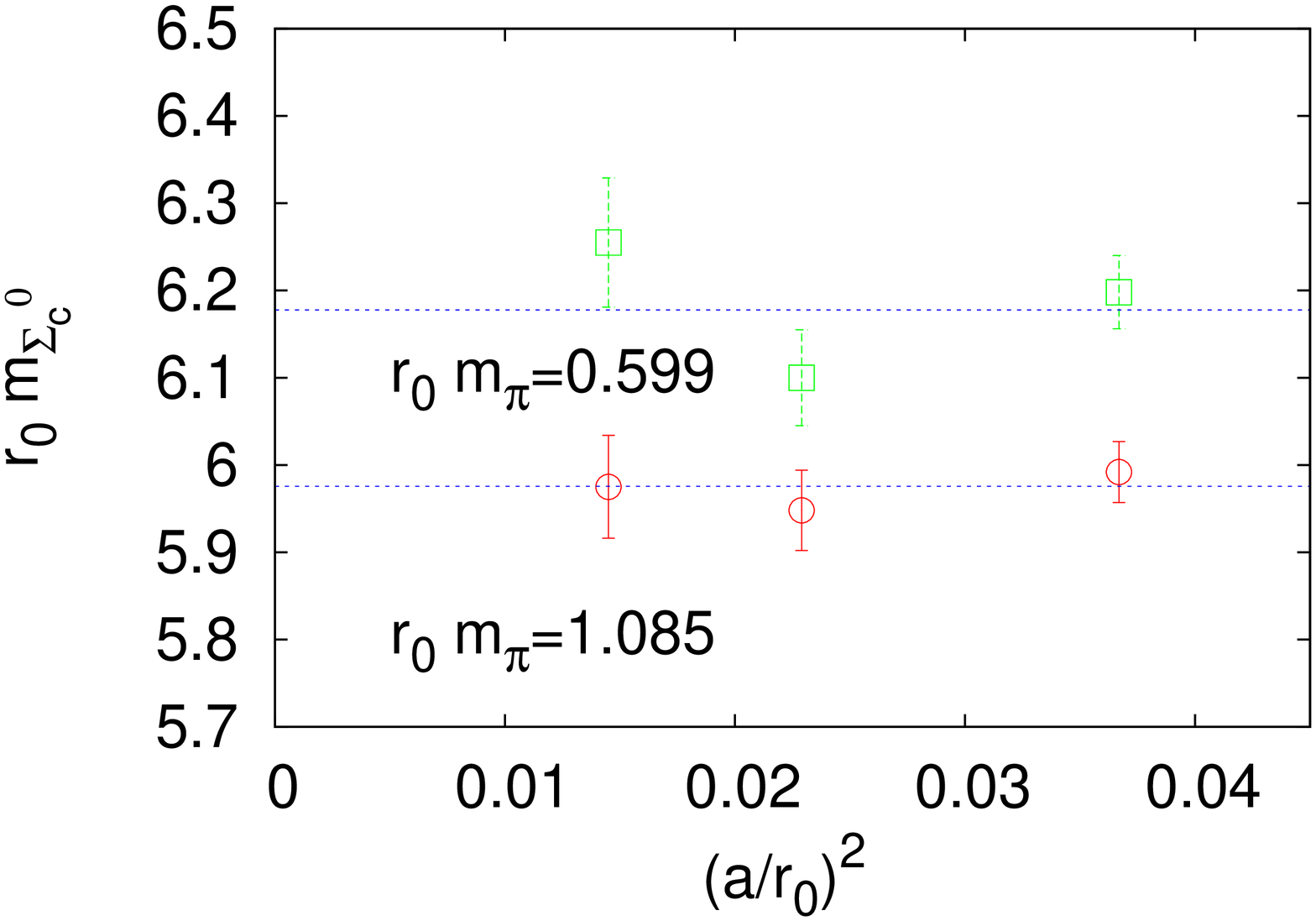} &
\includegraphics[width=.33\textwidth, angle=270]{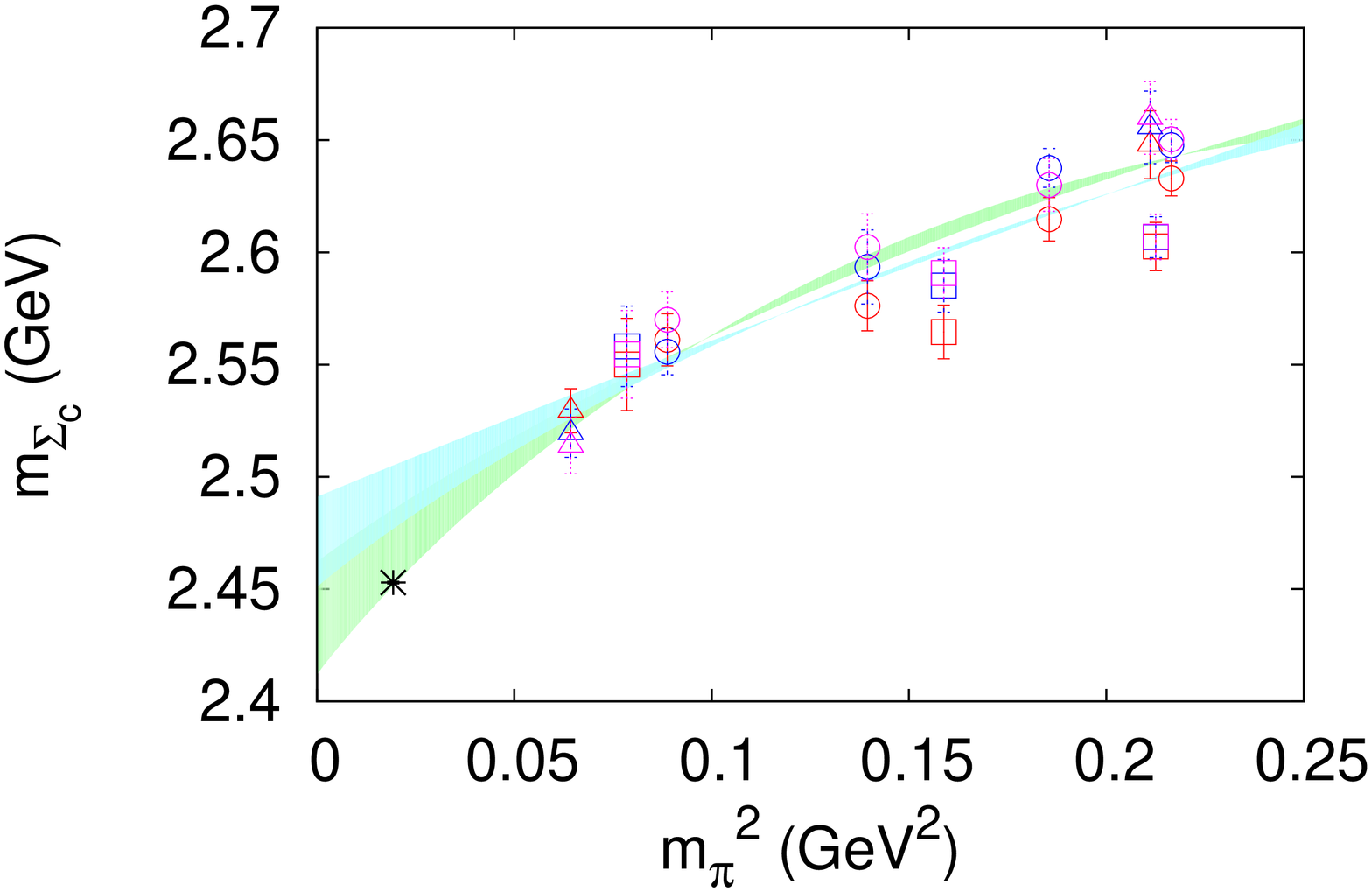} \\
\includegraphics[width=.33\textwidth, angle=270]{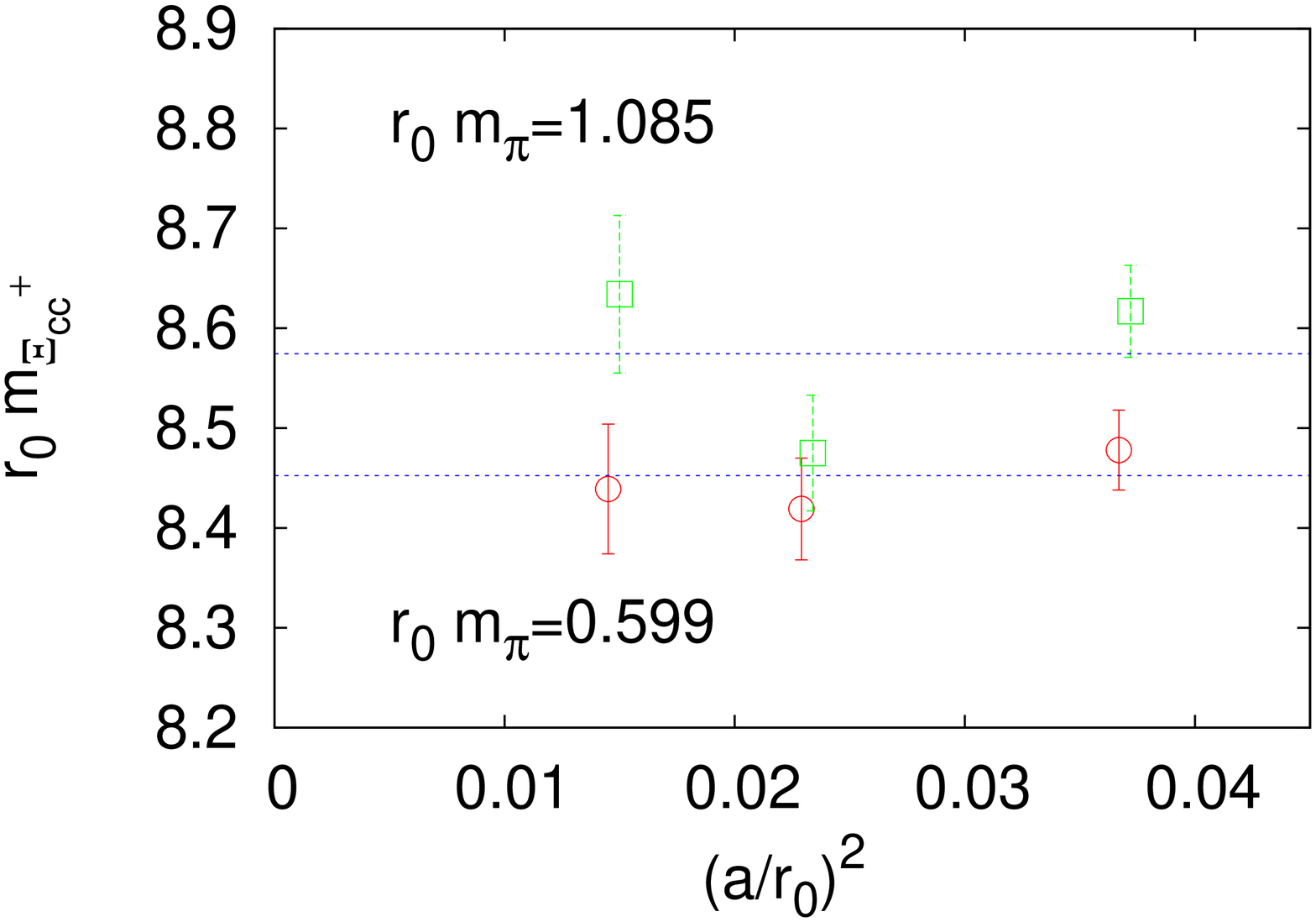} &
\includegraphics[width=.33\textwidth, angle=270]{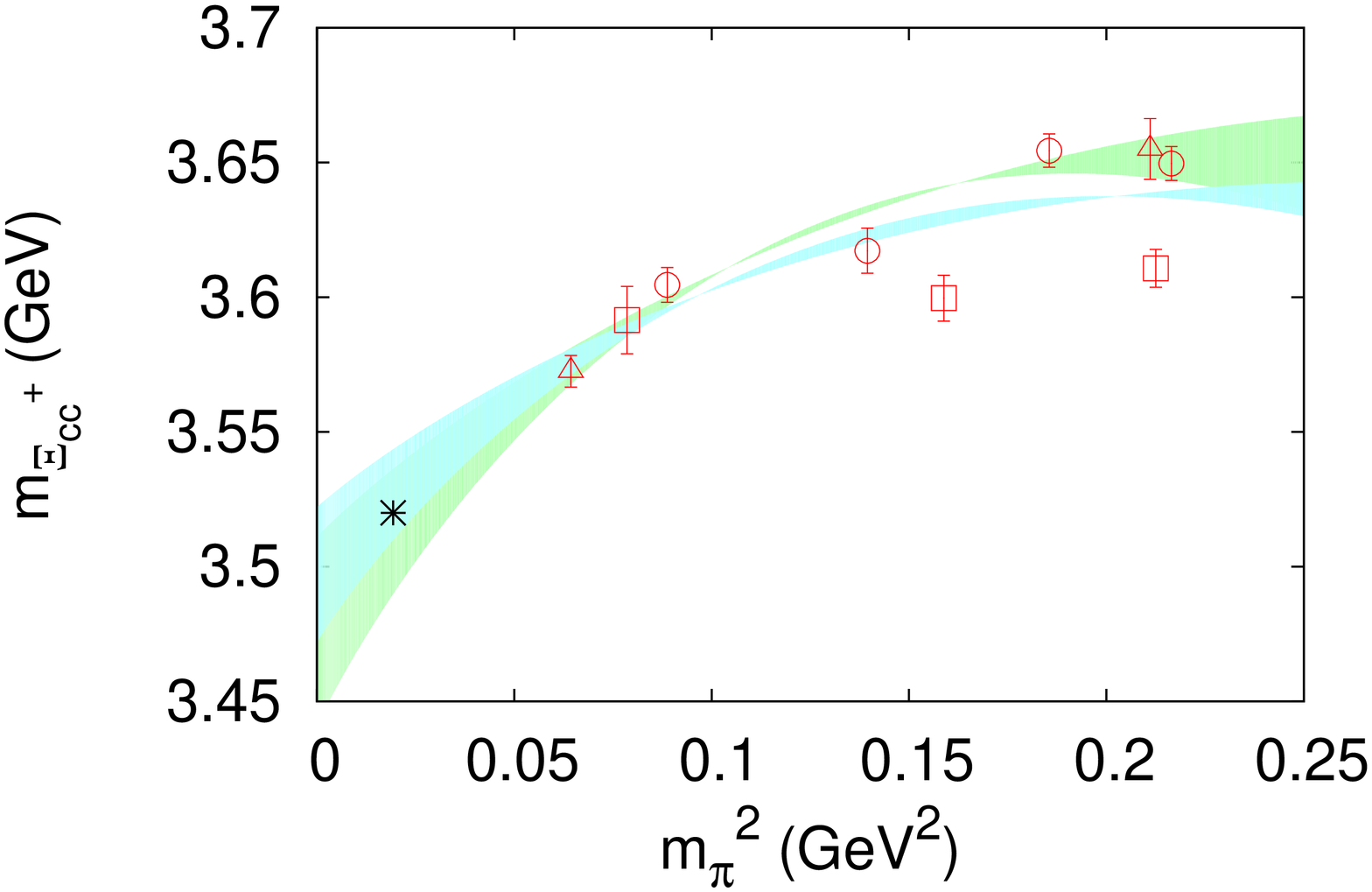} \\
\includegraphics[width=.33\textwidth, angle=270]{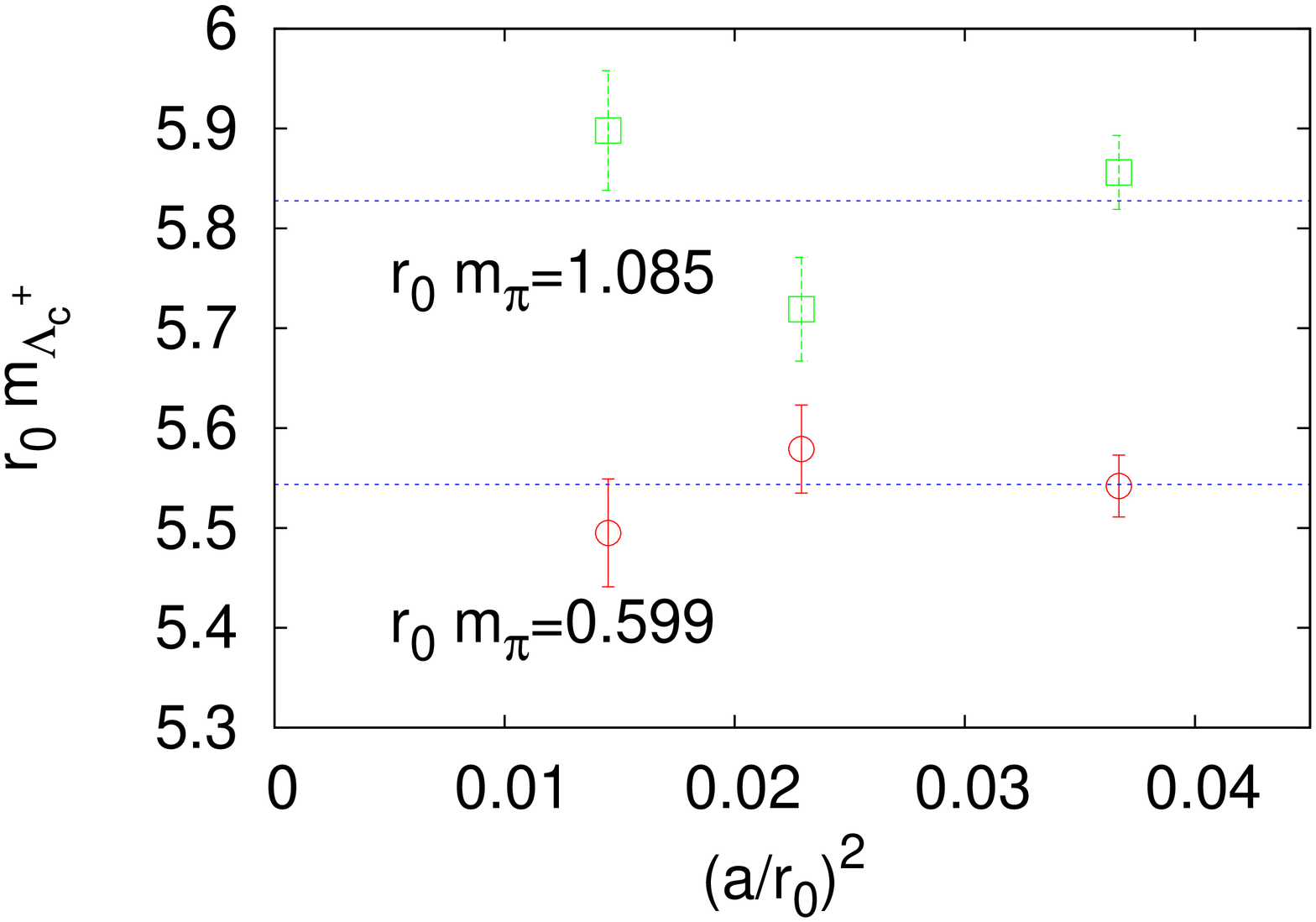} &
\includegraphics[width=.33\textwidth, angle=270]{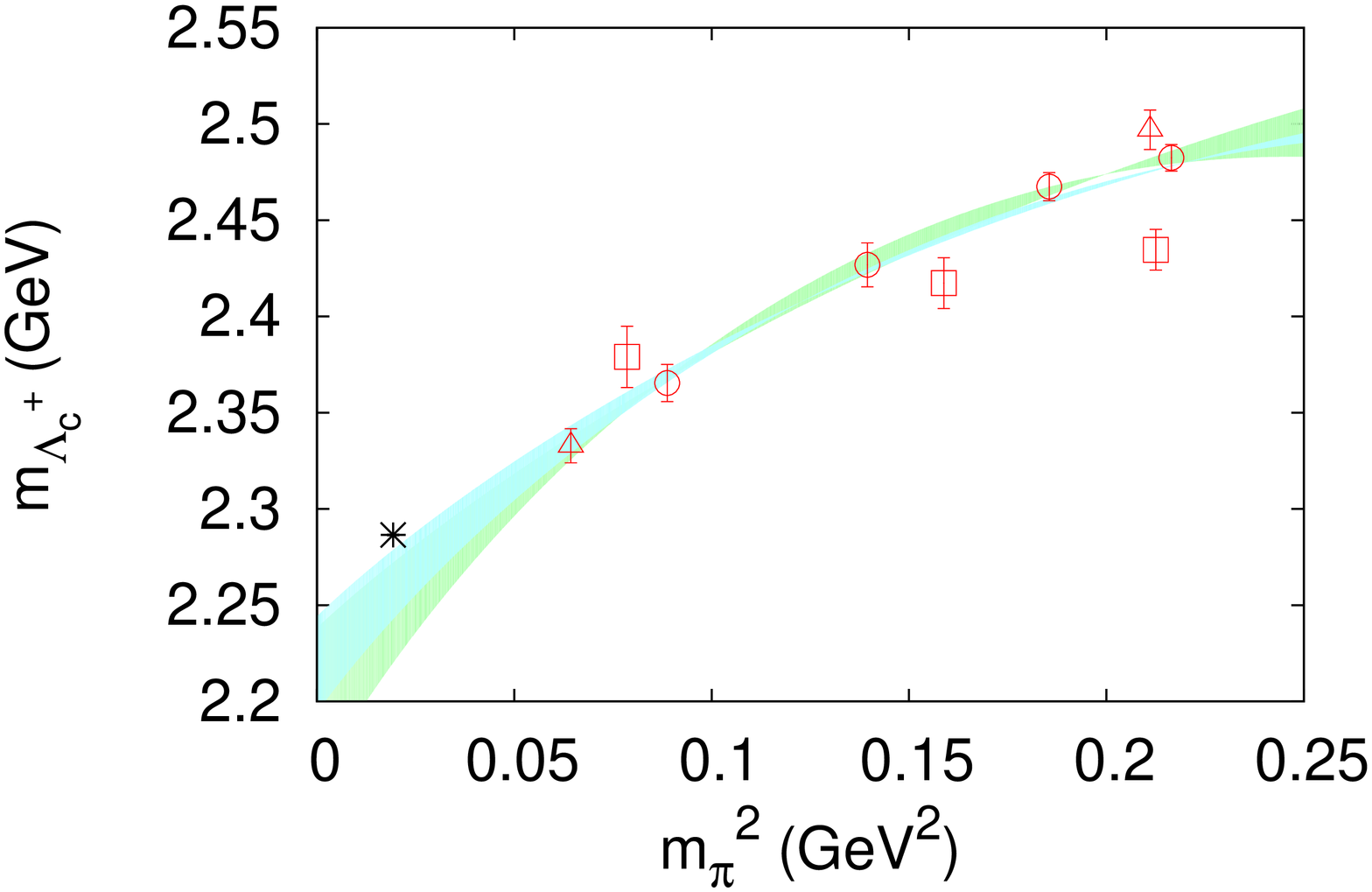} \\
\end{tabular}
\caption{Left: The mass of octet charm baryons in units of $r_0$ 
versus $(a/r_0)^2$ for several particles in the octet and decuplet at two
fixed values of the pion mass $m_\pi r_0$.
Right:  Chiral fits to  data at $\beta=3.9$ (circles) and at  $\beta=4.2$ (triangle) are shown with the green band and to all data with the lue band. 
For the $\Xi_{cc}$ we have extrapolated  $\Xi_{cc}^+$, whereas
 for $\Sigma_c$  we averaged over 
$\Sigma_c^{++}$ (purple), $\Sigma_c^{+}$ (blue) and $\Sigma_c^0$ (red). The spread of data at a given pion mass with the same symbol   indicates the level of isospin breaking for $\Sigma$.} 
\label{fig:continuum charm octet}
\end{figure}

\begin{figure}[h!]
%\centering
\begin{tabular}{cc}
\includegraphics[width=.33\textwidth, angle=270]{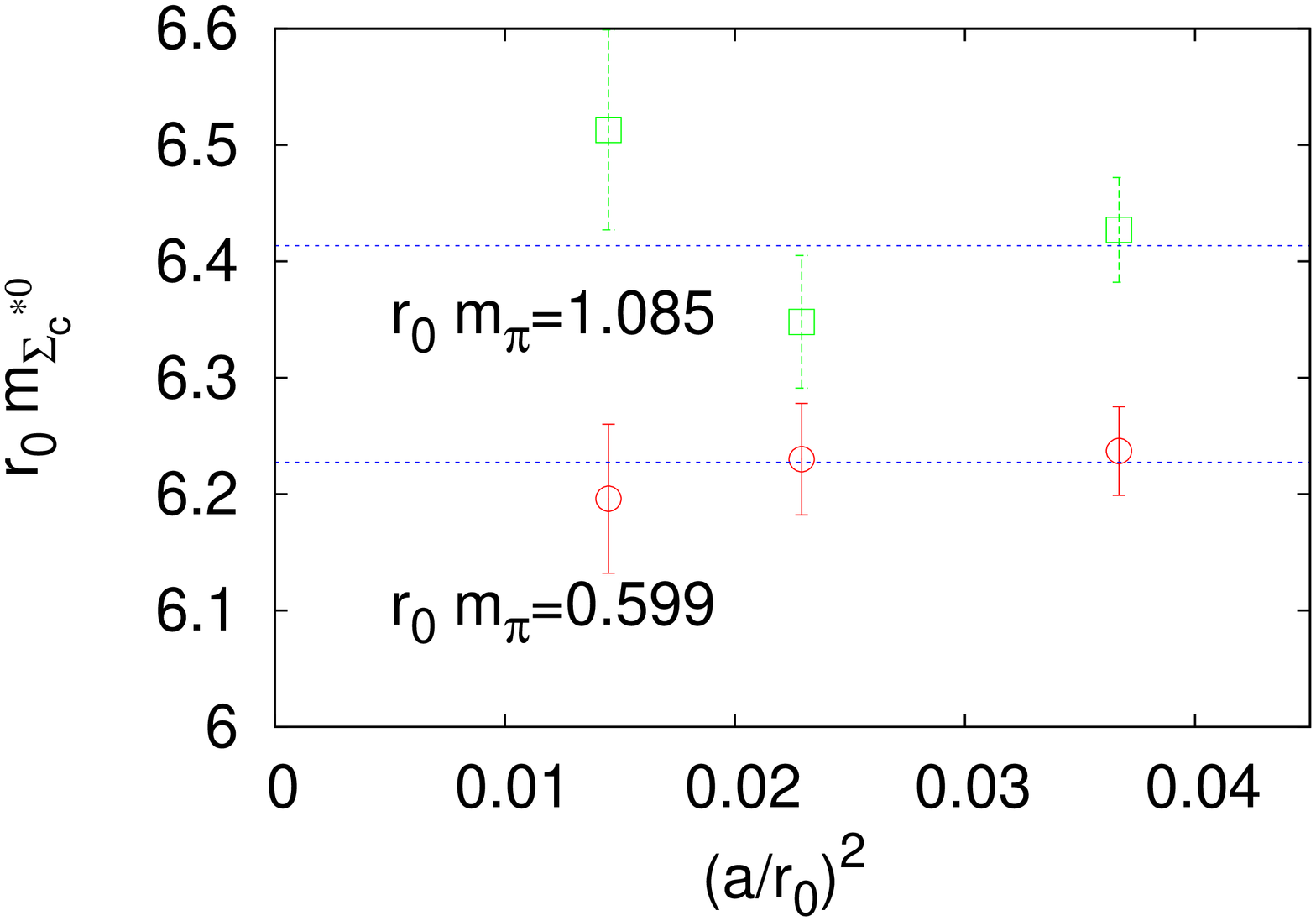} &
\includegraphics[width=.33\textwidth, angle=270]{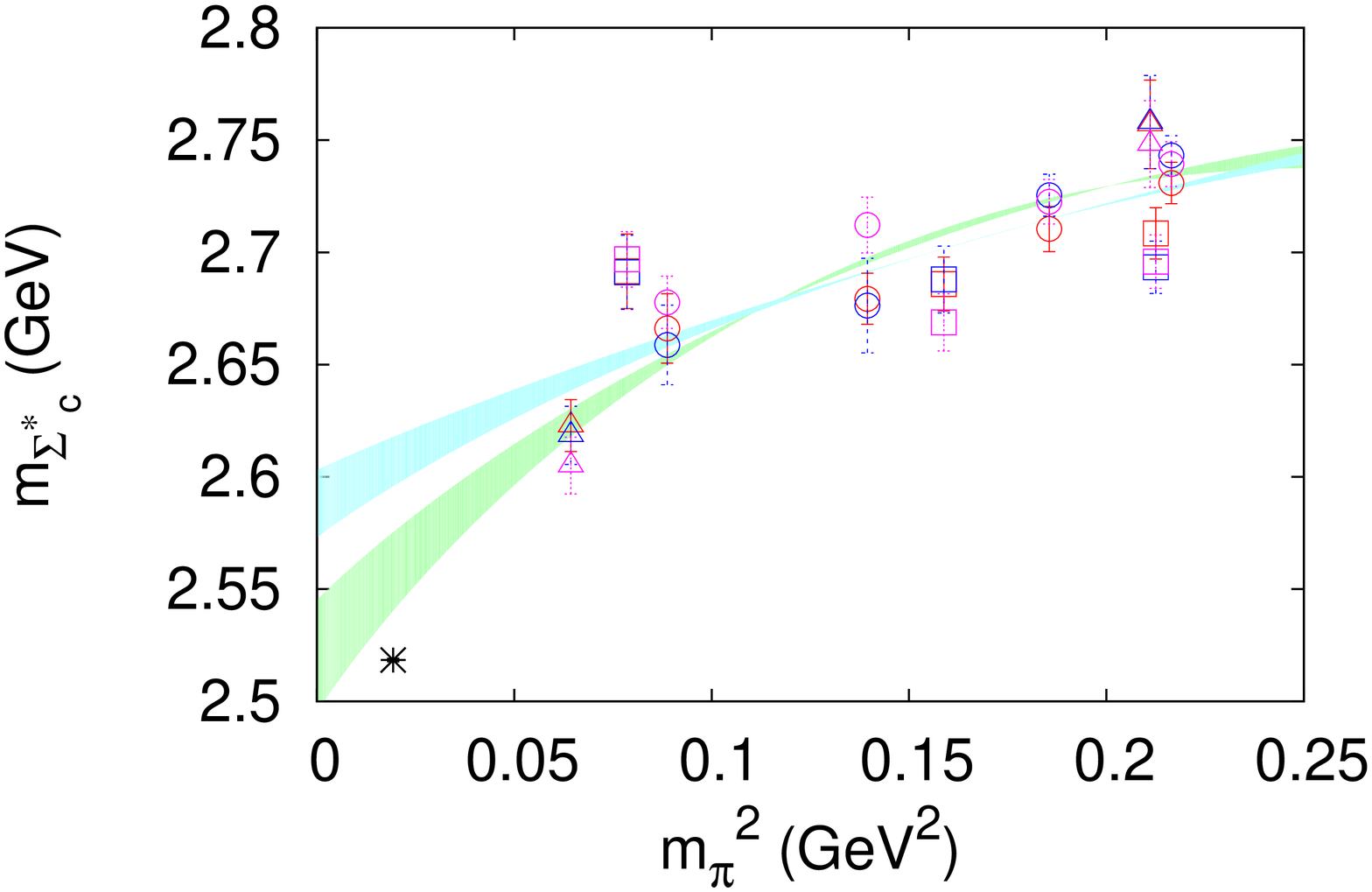} \\
\includegraphics[width=.33\textwidth, angle=270]{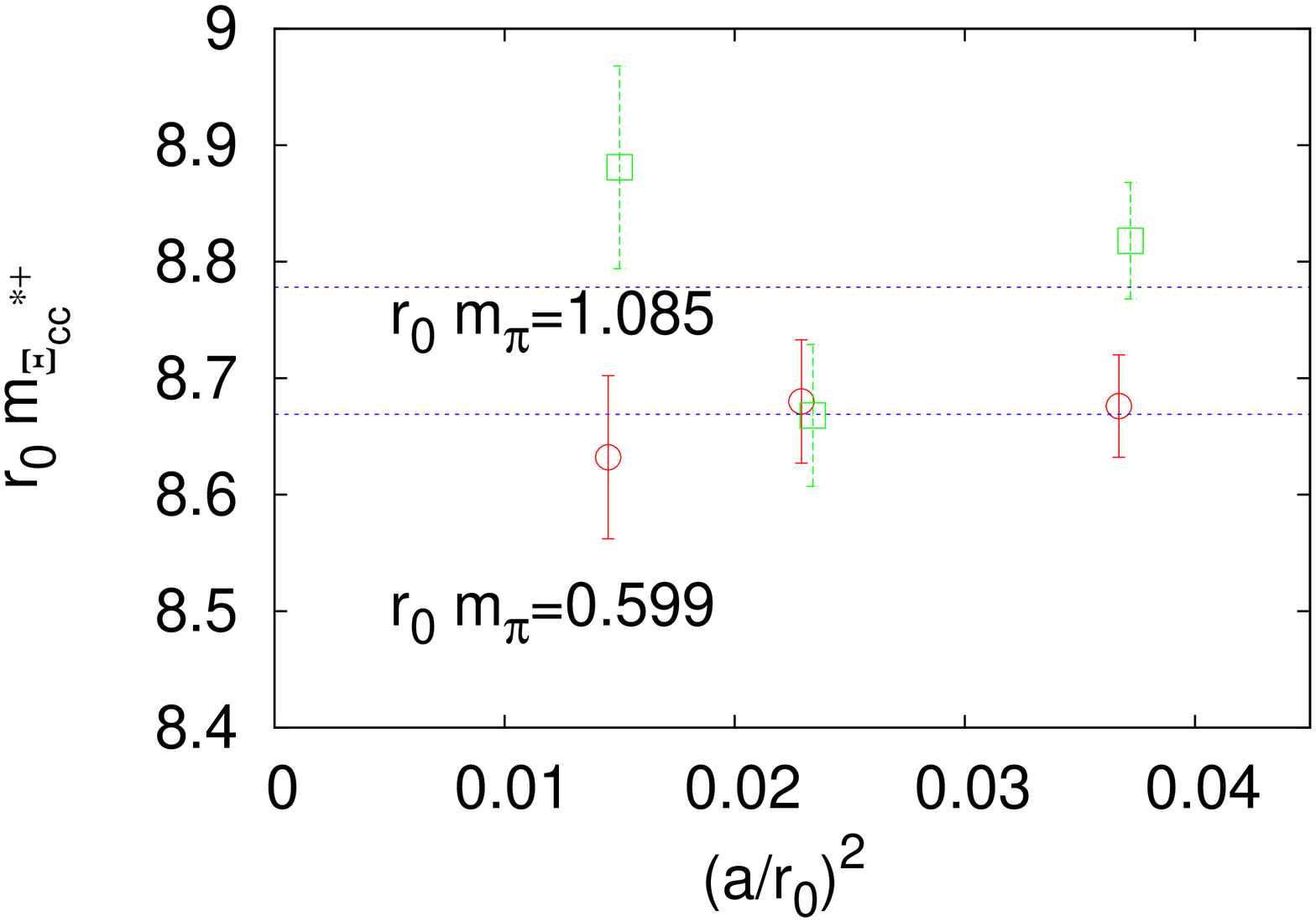} &
\includegraphics[width=.33\textwidth, angle=270]{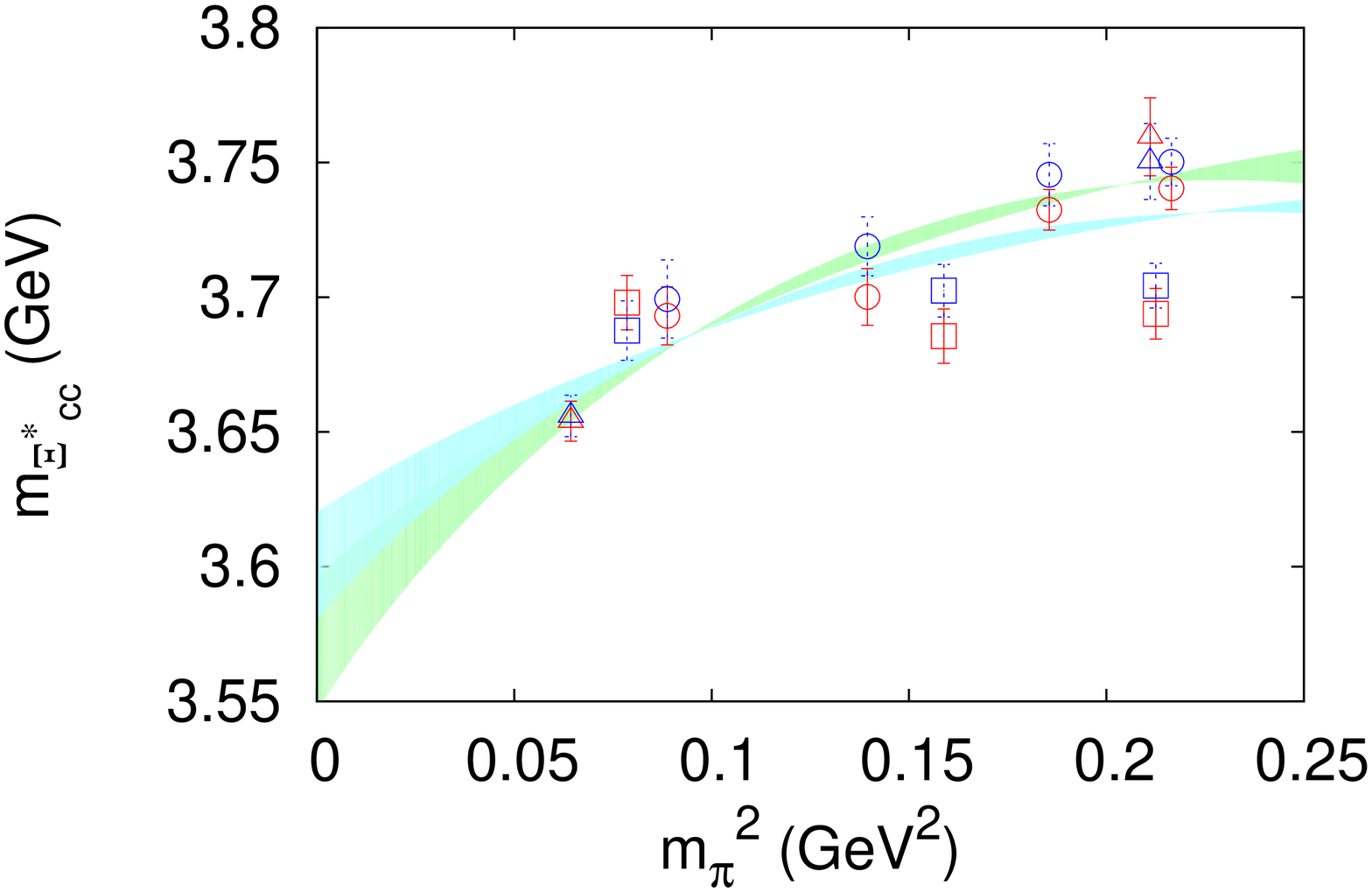} \\
\includegraphics[width=.33\textwidth, angle=270]{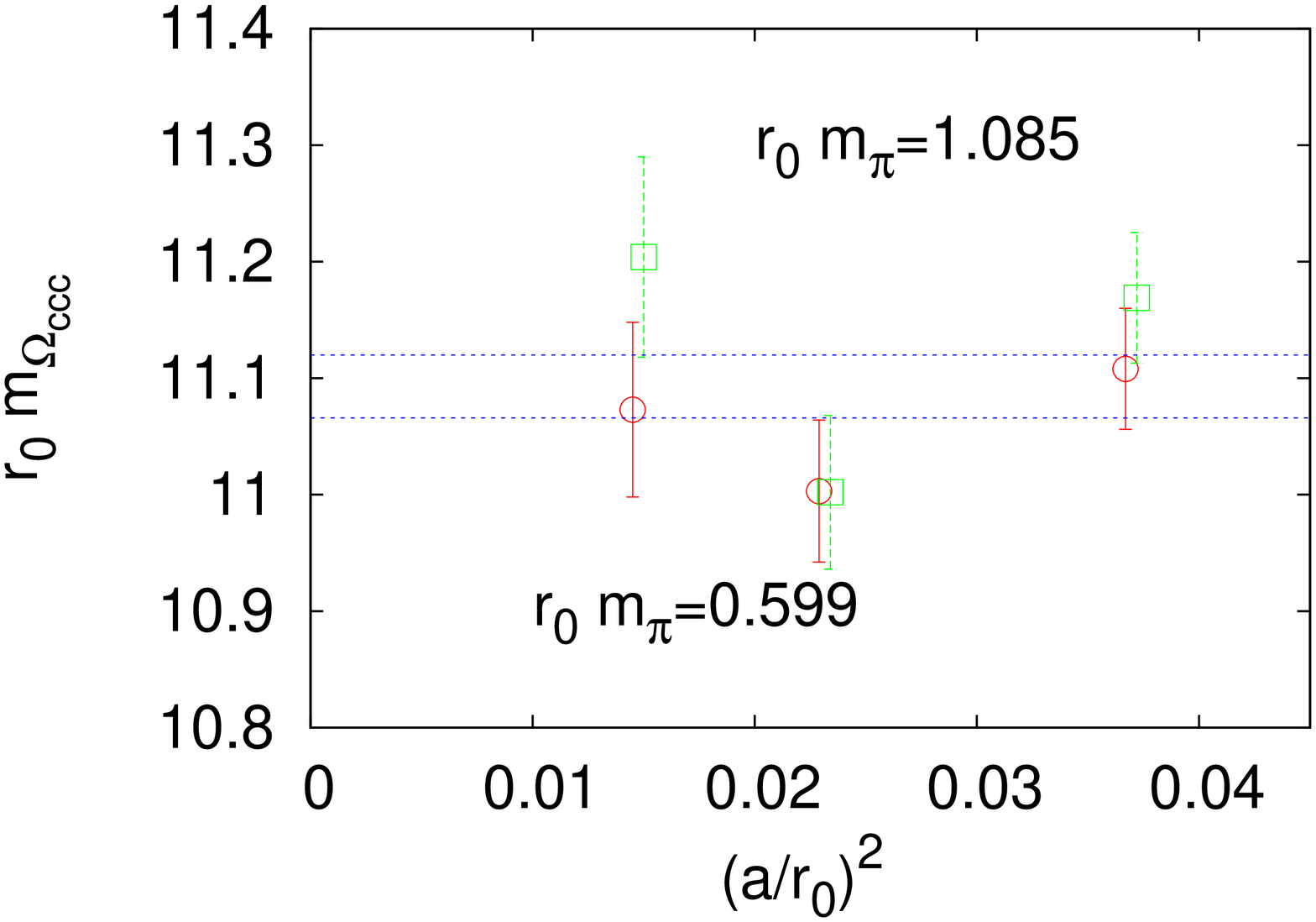} &
\includegraphics[width=.33\textwidth, angle=270]{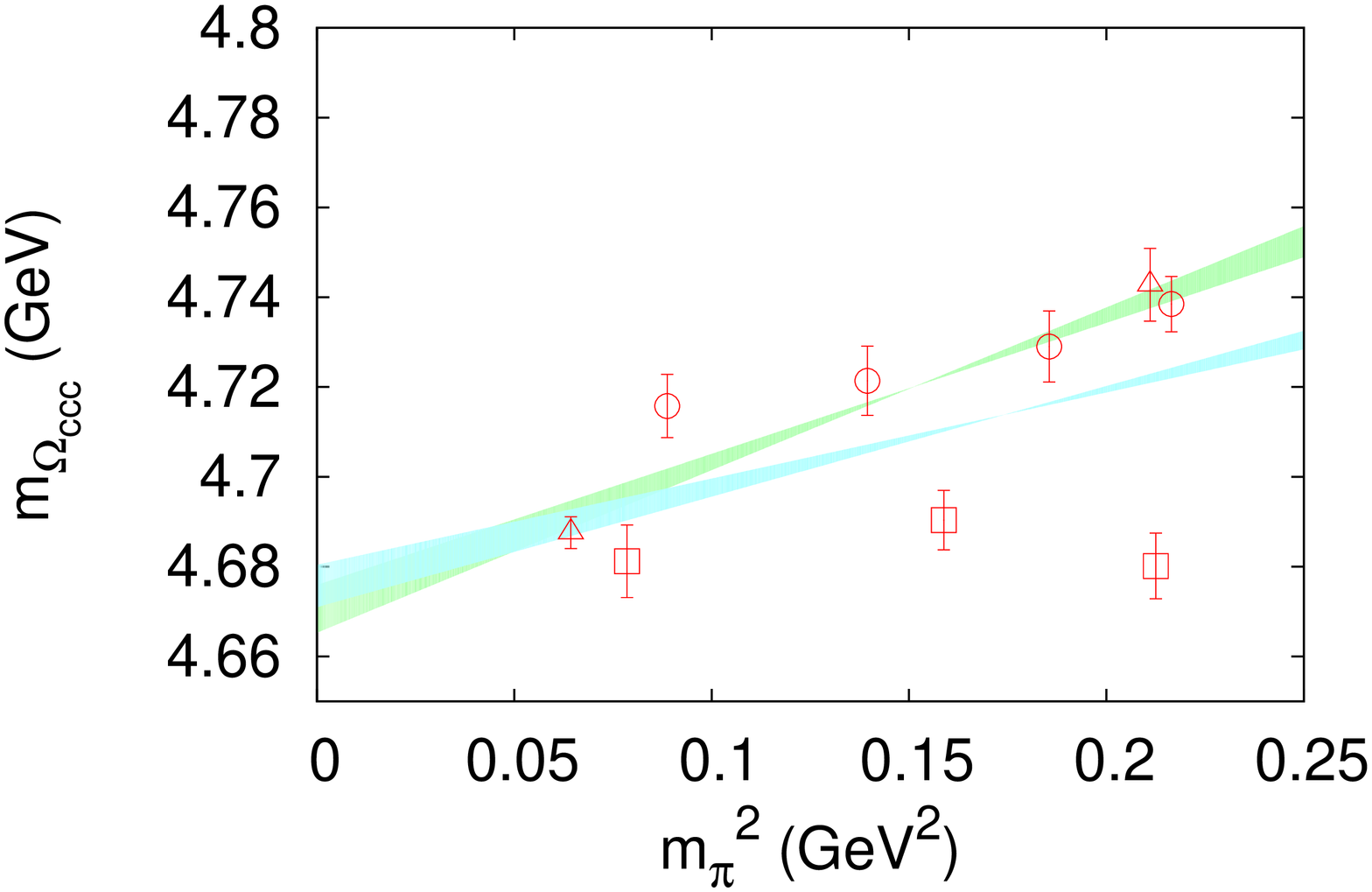} \\
\end{tabular}
\caption{Left: The mass of decuplet charm baryons in units of $r_0$ 
 versus $(a/r_0)^2$  at two
fixed values of $m_\pi r_0$.
Right: Chiral fits to the data at $\beta=3.9$ (circles) and  at $\beta=4.2$ (triangle) are shown with the green band and to all data with the blue band. 
For
 $\Xi_{cc}^*$ we averaged over 
$\Xi_{cc}^{*++}$ (blue) and $\Xi_{cc}^{*+}$ (red) and for $\Sigma_c^*$ we averaged over 
$\Sigma_c^{*++}$ (purple), $\Sigma_c^{*+}$ (blue) and $\Sigma_c^{*0}$ (red). The spread of data at a given pion mass with the same symbol 
  indicates the level of isospin breaking for the   $\Sigma^*_c$ and $\Xi^*_{cc}$ particles.} 
\label{fig:continuum charm decuplet}
\end{figure}

We follow the same analysis as in the case of the strange baryon sector.
In Fig.~\ref{fig:isospin charm} we show the mass difference between different 
charged states as a function of the lattice spacing at the smallest and
largest pion masses used in this work. As can be seen, the mass splittings
are zero at the smallest value of the lattice spacing for all
particles confirming restoration
of isospin symmetry in the continuum limit. 
Furthermore, except for the case of the $\Xi_{cc}$ mass, the mass splitting is
consistent with zero also at the other two $\beta$-values. Therefore,
for all particles except the $\Xi_{cc}$ one may average over the mass of different charge states.

In order to examine  the continuum limit we
interpolate our results at the three $\beta$-values at a given pion mass
in units of $r_0$.
In Figs.~\ref{fig:continuum charm octet} and \ref{fig:continuum charm decuplet}
we show the mass in  the octet and decuplet charm sector as a function
of lattice spacing for a given charge state, at the smallest and
largest value of the pion mass.  A linear fit in $(a/r_0)^2$ and a constant
fit yield consistent results at the continuum limit, albeit with large errors 
in the case of th elinear fit.
%%
%{\bf MARIO: also here mention the comparison between constant and linear 
%continuum limit. In the charm case the two limits always agree.}
%%
%As can be seen, at the smallest value
%of the pion mass results are consistent with a constant indicating
%small cut-off effects. 
%%
%{\bf MARIO: here refrase the sentence taking into account that 
%the referee is not happy about out interpretations involving
%cut-off effects.}
%%
We also note that at the largest pion mass, although 
results at $\beta=3.9$ are in agreement with those at $\beta=4.2$ indicating
neglegible ${\cal O}(a^2)$-dependence, at $\beta=4.05$ 
the results are systematically below. We note that we show only statistical
errors. Systematic errors due, for example, to the matching are not
shown. As discussed in the next section these are (5-10)\%.
Therefore, a reasonable way to extrapolate our results  in the charm
sector 
is to compare the chiral extrapolation
using all lattice data to those   using  results at $\beta=3.9$ and $\beta=4.2$.
We will take the different between the two values at the physical point as
an estimate of  a systematic error.

\section{Chiral extrapolation}

Having determined that ${\cal O}(a^2)$ effects are small for the lattice spacings
considered here we can combine  our lattice results at the various $\beta$-values to extrapolate to the physical pion mass (physical point). 

For the strange baryon sector, we consider SU(2) heavy baryon chiral perturbation, which was found to describe lattice data satisfactorily~\cite{Alexandrou:2009qu}. To leading one-loop one can described the pion mass dependence using  
\be
 m_B=m_B^{(0)}-4c_B^{(1)}m_{\pi}^2+c m_{\pi}^3
\label{chiral}
\ee
where $c$ is a known coefficient given in  Ref.~\cite{Alexandrou:2009qu}.
For completeness we give below the coefficients $c$~\cite{Nagels:1979xh,Nagels:1978sc}.
For the octet baryons $\Lambda$, $\Sigma$ and $\Xi$:
\be
c= - \frac{g_{\Lambda\Sigma}^2 }{16\pi f_\pi^2} \;, - \frac{2 g_{ \Sigma \Sigma}^2 + g_{\Lambda\Sigma}^2/3}{16\pi f_\pi^2}\; ,
  - \frac{3 g_{\Xi \Xi}^2}{16\pi f_\pi^2}  \quad,
\label{LO octet}
\ee
respectively, and for the decuplet baryons $\Sigma^*$, $\Xi^*$ and $\Omega$: 
\be
c=   - \frac{10}{9}\frac{g_{\Sigma^* \Sigma^*}^2 }{16\pi f_\pi^2}\;,
   - \frac{5}{3}\frac{g_{\Xi^* \Xi^*}^2 }{16\pi f_\pi^2}\;m_\pi^3 \; , 0\;.    \label{LO decuplet}
\ee

In addition we consider next to leading order SU(2) $\chi$PT results~\cite{Tiburzi:2008bk}. The expressions are included here for completeness:
\beq
m^{NLO}_\Lambda(m_\pi) &=& m^{(0)}_{\Lambda}-4c^{(1)}_{\Lambda}m_\pi^2  - \frac{g^2_{\Lambda\Sigma}}{(4\pi f_\pi)^2} \; {\cal F}(m_\pi,\Delta_{\Lambda \Sigma},\lambda)
                                                   - \frac{ 4g^2_{\Lambda\Sigma^*} } {(4\pi f_\pi)^2} \; {\cal F}(m_\pi,\Delta_{\Lambda \Sigma^*},\lambda) \nonumber \\
m^{NLO}_\Sigma(m_\pi)    &=&  m^{(0)}_{\Sigma}-4c^{(1)}_{\Sigma}m_\pi^2  - \frac{2 g_{ \Sigma \Sigma}^2}{16\pi f_\pi^2}\; m_\pi^3  -\frac{g^2_{\Lambda\Sigma}}{3(4\pi f_\pi)^2} \; {\cal F}(m_\pi,-\Delta_{\Lambda \Sigma},\lambda)
                                                   - \frac{4g^2_{\Lambda\Sigma^*}}{3(4\pi f_\pi)^2} \; {\cal F}(m_\pi,\Delta_{\Sigma\Sigma^*},\lambda) \nonumber \\
m^{NLO}_\Xi(m_\pi)            &=&  m^{(0)}_{\Xi}-4c^{(1)}_{\Xi}m_\pi^2 - \frac{3 g_{\Xi \Xi}^2}{16\pi f_\pi^2} \;m_\pi^3-\frac{2g_{\Xi^*\Xi}^2}{(4\pi f_\pi)^2} \; {\cal F}(m_\pi,\Delta_{\Xi\Xi^*},\lambda)    \label{NLO octet}
\eeq
and for the decuplet baryons:
\beq
m^{NLO}_{\Sigma^*}(m_\pi) &=& m^{(0)} _{\Sigma^* } - 4c^{(1)}_{\Sigma^*}m_\pi^2  - \frac{10}{9}\frac{g_{\Sigma^* \Sigma^*}^2 }{16\pi f_\pi^2}\;m_\pi^3  - \frac{2}{3(4\pi f_\pi)^2} \left[g_{\Sigma^*\Sigma} ^2
\; {\cal F}(m_\pi,-\Delta_{\Sigma\Sigma^*,\lambda}) + g_{\Lambda\Sigma^*}^2 \; {\cal F}(m_\pi,-\Delta_{\Lambda\Sigma^*,\lambda}) \right] \nonumber \\ 
m^{NLO}_{\Xi^*} (m_\pi)       &=& m^{(0)} _{\Xi^* } - 4c^{(1)}_{\Xi^*}m_\pi^2 - \frac{5}{3}\frac{g_{\Xi^* \Xi^*}^2 }{16\pi f_\pi^2}\;m_\pi^3- \frac{g_{\Xi^* \Xi}^2}{(4\pi f_\pi)^2}  \; {\cal F}(m_\pi,-\Delta_{\Xi\Xi^*,\lambda}) \nonumber \\
m^{NLO}_\Omega(m_\pi)     &=& m^{(0)} _{\Omega } - 4c^{(1)}_{\Omega}m_\pi^2
\label{NLO decuplet}
\eeq
with the non analytic function~\cite{Tiburzi:2005na}
\be\label{F}
{\cal F}(m,\Delta,\lambda) =(m^2-\Delta^2)\sqrt{\Delta^2-m^2+i\epsilon}
\;\log\left(\frac{\Delta-\sqrt{\Delta^2-m^2+i\epsilon}}{\Delta+\sqrt{\Delta^2-m^2+i\epsilon}}\right)
-\frac{3}{2}\Delta m^2\log\left(\frac{m^2}{\lambda^2}\right)-\Delta^3\log\left(\frac{4\Delta^2}{m^2}\right) \quad
\ee
depending on the threshold parameter $\Delta_{XY}= m^{(0)}_{Y}-m^{(0)}_X$ and on the 
scale $\lambda$ of chiral perturbation theory, fixed to
  $\lambda=1$~GeV.
For $\Delta>0$ the real part of the function ${\cal F}(m,\Delta,\lambda)$ has the property
\be
{\cal F}(m,-\Delta,\lambda) = \left \{\begin{array}{ll} -{\cal F}(m,\Delta,\lambda)& m<\Delta \\
-{\cal F}(m,\Delta,\lambda)+2\pi\left(m^2-\Delta^2\right)^{3/2} & m>\Delta\\
\end{array}\right.
\label{F symm}
\ee 
which corrects a typo in the sign of the
second term in Ref.~\cite{WalkerLoud:2008bp}.
We follow the procedure of Ref.~\cite{Alexandrou:2009qu} and fix the nucleon axial charge $g_A$ and pion decay constant $f_{\pi}$ to their experimental values
(we use the convention such that  $f_{\pi}=130.70$ MeV). The remaining pion-baryon axial coupling constants are taken from  SU(3) relations~\cite{Tiburzi:2008bk}.
The fit parameters extracted for fitting to the NLO  are given in Table~\ref{tab:chiral}. The deviation of the mean values obtained at the physical point 
when the results are fitted to leading order i.e. to Eq.(~\ref{chiral}) with 
$c=0$ and when they are fitted to the NLO expressions provide an estimate of the systematic error due to the chiral
extrapolation. We give this error in Table~\ref{tab:chiral}. In the case of the $\Omega$ there is no difference between leading order and next to leading order. 
Since the $\Omega$ contains three strange quarks any systematic error in the
tuning of the strange quark mass will be largest in this case. Having results at several values of the strange quark mass we can estimate the change in the $\Omega$ mass if the strange quark mass takes the maximum and minimum value
allowed by  the statistical error 
in the tuned strange quark mass. We take the
difference in  the mean values at the physical point obtained
by varying the strange quark mas to be the systematic error due to the tuning.
In Table~\ref{tab:chiral} we give the 
systematic error on the mass of $\Omega$ that we find following this procedure. This gives an
upper bound of the error expected from the uncertainty in the tuning. As can be
seen this is smaller as compared to the systematic error due to the chiral extrapolation and therefore it  is only taken into account for the case of the $\Omega$.

%%
%{\bf MARIO: This is the new table to estimate the systematic error at beta=3.9}
%%

\begin{table}
\begin{tabular}{|l|l|l|l||l|}
\hline
Particle(PDG) & $m_B^{(0)}$ (GeV) & $-4c_B^{(1)}$ (GeV$^{-1}$) & $\chi^2/$d.o.f. & $m$ (GeV)  \\ \hline
$\Sigma^{-}(1193)$  & 1.1368(70) & 3.560(40) & 2.7 & 1.1930(62)(660) \\ \hline
$\Xi^{-}(1315)$  & 1.3334(46) & 1.386(26) & 0.82 &  1.3538(41)(179) \\ \hline
$\Lambda(1116)$  & 1.0678(64) & 4.362(37) & 1.04 &  1.1276(57)(721) \\ \hline
$\Sigma^{*}_{av}(1384)$  & 1.4244(58) & 2.807(34) & 2.4 &  1.4757(51)(740) \\ \hline
$\Xi^{*-}(1531)$  & 1.4808(96) & 1.582(58) & 3.3 &  1.5113(89)(400) \\ \hline
$\Omega(1673)$ & 1.7522(76) & 0.361(45) & 2.0 & 1.7591(67)(200) \\
\hline
\end{tabular}
\caption{
The bare mass and $c_B$ (related to the $\sigma$-term by $\sigma_B=-4c_B m_\pi^2$) determined from fitting to the NLO expressions 
  for strange baryons at the tuned strange
quark mass. In the last column, we give the mass
 in GeV that we obtain at the physical point using the NLO expressions.
The second error given in the parenthesis is an estimate of the systematic
error coming from a comparison between the values obtained at the physical point using the leading order (LO) expressions  given in Eqs.~\ref{LO octet} and \ref{LO decuplet} and the NLO given by Eqs.~\ref{NLO octet} and \ref{NLO decuplet}. In the case of the $\Omega$
the systematic error is estimated by evaluating the impact of the error of the 
tuned strange quark mass on the extrapolated $\Omega$ mass.}
\label{tab:chiral}
\end{table}

\begin{table}
\begin{tabular}{|l|l|l|}
\hline
Particle(PDG) & $m$ (GeV)  & $\Delta m$  (GeV)  \\ \hline
$\Sigma_{c,av}(2.454)$  & 2.494(47) & 0.143   \\ \hline
$\Xi_{cc}^+$  & 3.563(25) & 0.397  \\ \hline
$\Lambda_{c}^+(2286)$  & 2.229(43) & 0.223   \\ \hline
$\Sigma^{*}_{c,av}(2.520)$ & 2.650(39) & 0.147    \\ \hline
$\Xi^{*}_{cc,av}$  & 3.672(42) & 0.274  \\ \hline
$\Omega_{ccc}$ & 4.702(11) & 0.308    \\
\hline
\end{tabular}
\caption{For each particle listed in the first column we give in the second column its mass at the physical pion mass using for the chiral extrpoaltion the masses computed at the
tuned value of the charm quark mass $m_c$. In the third column we give the mass difference between the baryon masses  obtained at the tuned value of $m_c$ and at the tuned value plus the error,  after extrapolation to the physical point.
This is done at $\beta=3.9$ where we have computed the masses at $m_c \pm$ error. }
\label{tab:charm mass 3.9}
\end{table}

\begin{table}
\begin{tabular}{|l|l|l|l|l||l|}
\hline
Particle(PDG) & $m_B^0$ (GeV) & $-4c_B$ (GeV$^{-1}$) & $c$ (GeV$^{-2}$) & $\chi^2/$d.o.f. & $m$ (GeV) \\ \hline
$\Sigma_{c,av}(2.454)$  & 2.437(25) & 1.92(54) & -2.09(91) & 1.1 & 2.468(17)(23) \\ \hline
$\Xi_{cc}^+$  & 3.476(35) & 2.39(83) & -3.39(1.5) & 2.7 & 3.513(23)(14) \\ \hline
$\Lambda_{c}^+(2286)$  & 2.198(40) & 2.99(96) & -3.6(1.7) & 0.10 & 2.246(27)(15) \\ \hline
$\Sigma^{*}_{c,av}(2.520)$  & 2.520(25) & 2.37(51) & -2.96(86) & 1.3 & 2.556(18)(51) \\ \hline
$\Xi^{*}_{cc,av}$  & 3.571(25) & 2.02(57) & -2.62(99) & 1.0 & 3.603(17)(21) \\ \hline
$\Omega_{ccc}$ & 4.6706(53) & 0.327(35) & 0. & 2.5 & 4.6769(46)(30) \\
% old $\Omega_{ccc}$ & 4.604(50) & 2.2(1.3) & -3.4(2.3) & 2.3 & 4.637(31) \\
\hline
\end{tabular}
\caption{Parameters of the chiral fit for charm baryons at the tuned charm
quark mass fitting results at $\beta=3.9$ and $\beta=4.2$. The last column is our prediction (in GeV) at the
physical point. The statistical error is given in the first paranthesis
and the systematic, computed by comparing the fit with all lattice data, in 
the second parenthesis.}
\label{tab:charm mass}
\end{table}

In Figs.~\ref{fig:continuum octet} and \ref{fig:continuum decuplet} we show
the chiral extrapolation for the octet and the decuplet. In the case of the $\Sigma$ and $\Xi^*$ the physical point is reproduced. However, for most
other particles the lattice results extrapolate to a higher value. The worse 
deviation is seen for the $\Omega$. Since this has three strange quarks
it may indicate that the tuning 
of the  strange quark mass performed using the kaon mass
introduces a systematic
error. One can study partial
quenching effects using  twisted mass fermion simulations
 with a dynamical strange quark. This will be considered in a future study.

In the charm baryon sector we 
use the Ansatz 
\be
m_B=a+bm_{\pi}^2+cm_{\pi}^3 ,
\label{chiral charm}
\ee
motivated by SU(2) HB$\chi$PT to leading one-loop 
order, with
  $c$ taken as a fit parameter. For the $\Omega_{ccc}$ we set $c=0$
since one does not expect a cubic term.

In order to assess the systematic error associated with the tuning of the charm
quark mass we consider our results at $\beta=3.9$. At this value of $\beta$ we have
computed the charm baryon masses at the tuned value of the charm quark
and at values of the charm quark shifted by the error on the tuned value.
Since these computations were performed at four different light quark masses
we can perform a chiral extrapolion using using the Ansatz of Eq.~(\ref{chiral charm} for the set of masses obtained
at the tuned value and at the value shifted by the error. The difference in the masses obtained at the physical pion mass is given in Tab.~\ref{tab:charm mass 3.9}.
As can be seen, this difference intorduces an error that varies between about 5\% and 10\%. This gives an estimate of the systemastic error due to the tuning
of the charm quark mass.  Since this analysis can only be done at $\beta=3.9$ 
we can only make a qualitative estimate of this error. Therefore in what follows
we will not quote this error on our values,. However, one has to bear in mind
that our final values can have a systematic error of about 10\% due to the
tunning.

In Figs.~\ref{fig:continuum charm octet} and \ref{fig:continuum charm decuplet} we show fits at for our three $\beta$ values.
We show fits using all data and fits using only data at $\beta=3.9$ and $\beta=4.2$. The latter case yields a better fit with a smaller value of $\chi/{\rm d.o.f.}$ and this is the value quoted in Table~\ref{tab:charm mass}. This is particularly noticeable for the case of $\Omega_{ccc}$ where
the results at $\beta=4.05$ are systematically lower. This maybe due to a small mismatch in
the tuned value of the charm quark mass, which for the $\Omega_{ccc}$ that
contains three charm quarks  would lead to the largest deviation. 
We take the difference in the extrapolated values at the physical points
when we exclude the $\beta=4.05$ data from the fit as a systematic error.

%{\bf Mario: Same as before: I wouldn't put the table}

%\begin{table}
%\begin{tabular}{|l|l|l|l|}
%\hline
%$m_\pi=0.254$ GeV & & & \\ \hline
%Particles & $\Delta m(3.9)$ & $\Delta m(4.05)$ & $\Delta m(4.2) $\\ \hline
%$\Sigma_c^++ - \Sigma_c^0$ & 0.014(11)  & 0.020(15) & -0.015(23) \\ \hline
%$ \Xi_{cc}^{++} - \Xi_{cc}^+$ & 0.0278(80) & 0.0204(97) & 0.005(12) \\ \hline
%$\Sigma_c^{*++} - \Sigma_c^{*0}$ & 0.009(11)  & 0.011(17) & -0.018(24) \\ \hlin%e
%$\Xi_{cc}^{*++} - \Xi_{cc}^{*+}$ & 0.011(10)  & -0.008(12) & 0.002(15) \\ \hline
%\hline
%$m_\pi=0.459$ GeV & & & \\ \hline
%Particles & $\Delta m(3.9)$ & $\Delta m(4.05)$ & $\Delta m(4.2) $\\ \hline
%$\Sigma_c^++ - \Sigma_c^0$ & 0.018(16)  & 0.010(22) & 0.012(32) \\ \hline
%$ \Xi_{cc}^{++} - \Xi_{cc}^+$ & 0.021(12) & 0.018(14) & 0.013(22) \\ \hline
%$\Sigma_c^{*++} - \Sigma_c^{*0}$ & 0.009(19)  & -0.017(24) & 0.009(39) \\ \hlin%e
%$\Xi_{cc}^{*++} - \Xi_{cc}^{*+}$ & 0.011(15)  & 0.015(18) & 0.009(29) \\ \hline
%\end{tabular}
%\caption{Difference of mass in GeV for particles with different isospin in 
%the charm baryon case.}
%\label{tab:charm}
%\end{table}

The extrapolation of the lattice data reproduce the mass of experimentally
measured charm baryon masses within a standard deviation, namely the mass
of the $\Sigma_c$, the $\Lambda_c$
and  the $\Sigma_c^*$. Therefore,  the extrapolated 
lattice value can be taken as a prediction for the
mass of the $\Xi^*_{cc}$ and the $\Omega_{ccc}$, within one standard deviation.  

\section{Comparison with the results of other lattice formulations}

%%
%{\bf MARIO: Here we should cite 1102.5300.}
%%

In this section we compare our results with those
using different discretization schemes by other collaborations. 
We also include a comparison for the nucleon and $\Delta$ masses although
they were not discussed in detail until now.

\begin{figure}[h!]
%\centering
\begin{tabular}{cc}
\includegraphics[width=.33\textwidth,angle=270]{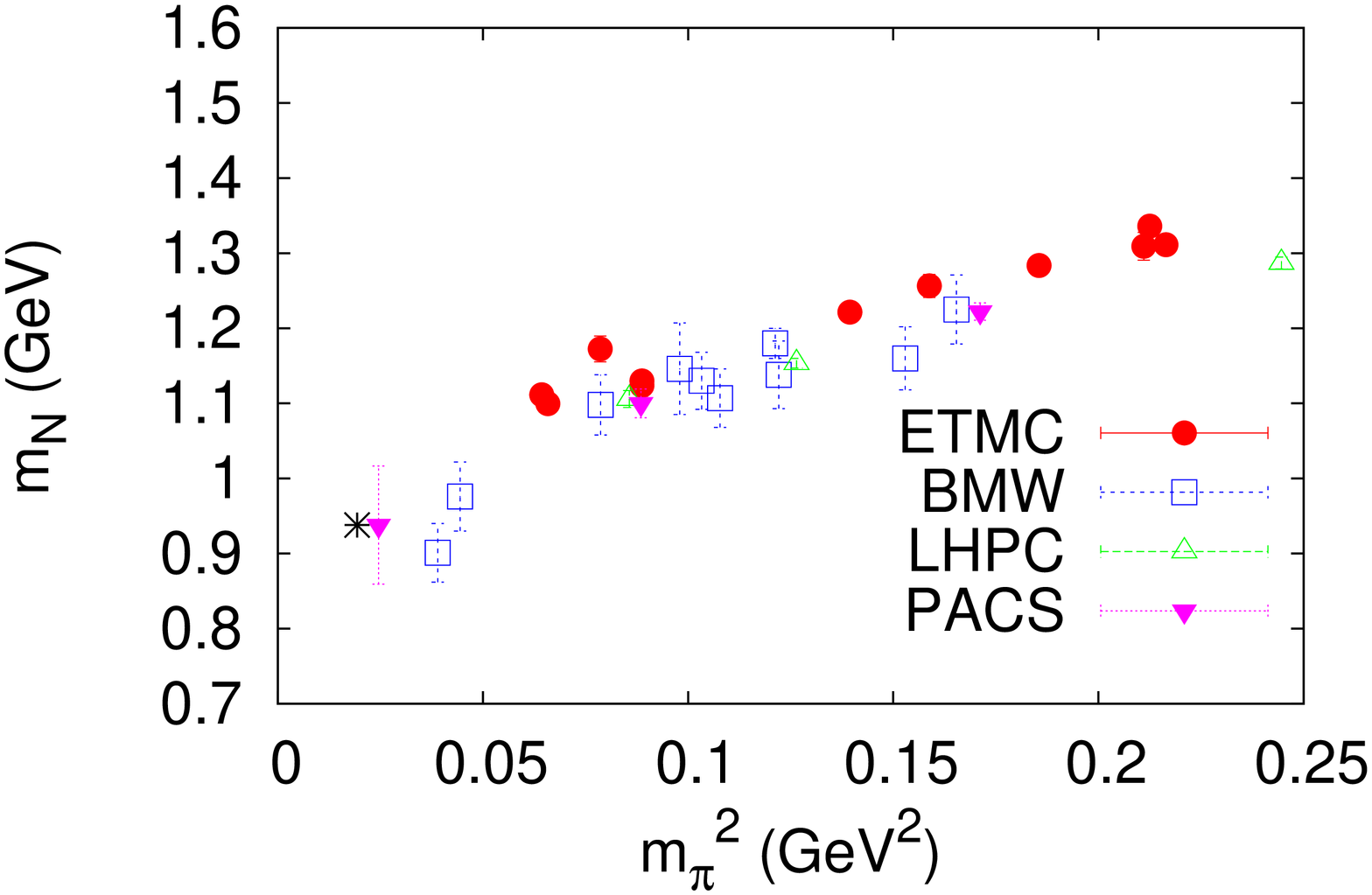} &
\includegraphics[width=.33\textwidth,angle=270]{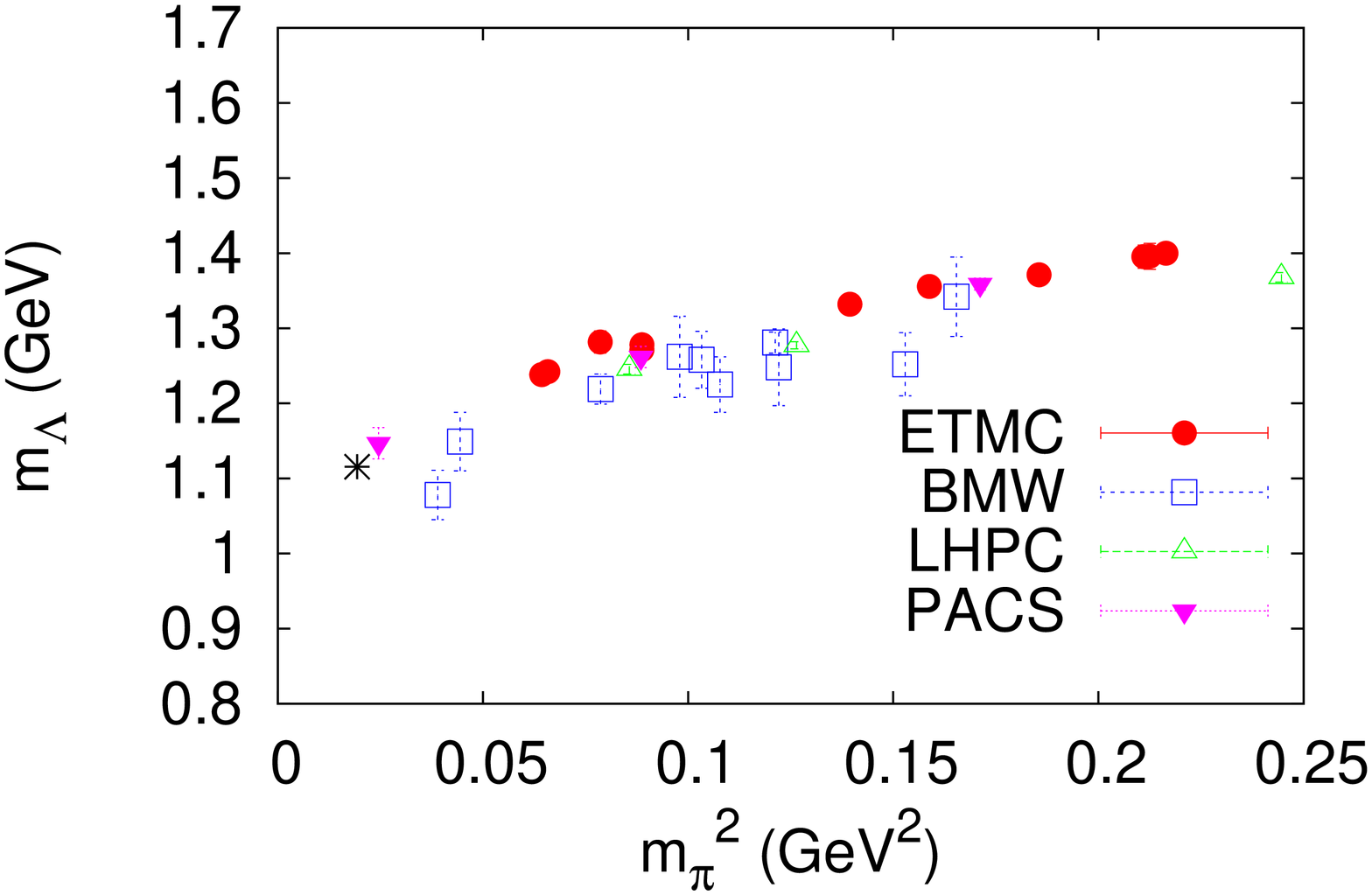} \\
\includegraphics[width=.33\textwidth,angle=270]{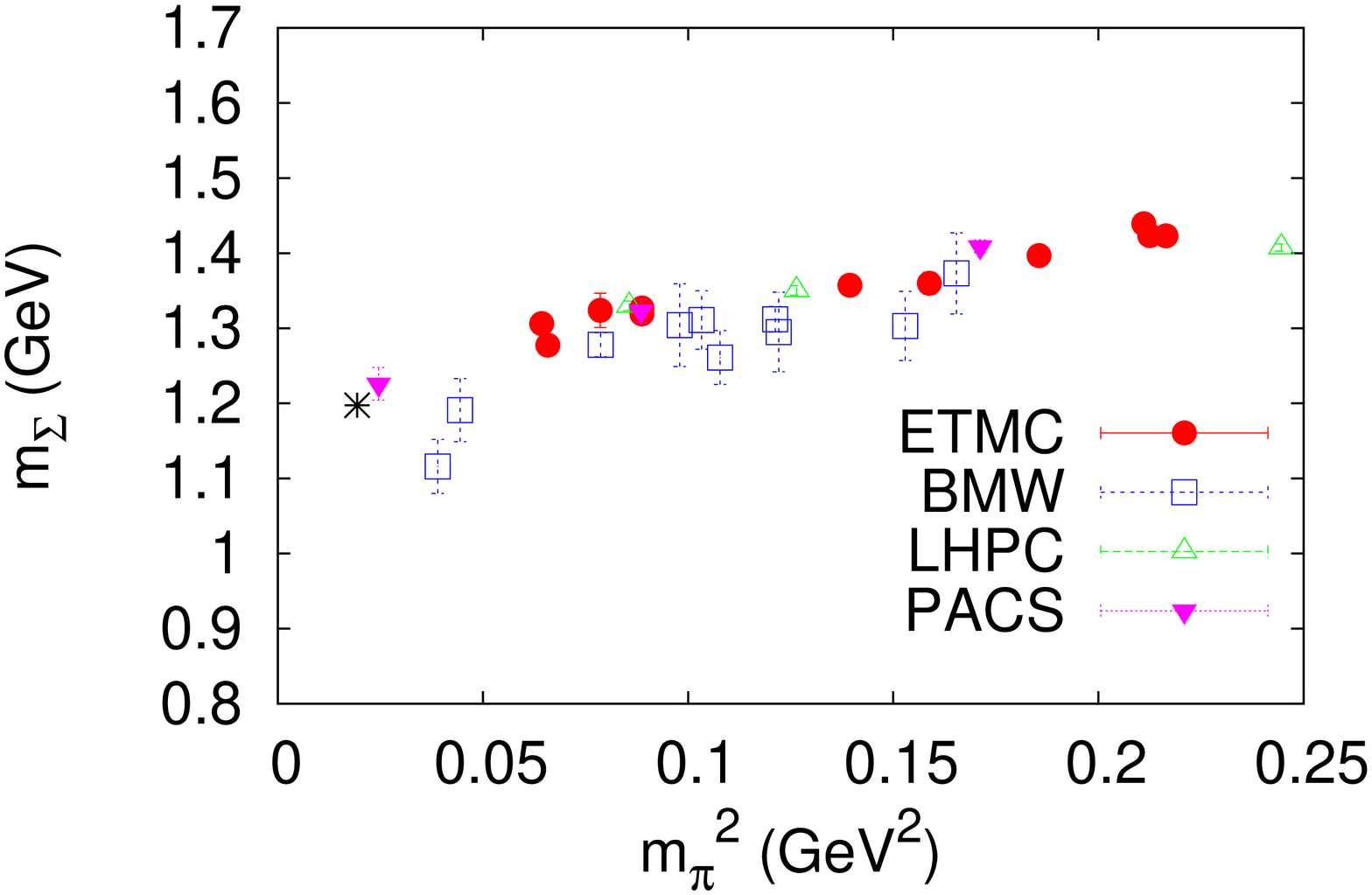} &
\includegraphics[width=.33\textwidth,angle=270]{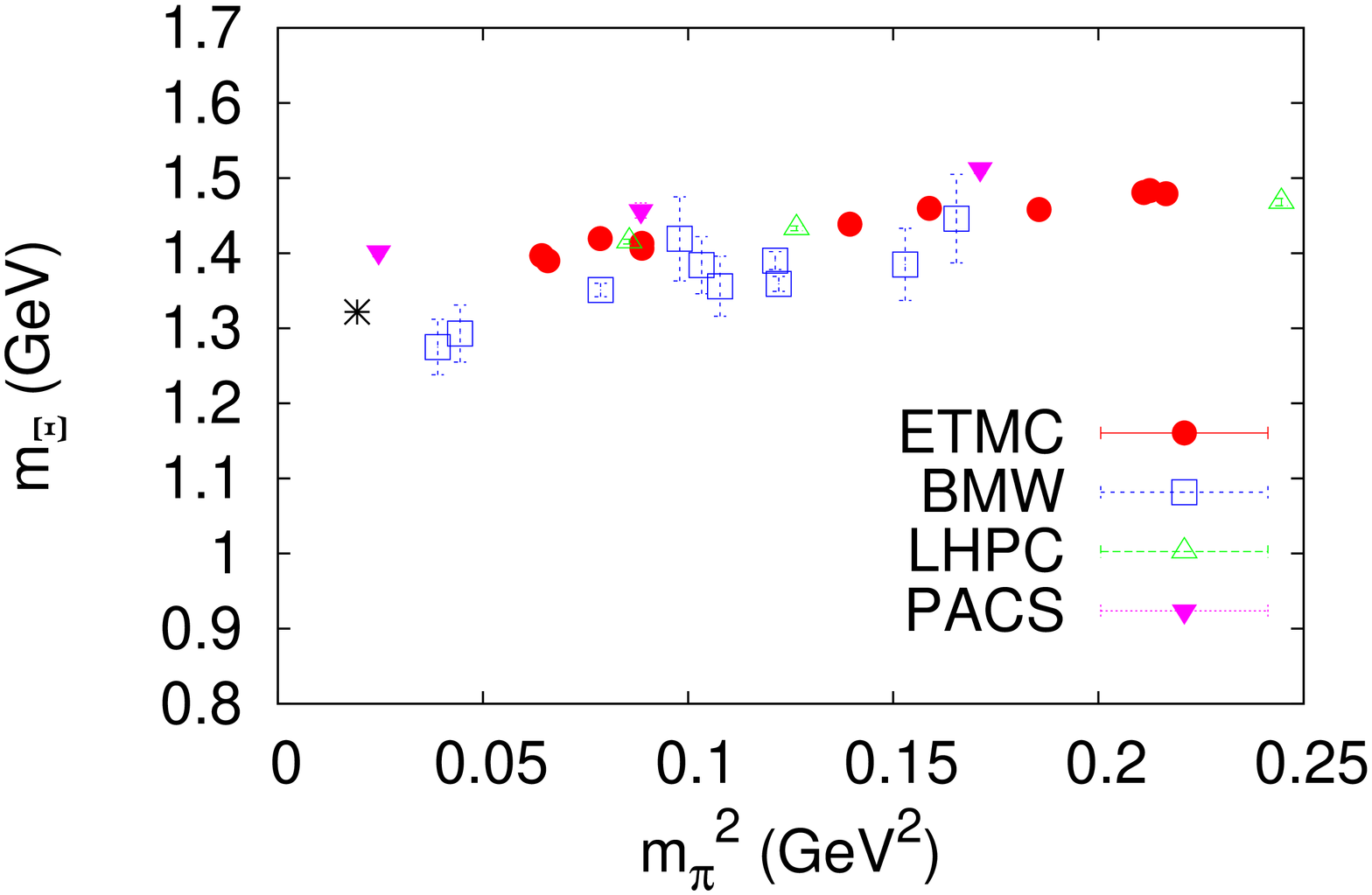} \\
\end{tabular}
\caption{The results of this work for the octet strange baryons are shown with the filled (red) circles,  results 
using clover fermions by the BMW Collaboration are shown with the open (blue) square~\cite{Durr:2008zz} and by the PACS-CS Collaboration with the filled (magenta) triangles~\cite{Aoki:2008sm}, and domain wall valence
quarks on a staggered sea by the LHPC with the open (green) triangles~\cite{WalkerLoud:2008bp}. The experimental value
is shown with the asterisk.}
\label{fig:comparison}
\end{figure}

\begin{figure}[h!]
%\centering
\begin{tabular}{cc}
\includegraphics[width=.33\textwidth,angle=270]{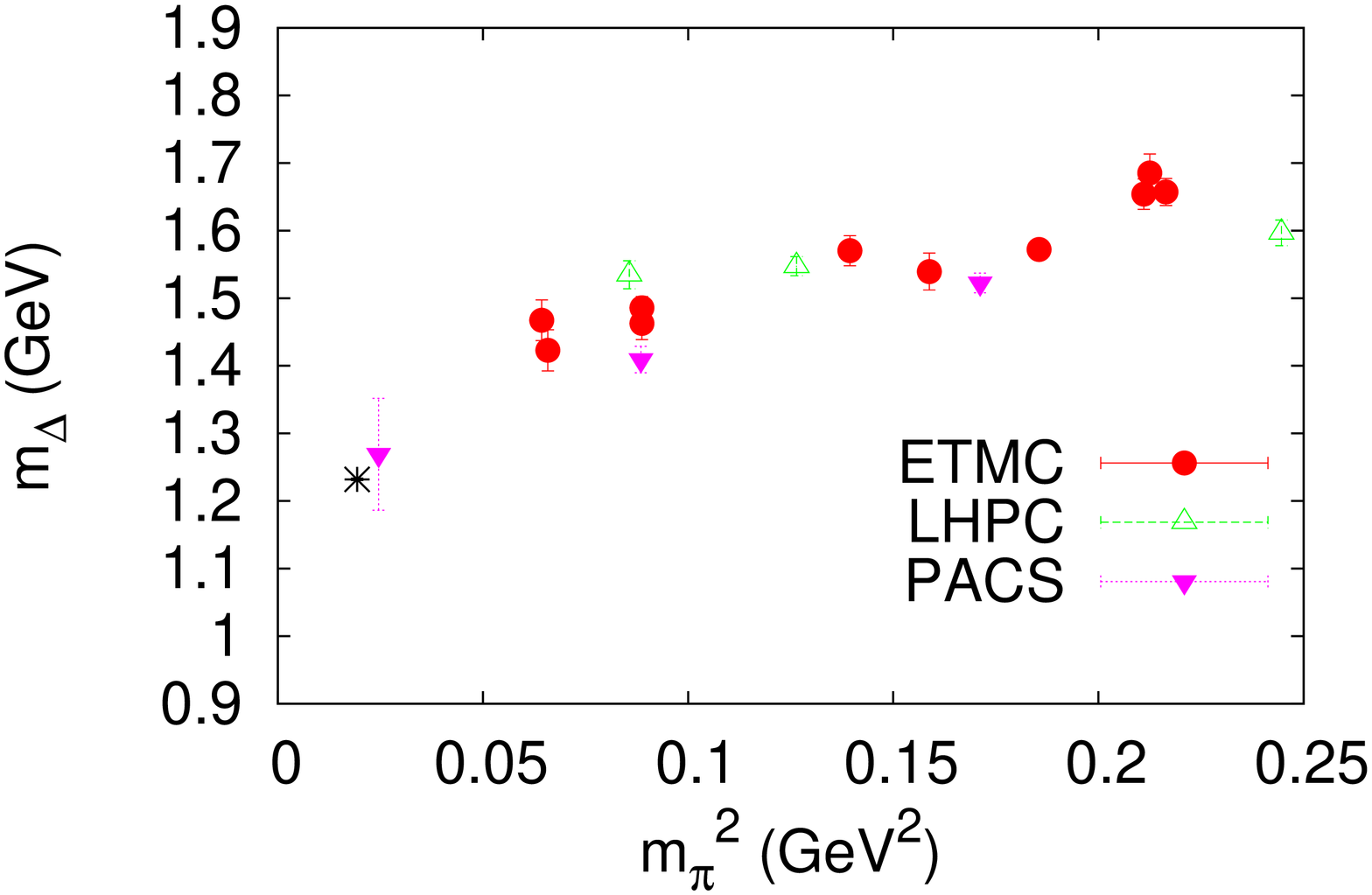} &
\includegraphics[width=.33\textwidth,angle=270]{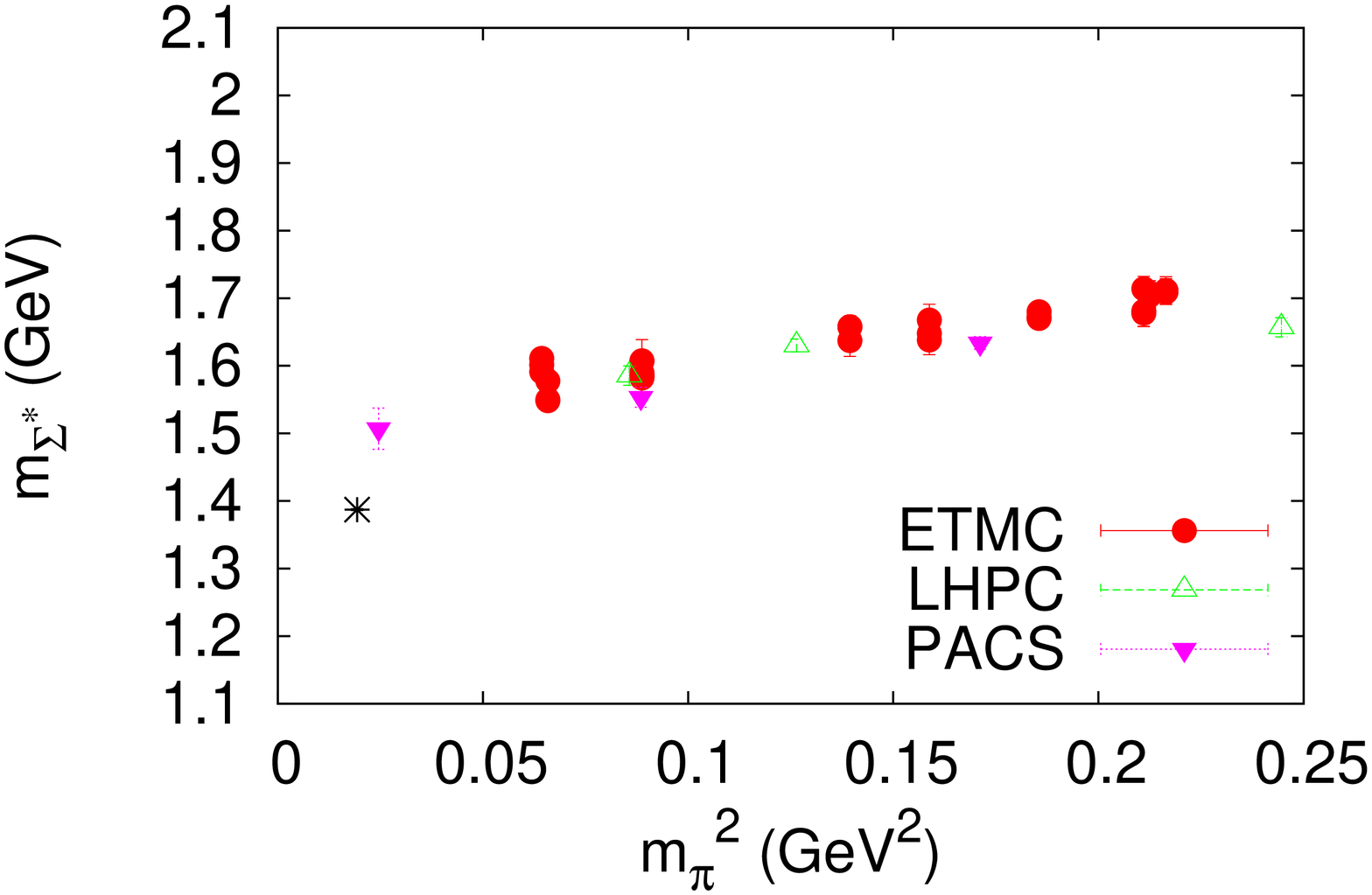} \\
\includegraphics[width=.33\textwidth,angle=270]{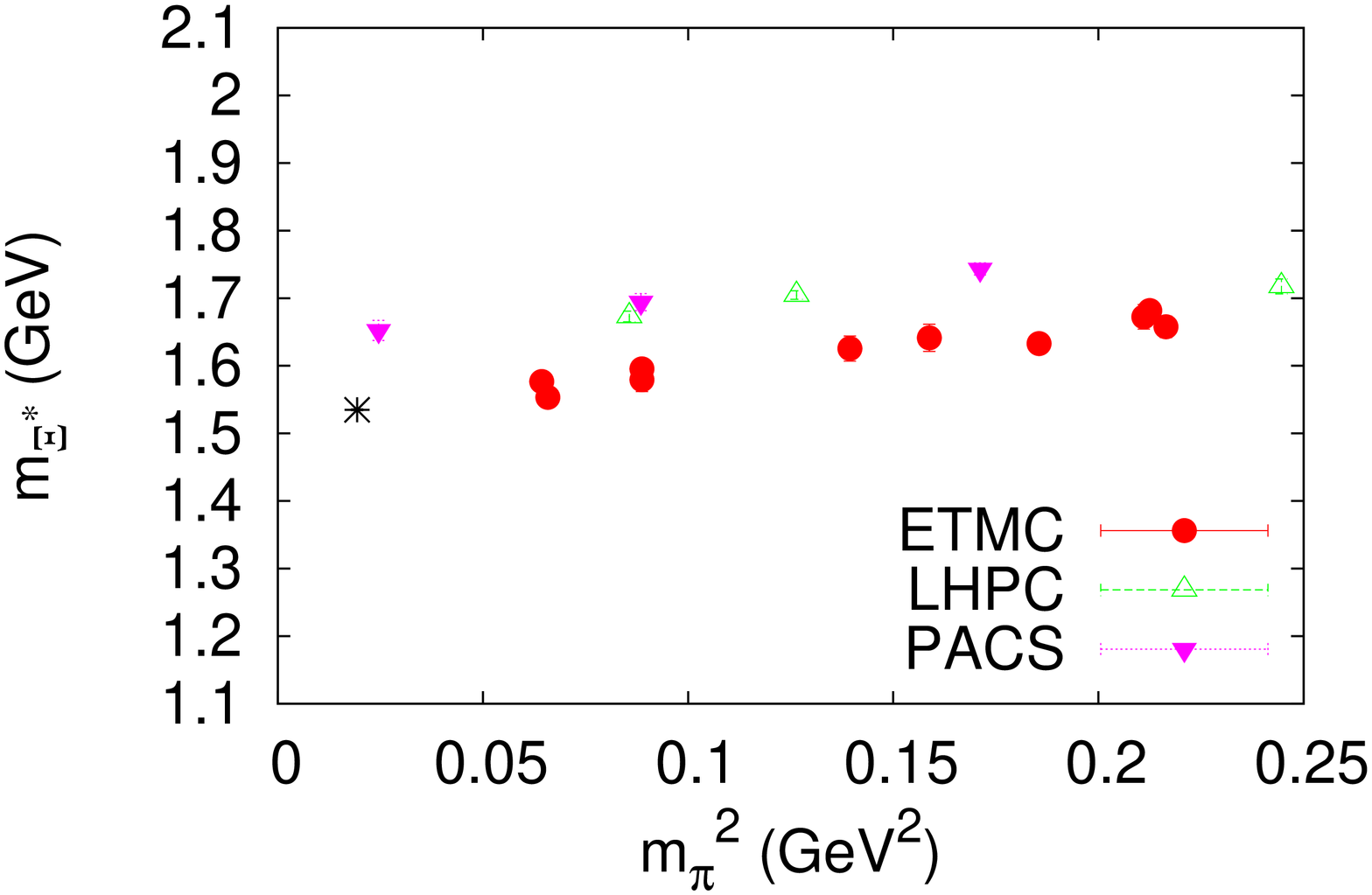} &
\includegraphics[width=.33\textwidth,angle=270]{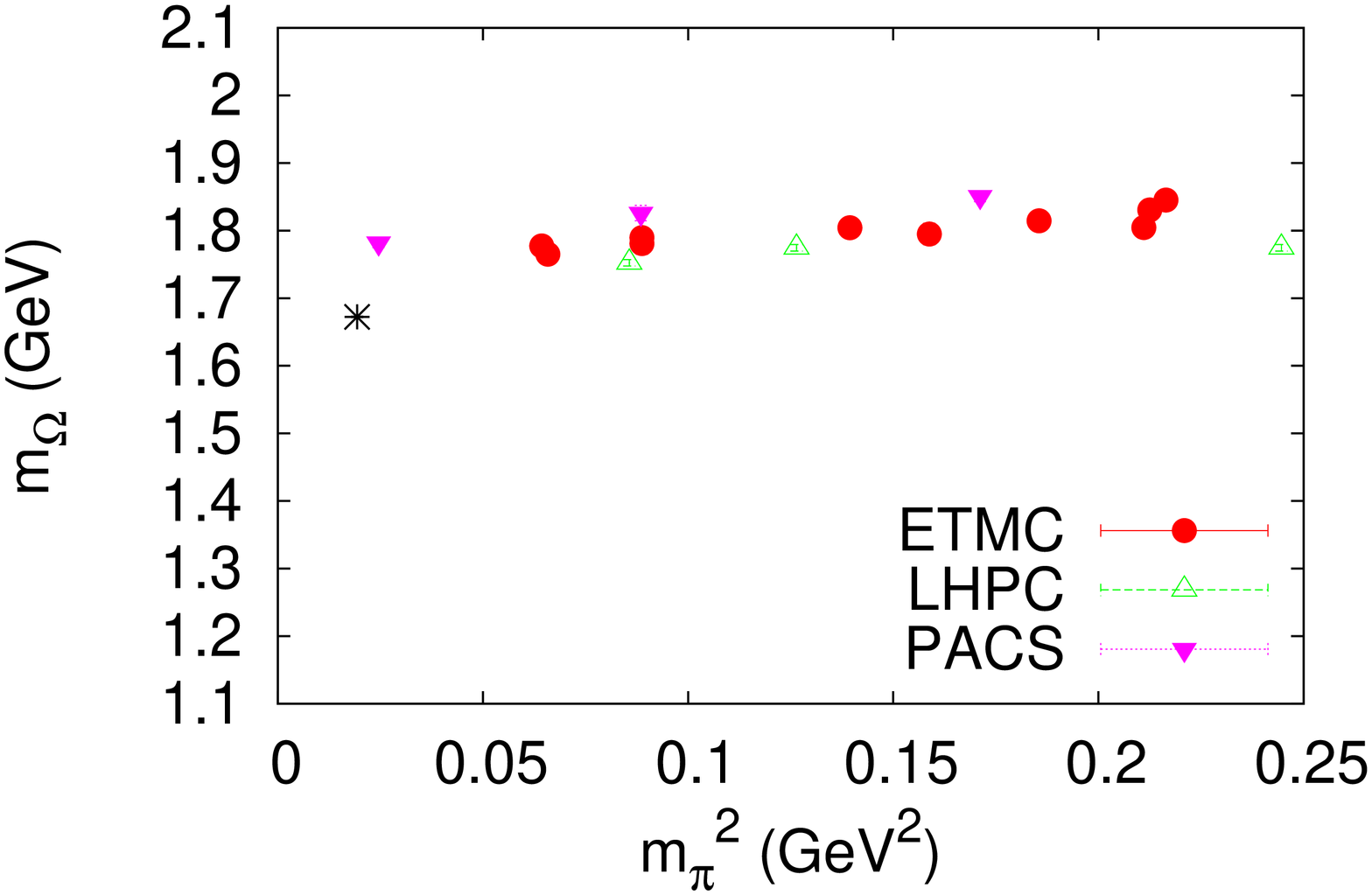} \\
\end{tabular}
\caption{Comparison of the results of this work for the strange decuplet baryons to those obtained
using domain wall valence
quarks on a staggered sea~\cite{WalkerLoud:2008bp}.The  notation is the same
as that in Fig.~\ref{fig:comparison}.}
\label{fig:comparison2}
\end{figure}

\begin{figure}[h!]
\includegraphics[width=.4\textwidth, angle=-90]{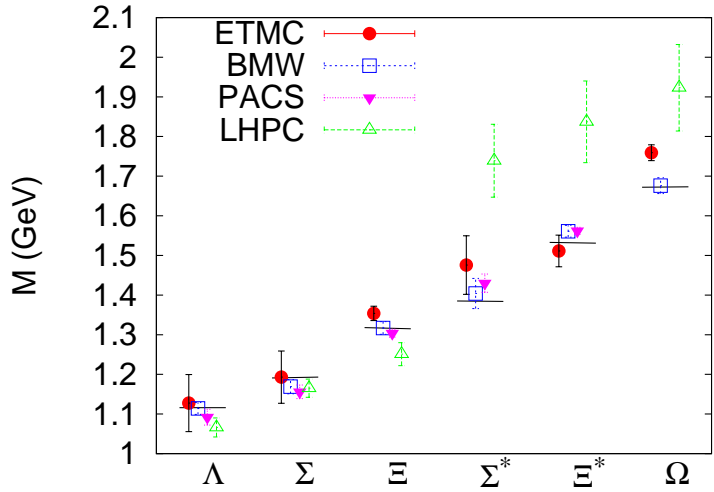}
\caption{Lattice data in the strange quark sector extrapolated to the physical pion mass and the experimental values 
shown by the horizontal lines. For the twisted mass results of this work the chiral extrapolation was carried out  
using NLO HB$\chi$PT.  We include in addition the results by the  QCDSF-UKQCD collaboration~\cite{Bietenholz:2011qq} (light blue crosses). The rest of the  notation is the same as in Fig.~\ref{fig:comparison}.}
\label{fig:spectrum-s}
\end{figure}

\begin{figure}[h!]
%\centering
\begin{tabular}{cc}
\includegraphics[width=.33\textwidth,angle=270]{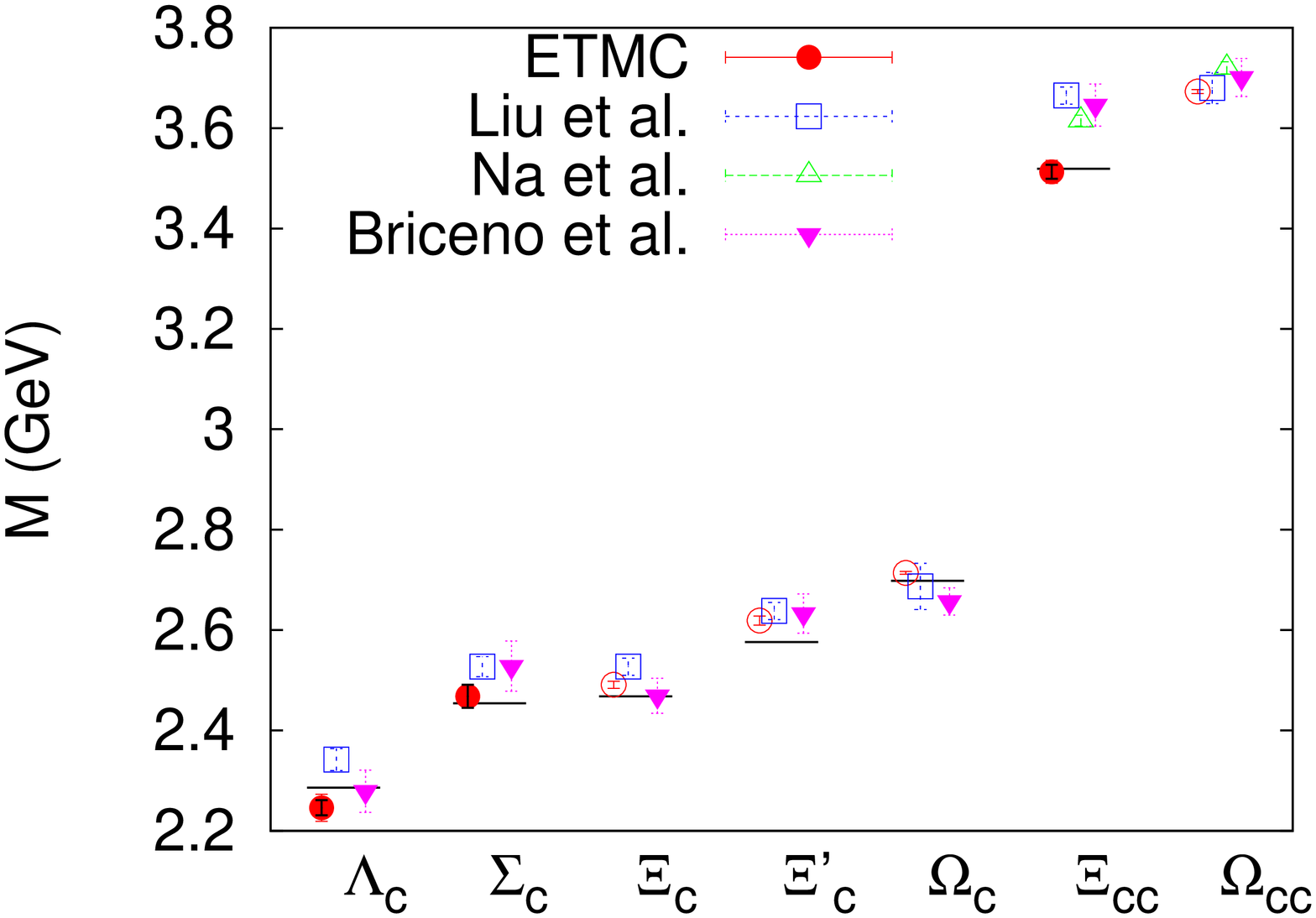} &
\includegraphics[width=.33\textwidth,angle=270]{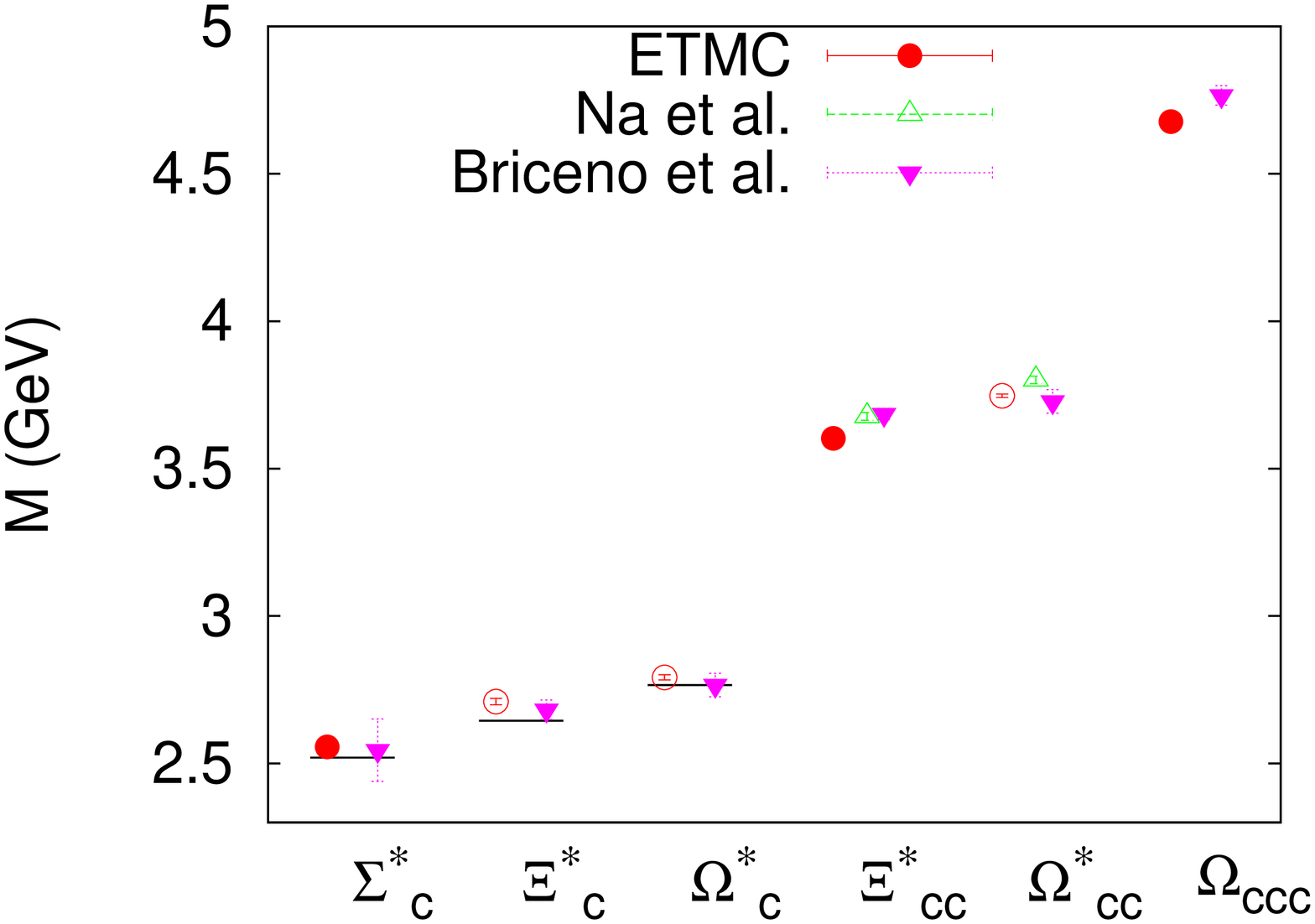} \\
\end{tabular}
\caption{Masses for charm baryons with spin 1/2 (left) and spin 3/2 (right)  computed within lattice QCD and the experimental values 
shown by the horizontal lines.  Our results for $\Lambda_c$, $\Sigma_c$, $\Xi_{cc}$, $\Sigma^*_c$, $\Xi^*_{cc}$ and $\Omega_{ccc}$  are extrapolated to the 
physical pion mass (shown with the red filled circles)
whereas for the rest we give the results obtained at $m_\pi=260$~MeV and $\beta=4.2$ (shown with open red circles). We include results obtained using a number of hybrid actions
with  staggered sea quarks from Refs.~\cite{Liu:2009jc} (open blue squares), 
\cite{Na:2007pv,Na:2008hz} (open green triangles) and \cite{Briceno:2011cb} (filled magenta triangles).}
\label{fig:spectrum-c}
\end{figure}

Several collaborations have
calculated the strange baryon spectrum. 
The Budapest-Marseille-Wuppertal (BMW) collaboration carried out simulations
using tree level
 improved 6-step stout smeared $N_f = 2 + 1$ clover fermions and a tree level
 Symanzik improved gauge action. Volume effects were studied using lattices of spatial extent of 2~fm to 4.1~fm.  
The continuum limit was taken using results produced on
three lattice spacings of $a=0.065$~fm,  0.085~fm and $a= 0.125$~fm.
Using pion masses  down to 190 MeV a polynomial was performed to extrapolate to the physical value of the pion mass~\cite{Durr:2008zz}.
The PACS-CS collaboration obtained results using $N_f = 2 + 1$
 non-perturbatively ${\cal O}(a)$ improved clover fermions 
on an Iwasaki gauge action on a lattice of spatial length of 2.9~fm and lattice
spacing $a = 0.09$~fm~\cite{Aoki:2008sm}.
The QCDSF-UKQCD collaboration~\cite{Bietenholz:2011qq} used $N_f=2+1$ Clover fermions with a single mild stout smearing and a lattice spacing $a=0.076 (2)$~fm.
Finally, the LPHC collaboration~\cite{WalkerLoud:2008bp} obtained results using a hybrid action
of domain wall valence quarks on a staggered sea on lattice
of spatial length 2.5~fm and 3.5~fm at lattice spacing $a=0.124$~fm.

 In Fig.~\ref{fig:comparison} we compare our
results on the strange octet baryons
 with those from the BMW, the PACS-CS, the QCDSF-UKQCD and the LHPC collaborations. 
Our results and the results by the PACS-CS and LHPC are not continuum
extrapolated. 
The BMW results 
are extrapolated to the continuum limit and have larger errors than the rest.
Nevertheless, there is 
an overall agreement, indicating that cut-off effects are small.
In Fig.~\ref{fig:comparison2} we compare our results on the strange decuplet baryons with the ones by PACS-CS and LHPC.
Lattice results are in agreement except for the case of the $\Xi^*$ where 
our results are consistently lower. Given the agreement of our results in the
case of the $\Omega$ this deviation cannot come from the mismatch in the strange 
quark mass. It is not clear what is the origin of this deviation for the $\Xi^*$, but the fact that the value obtained by PACS-CS at almost physical pion mass
is high than the experimental value may indicate that the strange quark
mass is larger than physical. In Fig.~\ref{fig:spectrum-s} we show
the masses for the strange baryons after extrapolating to the physical pion mass. Fort the results of this work we plot the values extracted using NLO HB$\chi$PT. Error shown on the twisted mass results is the estimate of the systematic
error due to the chiral extrapolation, whereas the statistical errors are equal to the size of the symbols and are not shown. As can be seen, our results are
in agreement with experiment except for $\Omega$, which is higher by 2\%, just like the value found by PACS-CS.

We also compare in Fig.~\ref{fig:spectrum-c} our results for the charm baryons
to those obtained using dynamical gauge configurations.
All previous  lattice computations of the mass of charm baryons 
used gauge configurations produced with staggered sea quarks with a number
of different actions for the valence quarks.
In Refs.~\cite{Na:2007pv,Na:2008hz} a Clover charm valence quark
was used  on 
MILC $N_f=2+1$ gauge configurations at three values of the lattice spacing, $a=0.09, 0.12, 0.15$~fm. In Ref.~\cite{Briceno:2011cb} $ N_f =2+1+1$ gauge configurations produced using the highly improved staggered quark (HISQ) action. The valence light-quark (up, down and strange) propagators are generated using the clover impoved Wilson action.
In order to reduce discretization artifacts a relativistic heavy-quark action 
was adopted for  the charm quark.
Finally in Ref.~\cite{Liu:2009jc} domain wall fermions are used for the up, the down and the strange quarks on $N_f=2+1$ improved Kogut-Susskind sea quarks
at one value of the lattice
spacing  $a=0.12$~fm. The
relativistic Fermilab action was employed for the charm quark. 
We show the comparison of our results to those obtained in the aforementioned
references in Fig.~\ref{fig:spectrum-c}.  Our results for $\Lambda_c$, $\Sigma_c$, $\Xi_{cc}$,
$\Sigma^*_c$, $\Xi^*_{cc}$ and $\Omega_{ccc}$  are extrapolated to the physical pion mass using a polynomial fit with up $m_\pi^3$-terms. For the spin 1/2 $\Xi_c$,
$\Xi_c^\prime$, $\Omega_c$ and $\Omega_{cc}$ we show the results
obtained at $m_\pi=260$~MeV, the smallest value of the pion mass considered in this work  on the lattice with the smallest lattice spacing at $\beta=4.2$, 
for which cut-off effects are smallest. 
   As can be seen, our results are in agreement  with the results of the other
studies except for the $\Xi_{cc}$ and with the experimental values.
Although for $\Xi_{cc}$ we find a value consistent with the result of the SELEX experiment, one has to study the pion mass dependence in order to reach
a final conclusion. In Fig.~\ref{fig:spectrum-c} we compare results for
the spin 3/2 charm baryons. Our results for $\Xi^*_c$, $\Omega^*_c$ and $\Omega^*_{cc}$ are obtained at $m_\pi=260$~MeV  at $\beta=4.2$.
There is good agreement among lattice results and with the known experimental
values for $\Sigma^*_c$, $\Xi^*_c$ and $\Omega^*_c$. Thus the
lattice results can be taken as a prediction for the masses of the other 
charm spin 3/2 baryons shown in the figure.

\section{Summary and Conclusions}
In this work we have computed the strange and charm baryon masses using 
$N_f=2$ twisted mass fermions.
 For the strange and charm sector we use an Osterwalder-Seiler 
valence  quarks.
 The bare strange and charm valence quark mass is tuned by requiring
that 
 the physical values of the mass of the  kaon and D-meson  are reproduced
after the lattice results are extrapolated at the physical value of the pion mass.

We analyze gauge configurations for three values of the lattice spacings
at the largest and smallest pion mass used in this study.
We find that cut-off effects are small even in case of the charm baryons.
This is a somewhat surprising result given that the Compton wave length
of the D-meson mass is same order as the lattice spacing. 

 Using simulations on two different volumes we obtained results
that are consistent showing that any volume effects are smaller
than our statistical accuracy.

Another artifact of our lattice formulation is 
isospin breaking at finite lattice spacing.
 We have found that isospin breaking decreases 
with the lattice spacing and it is 
consistent with zero for $a=0.056$~fm confirming the expected restoration of 
isospin symmetry.

Our results on the strange quark sector
are consistent with recent results using Clover improved fermions and domain
wall fermions on a staggered sea. There is an overall agreement
also in thee case of charm sector where we compare our results
to other studies that used staggered sea quarks.
The overall consistence among lattice results, despite the 
different discretizations used, provides a strong validation of 
lattice QCD computations.   
Our results on the charm baryons reproduce the experimentally known values 
and thus provide an estimate for the mass of the
$\Omega_{cc}$, $\Xi^*_{cc}$, $\Omega^*_{cc}$ and $\Omega_{ccc}$.

\section*{Acknowledgments}
We would like to thank all members of ETMC for a
very constructive and enjoyable collaboration.

Numerical calculations have been performed at GENCI/IDRIS
(Project i2011052271) and CC-IN2P3  computer centers. Computer time for this project was also 
made available to us by the
John von Neumann-Institute for Computing on the  Jugene BG/P system at the research center
in J\"ulich and by the Computation-based Science and Technology Research Center
of the Cyprus Institute
 through the infrastructure project Cy-Tera  co-funded by the European Regional Development Fund and the Republic of Cyprus through the Research Promotion Foundation (Project Cy-Tera NEA Y$\Pi$O$\Delta$OMH/$\Sigma$TPATH/0308/31).
%This work has been supported in part by  the DFG
%Sonder\-for\-schungs\-be\-reich/ Trans\-region SFB/TR9.
This work was partly supported by funding received  from the
 Cyprus Research Promotion Foundation under contracts PENEK/0609/17 and
KY-$\Gamma$A/0310/02. 
M.P.  acknowledges
 financial support by the Marie Curie European Reintegration
Grant of the 7th European Community Framework Programme
under contract number PERG05-GA-2009-249309.
M.G. was supported by the Marie-Curie European training network ITN STRONGnet 
grant PITN-GA-2009-238353.

\bibliography{paper_ref}

\newpage
\section{Appendix}

In Tables~\ref{tab:masses-strange-oct_3.9}, \ref{tab:masses-strange-oct_4.05} and 
\ref{tab:masses-strange-oct_4.2} the strange octet baryon masses are collected for 
$\beta=3.9$, $\beta=4.05$ and $\beta=4.2$,
respectively. In Tables~\ref{tab:masses-strange-dec_3.9}, \ref{tab:masses-strange-dec_4.05} and 
\ref{tab:masses-strange-dec_4.2} the strange decuplet baryon masses for the three beta are collected.
In Tables~\ref{tab:masses-charm-oct_3.9}, \ref{tab:masses-charm-oct_4.05} and 
\ref{tab:masses-charm-oct_4.2} we show the charm octet baryon masses for the three beta, while in
Tables~\ref{tab:masses-charm-dec_3.9}, \ref{tab:masses-charm-dec_4.05} and    
\ref{tab:masses-charm-dec_4.2} we show the charm decuplet baryon masses.
In Table~\ref{tab:masses-strange-charm_4.2} we show the masses for extra spin-1/2 and spin-3/2 
charm baryons at $\beta=4.2$ at the lightest pion mass we have.
%The errors are evaluated using jackknife and the Gamma method~\cite{Wolff:2004} to check consistency.

\begin{table}[!h]
\begin{tabular}{|lllllllll|}
\hline $L/a$ & $a \mu_h$ & $a \mu_l$ & $a m_{\Sigma^0}$ & $a m_{\Sigma^-}$ & $a m_{\Sigma^+}$ & $a m_{\Xi^0}$ & $a m_{\Xi^-}$ & $a m_\Lambda$ \\ \hline
32 & 0.0217 & 0.0030 & 0.597(3) & 0.577(4) & 0.607(7) & 0.657(4) & 0.628(3) & 0.561(3) \\
24 & 0.0217 & 0.0040 & 0.610(5) & 0.600(3) & 0.622(4) & 0.671(3) & 0.638(2) & 0.577(4) \\
32 & 0.0217 & 0.0040 & 0.616(4) & 0.596(5) & 0.628(7) & 0.675(4) & 0.635(3) & 0.574(4) \\
24 & 0.0217 & 0.0064 & 0.628(5) & 0.610(6) & 0.640(9) & 0.687(5) & 0.650(4) & 0.602(4) \\
24 & 0.0217 & 0.0085 & 0.649(3) & 0.631(2) & 0.668(4) & 0.697(3) & 0.659(2) & 0.619(3) \\
24 & 0.0217 & 0.0100 & 0.654(4) & 0.643(5) & 0.666(6) & 0.696(5) & 0.668(3) & 0.633(4) \\
24 & 0.015 & 0.0064 & 0.596(9) & 0.585(6) & 0.633(9) & 0.658(11) & 0.597(12) & 0.588(4) \\
24 & 0.015 & 0.0085 & 0.641(5) & 0.598(11) & 0.663(6) & 0.674(5) & 0.621(6) & 0.594(9) \\
24 & 0.015 & 0.0100 & 0.635(7) & 0.627(8) & 0.645(15) & 0.659(14) & 0.622(7) & 0.614(6) \\
32 & 0.025 & 0.0040 & 0.623(5) & 0.606(5) & 0.633(9) & 0.687(6) & 0.651(4) & 0.587(8) \\
32 & 0.030 & 0.0040 & 0.645(4) & 0.630(5) & 0.664(6) & 0.729(6) & 0.688(4) & 0.606(5) \\
24 & 0.030 & 0.0064 & 0.654(5) & 0.640(6) & 0.674(7) & 0.732(5) & 0.682(7) & 0.624(4) \\
24 & 0.030 & 0.0085 & 0.688(5) & 0.651(9) & 0.703(6) & 0.739(6) & 0.706(4) & 0.626(9) \\ \hline
\end{tabular}
\caption{Strange octet baryon masses at $\beta=3.9$.}
\label{tab:masses-strange-oct_3.9}
\end{table}

\begin{table}[!h]
\begin{tabular}{|lllllllll|}
\hline
$L/a$ & $a \mu_h$ & $a \mu_l$ & $a m_{\Sigma^0}$ & $a m_{\Sigma^-}$ & $a m_{\Sigma^+}$ & $a m_{\Xi^0}$ & $a m_{\Xi^-}$ & $a m_\Lambda$ \\ \hline
32 & 0.0178 & 0.0030&  0.477(5)&  0.470(8) & 0.488(9) & 0.519(6) & 0.504(4) & 0.455(5)\\
32 & 0.0178 & 0.0060&  0.500(5)&  0.483(5) & 0.503(8) & 0.529(5) & 0.519(4) & 0.482(5) \\
32 & 0.0178 & 0.0080&  0.512(4)&  0.506(5) & 0.522(4) & 0.541(4) & 0.527(3) & 0.496(5) \\
32 & 0.014  & 0.0030&  0.463(5)&  0.456(9) & 0.474(9) & 0.496(6) & 0.482(4) & 0.445(6) \\
32 & 0.014  & 0.0060&  0.495(4)&  0.473(6) & 0.497(8) & 0.513(6) & 0.500(4) & 0.479(4) \\
32 & 0.014  & 0.0080&  0.494(8)&  0.508(9) & 0.508(5) & 0.512(7) & 0.507(3) & 0.485(6) \\
32 & 0.0166 & 0.0030&  0.477(5)&  0.465(5) & 0.487(6) & 0.515(4) & 0.498(3) & 0.451(4) \\
32 & 0.0166 & 0.0060&  0.496(5)&  0.483(5) & 0.502(5) & 0.526(4) & 0.511(4) & 0.480(5) \\
32 & 0.0166 & 0.0080&  0.510(3)&  0.502(3) & 0.518(4) & 0.535(3) & 0.521(3) & 0.496(3) \\
32 & 0.019  & 0.0060&  0.501(6)&  0.495(8) & 0.519(13)& 0.532(6) & 0.524(6) & 0.483(5) \\
32 & 0.020  & 0.0030&  0.486(6)&  0.478(8) & 0.495(8) & 0.531(6) & 0.517(4) & 0.461(5) \\
32 & 0.020  & 0.0060&  0.510(6)&  0.493(5) & 0.515(7) & 0.544(6) & 0.534(4) & 0.492(5)\\
32 & 0.020  & 0.0080&  0.516(4)&  0.510(7) & 0.514(7) & 0.547(4) & 0.533(4) & 0.500(5)\\
32 & 0.025 & 0.0060&  0.522(5)&  0.501(6) & 0.531(7) & 0.572(5) & 0.556(4) & 0.501(5)\\ \hline
\end{tabular}
\caption{Strange octet baryon  masses at $\beta=4.05$.}
\label{tab:masses-strange-oct_4.05}
\end{table}

\begin{table}[!h]
\begin{tabular}{|lllllllll|} %3
\hline
$L/a$ & $a \mu_h$ & $a \mu_l$ & $a m_{\Sigma^0}$ & $a m_{\Sigma^-}$ & $a m_{\Sigma^+}$ & $a m_{\Xi^0}$ & $a m_{\Xi^-}$ & $a m_\Lambda$ \\ \hline
32 & 0.012 & 0.0065&  0.402(4) & 0.396(4) & 0.405(5) & 0.418(4) & 0.406(4) & 0.392(4) \\
48 & 0.012 & 0.0020&  0.362(3) & 0.360(3) & 0.368(4) & 0.390(2) & 0.381(2) & 0.344(3) \\
32 & 0.013 & 0.0065&  0.406(7) & 0.405(7) & 0.417(7) & 0.430(6) & 0.416(6) & 0.395(7) \\
32 & 0.015 & 0.0065&  0.411(4) & 0.409(4) & 0.413(4) & 0.431(4) & 0.421(4) & 0.397(4) \\
48 & 0.015 & 0.0020&  0.374(2) & 0.371(3) & 0.380(3) & 0.404(2) & 0.397(2) & 0.352(2) \\
32 & 0.016 & 0.0065&  0.417(7) & 0.409(4) & 0.413(4) & 0.442(5) & 0.421(4) & 0.397(4) \\ \hline
\end{tabular}
\caption{Strange octet baryon masses at $\beta=4.2$.}
\label{tab:masses-strange-oct_4.2}
\end{table}

\begin{table}[!h]
\begin{tabular}{|lllllllll|}
\hline
 $L/a$ & $a \mu_h$ & $a \mu_l$ & $a m_{{\Sigma^*}^0}$ & $a m_{{\Sigma^*}^-}$ & $a m_{{\Sigma^*}^+}$ & $a m_{{\Xi^*}^0}$ & $a m_{{\Xi^*}^-}$ & $a m_\Omega$ \\ \hline
32 & 0.0217 & 0.0030&  0.700(7) & 0.713(5) & 0.700(6) & 0.710(6) & 0.702(5) & 0.798(3) \\
32 & 0.0217 & 0.0040&  0.726(4) & 0.718(6) & 0.715(6) & 0.739(4) & 0.721(4) & 0.809(3) \\
24 & 0.0217 & 0.0040&  0.716(12)& 0.726(14)& 0.717(17)& 0.737(8) & 0.714(8) & 0.805(7) \\
24 & 0.0217 & 0.0064&  0.749(7) & 0.749(7) & 0.740(11)& 0.754(8) & 0.734(8) & 0.815(6) \\
24 & 0.0217 & 0.0085&  0.759(5) & 0.755(4) & 0.755(5) & 0.767(3) & 0.738(5) & 0.820(3) \\
24 & 0.0217 & 0.0100&  0.773(8) & 0.772(8) & 0.773(9) & 0.784(5) & 0.749(7) & 0.834(4) \\
24 & 0.015 & 0.0064 & 0.737(24) & 0.736(7) & 0.746(7) & 0.728(15)& 0.708(10)& 0.772(16)\\
24 & 0.015 & 0.0085 & 0.751(10) & 0.736(20)& 0.755(7) & 0.723(11)& 0.659(18)& 0.771(9) \\
24 & 0.015 & 0.0100 & 0.730(18) & 0.757(19)& 0.746(16)& 0.720(15)& 0.667(10)& 0.771(11) \\
32 & 0.025 & 0.0040 & 0.730(9)  & 0.723(12)& 0.710(14)& 0.747(12)& 0.738(8) & 0.827(7) \\
32 & 0.030 & 0.0040 & 0.724(11) & 0.743(10)& 0.739(9) & 0.782(7) & 0.765(8) & 0.870(6)\\
24 & 0.030 & 0.0064 & 0.772(6)  & 0.769(6) & 0.760(11)& 0.797(7) & 0.781(6) & 0.874(6)\\
24 & 0.030 & 0.0085 & 0.782(11) &  0.736(7)& 0.746(7) & 0.812(7) & 0.751(11)& 0.771(9) \\ \hline
\end{tabular}
\caption{Strange decuplet baryon  masses at $\beta=3.9$.}
\label{tab:masses-strange-dec_3.9}
\end{table}

\begin{table}[!h]
\begin{tabular}{|lllllllll|} %5
\hline $L/a$ & $a \mu_h$ & $a \mu_l$ & $a m_{{\Sigma^*}^0}$ & $a m_{{\Sigma^*}^-}$ & $a m_{{\Sigma^*}^+}$ & $a m_{{\Xi^*}^0}$ & $a m_{{\Xi^*}^-}$ & $a m_\Omega$ \\ \hline
32 & 0.0178&  0.0030 & 0.597(11) & 0.589(10)& 0.590(13)& 0.596(8) & 0.588(6) & 0.661(7) \\
32 & 0.0178&  0.0060 & 0.593(8)  & 0.586(8) & 0.582(8) & 0.582(7) & 0.583(7) & 0.638(6) \\
32 & 0.0178&  0.0080 & 0.606(6)  & 0.607(7) & 0.606(6) & 0.609(6) & 0.598(5) & 0.651(6) \\
32 & 0.014 & 0.0030  & 0.589(12) & 0.595(8) & 0.603(6) & 0.578(8) & 0.570(6) & 0.627(8) \\
32 & 0.014 & 0.0060  & 0.586(11) & 0.574(13)& 0.582(12)& 0.568(10)& 0.559(11)& 0.623(7) \\
32 & 0.014 & 0.0080  & 0.588(10) & 0.606(7) & 0.592(6) & 0.574(9) & 0.574(6) & 0.614(8) \\
32 & 0.0166&  0.0030 & 0.570(12) & 0.578(7) & 0.561(8) & 0.573(6) & 0.567(6) & 0.630(5) \\
32 & 0.0166&  0.0060 & 0.582(12) & 0.578(10)& 0.568(14)& 0.579(6) & 0.577(7) & 0.622(8) \\
32 & 0.0166& 0.0080  & 0.615(4)  & 0.609(6) & 0.604(7) & 0.601(5) & 0.586(5) & 0.648(4) \\
32 & 0.019 & 0.0060  & 0.592(16) & 0.576(13)& 0.551(19)& 0.579(8) & 0.583(9) & 0.648(5) \\
32 & 0.020 & 0.0030  & 0.611(7)  & 0.605(8) & 0.610(7) & 0.606(7) & 0.596(8) & 0.671(5) \\
32 & 0.020 & 0.0060  & 0.599(10) & 0.591(9) & 0.595(11)& 0.594(11)& 0.591(9) & 0.657(6) \\
32 & 0.020 & 0.0080  & 0.610(6)  & 0.619(6) & 0.605(6) & 0.613(8) & 0.597(6) & 0.660(5) \\
32 & 0.025 & 0.0060  & 0.598(11) & 0.589(12)& 0.577(15)& 0.613(7) & 0.613(8) & 0.674(8) \\ \hline
\end{tabular}
\caption{Strange decuplet baryon  masses at $\beta=4.05$.}
\label{tab:masses-strange-dec_4.05}
\end{table}

\begin{table}[!h]
\begin{tabular}{|lllllllll|} %6
\hline $L/a$ & $a \mu_h$ & $a \mu_l$ & $a m_{{\Sigma^*}^0}$ & $a m_{{\Sigma^*}^-}$ & $a m_{{\Sigma^*}^+}$ & $a m_{{\Xi^*}^0}$ & $a m_{{\Xi^*}^-}$ & $a m_\Omega$ \\ \hline
32 & 0.012&  0.0065&  0.480(7) & 0.482(6) & 0.476(8) & 0.468(6) & 0.464(6) & 0.498(6) \\
48 & 0.012&  0.0020&  0.453(4) & 0.451(4) & 0.451(4) & 0.444(3) & 0.440(3) & 0.487(3) \\
32 &0.013&  0.0065&  0.501(11)& 0.500(10)& 0.495(14)& 0.494(9) & 0.489(9) & 0.522(10)\\
32 & 0.015&  0.0065&  0.478(6) & 0.487(5) & 0.477(6) & 0.478(5) & 0.475(5) & 0.513(4) \\
48 & 0.015&  0.0020&  0.458(4) & 0.455(3) & 0.452(2) & 0.440(4) & 0.448(4) & 0.505(3) \\
32 & 0.016&  0.0065&  0.507(11)& 0.487(5) & 0.503(12)& 0.504(10)& 0.505(9) & 0.540(10)\\ \hline
\end{tabular}
\caption{Strange  decuplet baryon masses at $\beta=4.2$.}
\label{tab:masses-strange-dec_4.2}
\end{table}

\begin{table}[!h]
\begin{tabular}{|lllllllll|}
\hline
$L/a$ &  $a \mu_h$ & $a \mu_l$ & $a m_{\Sigma_c^+}$ & $a m_{\Sigma_c^0}$ & $a m_{\Sigma_c^{++}}$ & $a m_{\Xi_{cc}^{++}}$ & $a m_{\Xi_{cc}^+}$ & $a m_{\Lambda_c^+}$ \\ \hline
24 & 0.240 & 0.0040 &  1.100(6) & 1.105(13) & 1.102(14) &  1.532(5) & 1.528(4) & 1.015(8)\\
24 &  0.240 & 0.0064 &  1.117(6) & 1.059(20) & 1.122(6) &  1.552(3) & 1.533(4) & 1.045(5)\\
24 &   0.240 & 0.0085 &  1.131(6) & 1.125(5) & 1.135(5) &  1.555(4) & 1.541(5) & 1.063(3)\\
24 &  0.240 & 0.0100 &  1.139(4) & 1.131(5) & 1.138(5) &  1.559(3) & 1.551(2) & 1.070(3)\\
24 &  0.250 & 0.0040 &  1.128(7) & 1.120(9) & 1.099(14) &  1.591(5) & 1.575(5) & 1.055(6)\\
24 &  0.270 & 0.0040 &  1.154(5) & 1.157(5) & 1.161(6) &  1.640(4) & 1.628(3) & 1.069(4)\\
24 &  0.270 & 0.0064 &  1.172(7) & 1.164(5) & 1.176(7) &  1.648(5) & 1.634(4) & 1.096(5)\\
24 &  0.270 & 0.0085 &  1.192(4) & 1.181(4) & 1.188(5) &  1.658(3) & 1.651(3) & 1.115(3)\\
24 &  0.270 & 0.0100 &  1.196(4) & 1.189(4) & 1.197(4) &  1.660(3) & 1.649(25) & 1.122(3)\\
24 &  0.300 & 0.0040 &  1.203(9) & 1.214(5) & 1.221(5) &  1.736(6) & 1.731(4) & 1.135(4)\\
24 &  0.300 & 0.0064 &  1.227(4) & 1.162(22) & 1.236(5) &  1.747(4) & 1.722(6) & 1.148(5) \\
24 & 0.300 & 0.0085 &  1.244(4) & 1.236(4) & 1.244(4) &  1.759(3) & 1.736(6) & 1.169(3)\\
24 &  0.300 & 0.0100 &  1.251(3) & 1.237(5) & 1.249(4) &  1.762(2) & 1.751(3) & 1.172(3)\\ \hline
\end{tabular}
\caption{Charm spin-1/2 baryon  masses at $\beta=3.9$.}
\label{tab:masses-charm-oct_3.9}
\end{table}

\begin{table}[!h]
\begin{tabular}{|lllllllll|} %8
\hline
$L/a$ & $a \mu_h$ & $a \mu_l$ & $a m_{\Sigma_c^+}$ & $a m_{\Sigma_c^0}$ & $a m_{\Sigma_c^{++}}$ & $a m_{\Xi_{cc}^{++}}$ & $a m_{\Xi_{cc}^+}$ & $a m_{\Lambda_c^+}$ \\ \hline
32 &   0.170 & 0.0030 &  0.843(4) & 0.841(5) & 0.839(6) &  1.134(5) & 1.137(4) & 0.774(5)\\
32 &   0.170 & 0.0060 &  0.844(4) & 0.831(9) & 0.836(8) &  1.149(3) & 1.139(4) & 0.791(5)\\
32 &   0.170 & 0.0080 &  0.852(4) & 0.852(5) & 0.852(7) &  1.146(2) & 1.142(3) & 0.791(6)\\
32 &   0.200 & 0.0030 &  0.900(4) & 0.888(7) & 0.894(6) &  1.244(3) & 1.244(5) & 0.828(6)\\
32 &   0.200 & 0.0060 &  0.903(4) & 0.886(10) & 0.891(8) &  1.255(2) & 1.250(3) & 0.843(5)\\
32 &   0.200 & 0.0080 &  0.905(5) & 0.907(5) & 0.908(8) &  1.250(2) & 1.247(3) & 0.848(5)\\
32 &   0.210 & 0.0030 &  0.917(4) & 0.912(6) & 0.908(7) &  1.282(3) & 1.276(4) & 0.845(6)\\
32 &   0.210 & 0.0060 &  0.919(4) & 0.911(4) & 0.921(4) &  1.286(3) & 1.279(3) & 0.859(5)\\
32 &   0.210 & 0.0080 &  0.926(3) & 0.925(4) & 0.926(4) &  1.288(2) & 1.283(2) & 0.865(4)\\
32 &   0.230 & 0.0030 &  0.952(5) & 0.942(7) & 0.948(6) &  1.345(3) & 1.342(5) & 0.883(5)\\
32 &   0.230 & 0.0060 &  0.952(4) & 0.940(11) & 0.945(8) &  1.355(3) & 1.353(3) & 0.895(5) \\
32 &   0.230 & 0.0080 &  0.959(5) & 0.960(5) & 0.959(5) &  1.349(3) & 1.349(3) & 0.899(5)\\
32 &   0.260 & 0.0030 &  1.004(5) & 1.005(10) & 1.003(5) &  1.444(3) & 1.441(5) & 0.935(5)\\
32 &   0.260 & 0.0060 &  1.001(5) & 0.992(11) & 0.997(8) &  1.457(3) & 1.451(4) & 0.946(5)\\
32 &   0.260 & 0.0080 &  1.010(6) & 1.011(7) & 1.012(8) &  1.449(3) & 1.446(5) & 0.955(3)\\  \hline
\end{tabular}
\caption{Charm spin-1/2 baryon masses at $\beta=4.05$.}
\label{tab:masses-charm-oct_4.05}
\end{table}

\begin{table}[!h]
\begin{tabular}{|lllllllll|} %9
\hline
$L/a$ & $a \mu_h$ & $a \mu_l$ & $a m_{\Sigma_c^+}$ & $a m_{\Sigma_c^0}$ & $a m_{\Sigma_c^{++}}$ & $a m_{\Xi_{cc}^{++}}$ & $a m_{\Xi_{cc}^+}$ & $a m_{\Lambda_c^+}$ \\ \hline
32 & 0.130 & 0.0065 &  0.696(6) & 0.694(6) & 0.698(6) &  0.932(4) & 0.927(5) & 0.653(4)\\
32 & 0.160 & 0.0065 &  0.733(6) & 0.731(6) & 0.734(7) &  0.999(4) & 0.997(5) & 0.688(4)\\
32 & 0.185 & 0.0065 &  0.778(7) & 0.776(6) & 0.779(7) &  1.085(4) & 1.082(5) & 0.731(4)\\
32 & 0.210 & 0.0065 &  0.821(7) & 0.819(7) & 0.822(7) &  1.168(4) & 1.156(5) & 0.774(4)\\
48 & 0.136 & 0.0020 &  0.653(3) & 0.656(3) & 0.652(3) &  0.899(2) & 0.898(2) & 0.603(2)\\
48 & 0.170 & 0.0020 &  0.716(3) & 0.719(3) & 0.715(4) &  1.017(2) & 1.016(2) & 0.663(3) \\
\hline
\end{tabular}
\caption{Charm spin-1/2 baryon  masses at $\beta=4.2$.}
\label{tab:masses-charm-oct_4.2}
\end{table}

\begin{table}[!h]
\begin{tabular}{|lllllllll|} %10
\hline
$L/a$ &  $a \mu_h$ & $a \mu_l$ & $a m_{{\Sigma_c^*}^+}$ & $a m_{{\Sigma_c^*}^0}$ & $a m_{{\Sigma_c^*}^{++}}$ & $a m_{{\Xi_{cc}^*}^{++}}$ & $a m_{{\Xi_{cc}^*}^+}$ & $a m_{\Omega_{ccc}}$ \\ \hline
24 &    0.240 & 0.0040 &  1.148(10) & 1.142(16) & 1.147(15) &  1.572(6) & 1.564(6) & 1.989(3)\\
24 &    0.240 & 0.0064 &  1.159(10) & 1.151(9) & 1.166(11) &  1.580(6) & 1.572(5) & 1.991(4)\\
24 &    0.240 & 0.0085 &  1.175(8) & 1.164(8) & 1.174(10) &  1.594(5) & 1.578(7) & 1.997(4)\\
24 &    0.240 & 0.0100 &  1.184(6) & 1.184(4) & 1.181(6) &  1.599(3) & 1.591(3) & 1.999(3) \\
24 &    0.250 & 0.0040 &  1.173(8) & 1.182(9) & 1.173(11) &  1.602(7) & 1.606(8) & 2.043(4)\\
24 &    0.270 & 0.0040 &  1.201(8) & 1.204(7) & 1.210(5) &  1.671(7) & 1.668(5) & 2.130(3) \\
24 &    0.270 & 0.0064 &  1.209(10) & 1.210(5) & 1.225(6) &  1.680(5) & 1.672(5) & 2.133(3)\\
24 &    0.270 & 0.0085 &  1.231(4) & 1.224(5) & 1.230(4) &  1.692(5) & 1.686(3) & 2.136(4) \\
24 &   0.270 & 0.0100 &  1.239(4) & 1.234(4) & 1.238(5) &  1.694(4) & 1.690(4) & 2.141(3) \\
24 &    0.300 & 0.0040 &  1.263(5) & 1.245(15) & 1.260(6) &  1.775(6) & 1.764(5) & 2.269(3)\\
24 &    0.300 & 0.0064 &  1.267(5) & 1.260(5) & 1.275(6) &  1.779(5) & 1.768(5) & 2.270(3) \\
24 &    0.300 & 0.0085 &  1.281(4) & 1.274(5) & 1.280(4) &  1.794(3) & 1.770(8) & 2.274(4) \\
24 &    0.300 & 0.0100 &  1.292(3) & 1.283(4) & 1.282(6) &  1.795(3) & 1.789(3) & 2.277(2) \\ \hline
\end{tabular}
\caption{Charm spin-3/2 baryon masses at $\beta=3.9$.}
\label{tab:masses-charm-dec_3.9}
\end{table}

\begin{table}[!h]
\begin{tabular}{|lllllllll|} %11
\hline
$L/a$ & $a \mu_h$ & $a \mu_l$ & $a m_{{\Sigma_c^*}^+}$ & $a m_{{\Sigma_c^*}^0}$ & $a m_{{\Sigma_c^*}^{++}}$ & $a m_{{\Xi_{cc}^*}^{++}}$ & $a m_{{\Xi_{cc}^*}^+}$ & $a m_{\Omega_{ccc}}$ \\ \hline
32 &   0.170 & 0.0030 &  0.889(6) & 0.889(6) & 0.890(4) &  1.170(6) & 1.174(5) & 1.468(3) \\
32 &   0.170 & 0.0060 &  0.887(4) & 0.883(5) & 0.868(12) &  1.184(3) & 1.172(6) & 1.476(2)\\
32 &   0.170 & 0.0080 &  0.885(5) & 0.890(7) & 0.884(5) &  1.176(3) & 1.175(4) & 1.468(3) \\
32 &   0.200 & 0.0030 &  0.940(6) & 0.940(6) & 0.941(4) &  1.277(4) & 1.281(4) & 1.616(3) \\
32 &   0.200 & 0.0060 &  0.928(6) & 0.937(4) & 0.918(11) &  1.288(3) & 1.281(3) & 1.623(2)\\
32 &   0.200 & 0.0080 &  0.936(5) & 0.945(6) & 0.935(5) &  1.279(3) & 1.281(3) & 1.612(3) \\
32 &   0.210 & 0.0030 &  0.956(6) & 0.956(6) & 0.958(4) &  1.310(4) & 1.314(4) & 1.663(3) \\
32 &   0.210 & 0.0060 &  0.955(5) & 0.954(4) & 0.948(5) &  1.316(3) & 1.310(4) & 1.667(2) \\
32 &   0.210 & 0.0080 &  0.957(4) & 0.962(4) & 0.958(4) &  1.316(3) & 1.313(3) & 1.663(3) \\
32 &   0.230 & 0.0030 &  0.989(6) & 0.989(6) & 0.992(4) &  1.376(4) & 1.379(4) & 1.761(3) \\
32 &   0.230 & 0.0060 &  0.984(4) & 0.987(4) & 0.967(11) &  1.387(3) & 1.380(4) & 1.768(2)\\
32 &   0.230 & 0.0080 &  0.986(5) & 0.995(6) & 0.986(5) &  1.378(3) & 1.380(3) & 1.758(3) \\
32 &   0.260 & 0.0030 &  1.039(6) & 1.038(6) & 1.036(7) &  1.473(4) & 1.476(4) & 1.905(3) \\
32 &   0.260 & 0.0060 &  1.033(5) & 1.032(5) & 1.016(11) &  1.485(3) & 1.478(4) & 1.910(3)\\
32 &   0.260 & 0.0080 &  1.034(5) & 1.039(6) & 1.032(10) &  1.472(3) & 1.472(5) & 1.898(4) \\ \hline
\end{tabular}
\caption{Charm spin-3/2 baryon masses at $\beta=4.05$.}
\label{tab:masses-charm-dec_4.05}
\end{table}

\begin{table}[!h]
\begin{tabular}{|lllllllll|} %12
\hline
$L/a$ &  $a \mu_h$ & $a \mu_l$ & $a m_{{\Sigma_c^*}^+}$ & $a m_{{\Sigma_c^*}^0}$ & $a m_{{\Sigma_c^*}^{++}}$ & $a m_{{\Xi_{cc}^*}^{++}}$ & $a m_{{\Xi_{cc}^*}^+}$ & $a m_{\Omega_{ccc}}$ \\ \hline
32 & 0.130 & 0.0065 &  0.730(8) & 0.730(8) & 0.727(8) &  0.958(6) & 0.963(6) & 1.191(3) \\
32 & 0.160 & 0.0065 &  0.763(8) & 0.763(8) & 0.761(8) &  1.025(6) & 1.029(6) & 1.287(3) \\
32 & 0.185 & 0.0065 &  0.805(8) & 0.804(8) & 0.802(8) &  1.107(6) & 1.109(6) & 1.408(3) \\
32 & 0.210 & 0.0065 &  0.845(9) & 0.845(8) & 0.843(8) &  1.188(6) & 1.193(6) & 1.526(3) \\
48 & 0.136 & 0.0020 &  0.686(4) & 0.688(3) & 0.683(4) &  0.925(2) & 0.926(2) & 1.166(1) \\
48 & 0.170 & 0.0020 &  0.744(4) & 0.746(3) & 0.741(4) &  1.039(2) & 1.039(2) & 1.333(1) \\ \hline
\end{tabular}
\caption{Charm spin-3/2 baryon masses at $\beta=4.2$.}
\label{tab:masses-charm-dec_4.2}
\end{table}

\begin{table}[!h]
\begin{tabular}{|lllllllllll|} %12
\hline
$L/a$ &  $a \mu_s$ & $a \mu_c$ & $a \mu_l$ & $a m_{{\Xi_c}}$ & $a m_{{\Xi_c'}}$ & $a m_{{\Omega_c}}$ & $a m_{{\Omega_{cc}}}$ & $a m_{{\Xi_{c}^*}}$ & $a m_{\Omega_{c}^*}$ & $a m_{\Omega_{cc}^*}$ \\ \hline
48 & 0.015 & 0.17 & 0.0020 &  0.708(2) & 0.745(3) & 0.771(2) &  1.044(1) & 0.770(3) & 0.794(2) & 1.065(2)\\ \hline
\end{tabular}
\caption{Strange-charm spin-1/2 and 3/2 baryon masses at $\beta=4.2$
at the tuned heavy quark masses.}
\label{tab:masses-strange-charm_4.2}
\end{table}

\end{document}